\documentclass{SciPost}

\hypersetup{
    colorlinks,
    linkcolor={red!50!black},
    citecolor={blue!50!black},
    urlcolor={blue!80!black}
}

\usepackage[bitstream-charter]{mathdesign}
\DeclareSymbolFont{usualmathcal}{OMS}{cmsy}{m}{n}
\DeclareSymbolFontAlphabet{\mathcal}{usualmathcal}

\fancypagestyle{SPstyle}{
\fancyhf{}
\lhead{\colorbox{scipostblue}{\bf \color{white} ~SciPost Physics }}
\rhead{{\bf \color{scipostdeepblue} ~Submission }}

\fancyfoot[C]{\textbf{\thepage}}
}

\usepackage{float}
\usepackage{graphicx}
\usepackage{bm}
\usepackage{physics}
\usepackage{amsmath}
\usepackage{appendix}
\usepackage{color}
\usepackage{placeins}

\newcommand{\up}{\uparrow}
\newcommand{\dw}{\downarrow}
\newcommand{\halfbox}{
  \raisebox{0.5ex}{\rule{0.4pt}{0.7ex}}
  \raisebox{0.3ex}{\rule{1.5ex}{0.4pt}}
  \raisebox{0.5ex}{\rule{0.4pt}{0.7ex}}
}

\begin{document}

\pagestyle{SPstyle}

\begin{center}{\Large \textbf{\color{scipostdeepblue}{
Effects of electron-electron interaction and spin-orbit coupling on Andreev pair qubits in quantum dot Josephson junctions
}}}\end{center}

\begin{center}\textbf{
Teodor Iličin\textsuperscript{1,2,$\star$}, Rok \v Zitko\textsuperscript{1,$\dagger$} 
}\end{center}

\begin{center}
{\bf 1} Jo\v zef Stefan Institute, Jamova 39, SI-1000 Ljubljana, Slovenia\\
{\bf 2} Faculty of mathematics and physics, University of Ljubljana, Jadranska 19, SI-1000 Ljubljana, Slovenia
\\[\baselineskip]
$\star$ \href{mailto:teodor.ilicin@ijs.si}{\small teodor.ilicin@ijs.si}
$\dagger$ \href{mailto:rok.zitko@ijs.si}{\small rok.zitko@ijs.si}
\end{center}

\section*{\color{scipostdeepblue}{Abstract}}
\textbf{\boldmath{%
We investigate the superconducting Anderson impurity model for interacting quantum dot Josephson junctions with spin–orbit coupling and a term accounting for tunnelling through higher-energy orbitals. These elements establish the conditions required for spin polarization in the absence of external magnetic field at finite superconducting phase bias. This Hamiltonian has been previously used to model the Andreev spin qubit, where quantum information is encoded in spinful odd-parity subgap states. Here we instead analyse the even-parity sector, i.e., the Andreev pair qubit based on Andreev bound states (ABS). The model is solved using the zero-bandwidth approximation and the numerical renormalization group, with further insight from variational calculations. Electron–electron interaction admixes single-occupancy Yu–Shiba–Rusinov (YSR) components into the ABS states, thereby strongly enhancing spin transitions in the presence of spin-orbit coupling. The ABS states can thus become sensitive to local magnetic field fluctuations, which has implications for decoherence in Andreev pair qubits. For strong interaction $U$, especially in the cross-over region between the ABS and YSR regimes for $U \sim 2\Delta$, charge, spin, and inductive transitions can all become strong, offering avenues for spin control and quantum transduction.
}}

\vspace{\baselineskip}

\noindent\textcolor{white!90!black}{%
\fbox{\parbox{0.975\linewidth}{%
\textcolor{white!40!black}{\begin{tabular}{lr}%
  \begin{minipage}{0.6\textwidth}%
    {\small Copyright attribution to authors. \newline
    This work is a submission to SciPost Physics. \newline
    License information to appear upon publication. \newline
    Publication information to appear upon publication.}
  \end{minipage} & \begin{minipage}{0.4\textwidth}
    {\small Received Date \newline Accepted Date \newline Published Date}%
  \end{minipage}
\end{tabular}}
}}
}

\vspace{10pt}
\noindent\rule{\textwidth}{1pt}  
\tableofcontents
\noindent\rule{\textwidth}{1pt}
\vspace{10pt}

\section{Introduction}

Andreev level qubits store quantum information in the discrete subgap states localized at Josephson junctions between superconducting leads. The information can be encoded in different ways, either in the odd-occupancy subspace (Andreev spin qubits) 
\cite{chtchelkatchev2003,padurariu2010,hays2021,bargerbos2022,Bargerbos2023,pitavidal2023,PitaVidal2025,Lu2025} or in the even-occupancy subspace (Andreev pair qubits) \cite{Ivanov1999,Desposito2001,Lantz2002,Zazunov2003,Zazunov2005,bretheau2013nature,bretheau2013prx,Bretheau2014,janvier2015,Hays2018,tosi2019,metzger2021,Cheung2024,shvetsov2025}. The chosen computational basis states are not necessarily those with the absolute lowest energy, and it was proved possible to control the occupancy of the subgap state manifold (its odd or even parity) using microwave pumping, reaching high (90\%) parity polarization \cite{Wesdorp2023}. Provided that the parity lifetime is long enough, this allows parity-selective spectroscopy ``in hardware'' without heralding or post-selection, as well as state manipulation in what would otherwise be an excited-state manifold.

\begin{figure}[hbt]
\centering
\includegraphics[width=8cm]{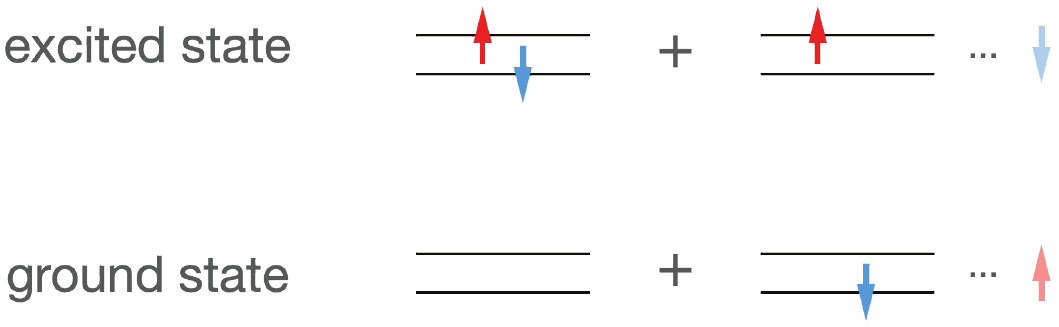}
\caption{Schematic representation of the sub-gap states in the even-parity subsector. The lines represent the QD levels,
the arrows represent Bogoliubov quasiparticles located either inside the junction (dark) or in the bath (light). The ABS components (left) 
differ in the number (zero or two) of trapped quasiparticles, the YSR components (right) differ in the wavefunction of the quasiparticle
in the bath.
} 
\label{schematic}
\end{figure}

In the case of Andreev pair qubits based on the Andreev bound states (ABS), the two basis states are distinguished by the relative phase of the empty-orbital and two-electron parts of the local wavefunction, $(\ket{0} \pm \ket{2})/\sqrt{2}$, or equivalently, in the basis of Bogoliubov excitations, by the absence or presence of two trapped quasiparticles (QPs), see Fig.~\ref{schematic}. One possible implementation of Andreev level qubits uses hybrid semiconductor-superconductor structures to build quantum-dot (QD) Josephson junctions that can trap QPs, for instance using InAs nanowires or Ge quantum wells \cite{deLange2015,Larsen2015,Zellekens2022,shvetsov2025,tenKate2025FiniteLengthGeJJ, hybrid2010,meden2019review}. Such structures can be modelled using the superconducting Anderson impurity model \cite{pillet2013,lee2017prb}. Electrons confined in the QD experience Coulomb interaction, the strength of which, $U$, depends strongly on the material properties, the device geometry, and the tuning of gate voltages. It can range from very small values, in which case the interaction effects require some effort to be detected \cite{Fatemi2022}, to values where the nature of the subgap states changes completely and the QD behaves as a local magnetic moment \cite{bargerbos2022}; this is the regime of Yu-Shiba-Rusinov (YSR) states  \cite{Kirsanskas2015,jellinggaard2016,meden2019review,ilicin2025}. The change occurs for values $U \approx 2\Delta$, where $\Delta$ is the superconducting gap \cite{ilicin2025}. In the limit of small hybridization strength $\Gamma$, the transition between the two limits is sharp; however, for experimentally relevant couplings it becomes continuous, accompanied by a strong mixing between the ABS and YSR components of the wavefunction \cite{ilicin2025}, as schematically represented in Fig.~\ref{schematic}. 

As a result, in presence of electron-electron (e-e) interaction the ABS states generally acquire some local-moment character and are therefore sensitive to magnetic fields. The implications of this mixed ABS/YSR character in realistic devices remain little explored, although the e-e interaction clearly affects the device response \cite{Kurilovich2021MicrowaveResponseABS,Fatemi2022,MatuteCanadas2022,bargerbos2022,Bargerbos2023,Pavesic2023ImpurityKnightShift,Kurilovich2024,Wesdorp2024,Bordin2025ImpactABSLeads,tenKate2025FiniteLengthGeJJ}. In particular, the magnetic-field dependence of the ABS energy can induce unexpectedly large dephasing through local magnetic field fluctuations. Conversely, the coupling of even-parity subgap states to the magnetic field may prove advantageous, enabling magnetic state manipulation—akin to that in Andreev spin qubits—or supporting sensing and quantum transduction applications.

As key results, we will demonstrate that 1) because ABS acquire YSR admixture due to interaction, even-parity Andreev pair qubits are not purely charge qubits, but carry some local-moment character, 2) spin-orbit coupling in combination with chirality generates spin polarization and strong spin transitions without external magnetic field, and 3) in the cross-over regime around $U \approx 2\Delta$, charge, spin, and inductive transition matrix elements are large and tunable.

\section{Model}

We consider the situation where a single localized level in the QD embedded in the Josephson junction (JJ) dominates the low-energy physics. The Hamiltonian is then
\newcommand{\VN}{V^\mathrm{(N)}}
\newcommand{\VS}{V^\mathrm{(SF)}}
\newcommand{\VD}{V^\mathrm{(D)}}
\newcommand{\geff}{g_\mathrm{eff}}

\begin{equation}
\begin{split}
    H &= \frac{U}{2} (\hat{n}-\nu)^2 + \sum_{lk\sigma} \epsilon_k c^\dag_{lk\sigma} c_{lk\sigma} 
    - \sum_{lk} \left( \Delta e^{i\phi_l}  c^\dag_{l k\uparrow} c^\dag_{l\, -k\downarrow} + \text{H.c.} \right) \\
    &+ \sum_{k\sigma} \left( \frac{\VN_L}{\sqrt{N}}  c^\dag_{Lk\sigma} d_{\sigma} + \text{H.c.} \right) 
     + \sum_{k\sigma} \left( \frac{\VN_R}{\sqrt{N}}  d^\dag_{\sigma} c_{Rk\sigma} + \text{H.c.} \right) \\
    &+ \sum_{k\sigma} \left( \frac{\VS_L}{\sqrt{N}} i c^\dag_{Lk\sigma} d_{\bar \sigma} + \text{H.c.} \right) 
    + \sum_{k\sigma} \left( \frac{\VS_R}{\sqrt{N}} i d^\dag_{\sigma} c_{Rk\bar \sigma} + \text{H.c.} \right) \\
    &+ \sum_{kq\sigma} \left( \frac{\VD}{N} c^\dag_{Lk\sigma} c_{Rq\sigma} + \text{H.c.} \right) \\
    &+ \geff \mu_B (B_x S_x + B_y S_y + B_z S_z).
\end{split}
\end{equation}
Here $U$ is the electron--electron (e-e) repulsion, $\hat{n}=\sum_\sigma d^\dag_\sigma d_\sigma$ is the QD occupancy operator for
the orbital $d$ (the partially occupied active level),
$\nu$ is the gate voltage expressed in electron units (with $\nu=1$ corresponding to nominal half
filling of the QD). $l=L,R$ indexes the two superconducting leads, assumed to have equal dispersion $\epsilon_k$,
equal superconducting gap amplitude $\Delta$, but different phases $\phi_l$. We define $\phi_R=\phi/2$ and $\phi_L=-\phi/2$, so that $\phi=\phi_R-\phi_L$. The remaining terms describe electron transfer processes: $\VN$ and $\VS$ are the spin-conserving and spin-flipping tunneling amplitudes (the notation $\bar\sigma$ stands for spin inversion), while the $\VD$ term encodes all background tunnelling processes through the "inactive" orbitals away from the Fermi level (those that are nearly fully occupied or nearly empty). These orbitals are assumed to be integrated out, so that the only remaining contribution is this direct inter-lead hopping \cite{Bargerbos2023,pitavidal2025novel}. This contribution to the Hamiltonian is important as a means to include the hybridisation chirality, which is a prerequisite for "spontaneous" spin polarization in the absence of external magnetic field \cite{Beri2008,zazunov2009,Reynoso2012}. Note that in most circumstances a nanowire Josephson junction will have a significant number of "inactive" orbitals, hence the $\VD$ term is generically expected to be present and need not be small, as evidenced by the properties of experimental devices that show large mesoscopic variability with many tuning points exhibiting very large spin splitting \cite{Bargerbos2023}. We provide an estimate of the magnitude of $\VD$ in Appendix~\ref{app:VD}. $\VD$ processes are assumed to be spin-conserving (without loss of generality). We also remark that such Hamiltonian terms are relevant in problems where a QD JJ is embedded in a Bohm-Aharonov interferometer \cite{Karrasch2009Supercurrent,zalom2025}.
We will assume that the $g$-factor at the QD
is significantly larger than that in the SC, thus the Zeeman term is included only at the QD orbital; $\geff$ is then the effective
$g$-factor (essentially the difference of QD and SC $g$-factors). $S_{x,y,z}$ are local spin operator on the QD site, and $\mu_B$ is the Bohr magneton.

We assume $\VN_l$ to be real and to have equal value for both leads, $\VN_L=\VN_R=\VN$. $\VS_l$ will also be assumed to have equal values,
$\VS_L=\VS_R=\VS$; here the signs are particularly important, because the spin-flip tunnelling terms are "directional" (note the order of the operators in the spin-flip hopping terms in the Hamiltonian: from $c_L$ to $d$, then from $d$ to $c_R$). Equal values of $\VS$ correspond to the homogeneous case where the SOC effective field direction is constant along the electron path from one lead through the QD to the other lead.
In matrix notation, the tunnelling terms can be expressed in the form
\begin{equation}
H'=\left( c^\dag_{lk\uparrow}, c^\dag_{lk\downarrow} \right) 
\left( \VN+i \sigma_x \VS \right)
\left( d_\uparrow, d_\downarrow \right)^T + \text{H.c.},
\end{equation}
where $\sigma_x$ is the Pauli $X$ matrix; the effective SOC field thus points along the $x$-axis. This specific choice of the SOC direction has been made for notational and computational convenience and does not restrict generality; for a different orientation of the SOC field, one can choose the spin quantization axis so as to obtain this form.

To facilitate the discussion, we introduce the relative strength of the spin-flip terms (and thus the amplitude of the SOC effects) through a dimensionless ratio $\lambda$, so that
\begin{equation}
    \VN = V (1-\lambda), \quad \VS = V \lambda.
\end{equation}
The parameter $V$ quantifies the overall coupling; the total hybridization is 
\begin{equation}
    \Gamma=\pi \rho \left[ (\VN_L)^2+ (\VN_R)^2 +
(\VS_L)^2+(\VS_R)^2 \right] = 2\pi \rho V^2 \left[ (1-\lambda)^2 + \lambda^2 \right].
\end{equation}
The background tunnelling will be parametrized as 
\begin{equation}
\VD=\tau V,
\end{equation}
where $\tau$ quantifies the relative importance of tunnelling through higher-energy QD levels \cite{Bargerbos2023}. 

In many experimental devices multiple subgap modes (multiple ``channels'') are observed \cite{tosi2019,Hays2020,hays2021,MatuteCanadas2022,Wesdorp2024}. In this work we focus on the minimal situation where a single localized level dominates the low-energy physics; multi-level effects are neglected and could modify quantitative details, and certainly would be required to obtain transition lines involving several levels \cite{MatuteCanadas2022}.

\subsection{Parameter values}
\label{secparam}

\begin{figure}
\centering
    \includegraphics[width=0.45\textwidth]{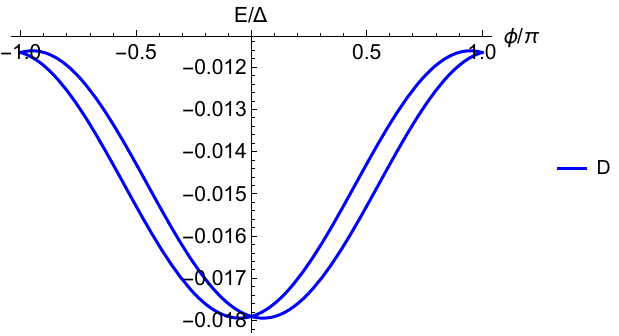}
    \includegraphics[width=0.45\textwidth]{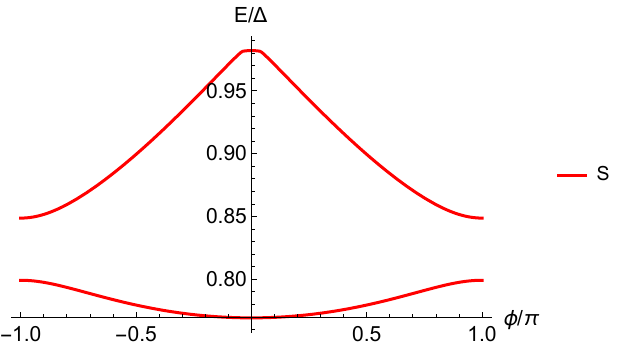}
\caption{Energy-phase relationships (dispersion curves) of the "doublet" (odd parity, the states relevant in the Andreev spin qubit) and "singlet" (even parity, the states relevant in the Andreev pair qubit) subgap states for the reference set of parameters. (ZBW results)}
\label{figdispersion}
\end{figure}

In a realistic hybrid nanowire JJ device operated at a generic tuning point, no parameter is particularly small and all terms in the Hamiltonian are relevant. This is especially important for the formation of local moments and spin polarization. Experiments (for example those described in Refs.~\cite{bargerbos2022,Bargerbos2023}) show a large mesoscopic variability of parameters, including the ratio of spin-preserving and spin-flip tunneling, implying that $\lambda$ can take essentially arbitrary value. Unless stated otherwise, we will use the following parameters (referred to as "reference values"): interaction strength $U=2\Delta$, overall hybridisation $V=0.2\Delta$, nominal filling $\nu=1$, dimensionless spin-orbit coupling $\lambda=0.3$, ratio of background tunnelling $\tau=0.4$, and phase bias $\phi=0.75\pi$. The corresponding energy-phase relationships are shown in Fig.~\ref{figdispersion}; in particular, the odd-parity states display the expected spin-splitting. Compared to experiments, this represents a rather modest splitting, thus the parameter set (including $V$ and $\tau$) is realistic and does not correspond to some extreme situation.

Parity switching is generally a complication for Andreev level qubits \cite{janvier2015}. Recent years have seen major progress in parity control and parity-selective spectroscopy, particularly using microwave techniques \cite{Blais2004,Schuster2005,bretheau2013nature,Bretheau2014,janvier2015,metzger2021,kurilovich2021,Zellekens2022,Hinderling2023}. For example, dispersive readout based on parity- and state-dependent resonator-frequency shifts can be used as a spectroscopic tool, and quantum capacitance can be measured by reflectometry \cite{Petersson2010,Colless2013,GonzalezZalba2015,Crippa2019,vanVeen2019,Razmadze2019,Menard2019,malinowski2022,aghaee2025,Vigneau2023}. Unlike conventional tunnelling spectroscopy, which involves parity-changing processes through electron injection into the device, these approaches are non-invasive and conserve parity. 

In experiments targeting a selected parity subspace, microwave spectroscopy and control are performed within the parity lifetime of the relevant states \cite{Hays2018,tosi2019}, even when these states belong to an excited manifold. This was the case, for example, in the first demonstration of coherent control of Andreev spin qubits in a low-$U$ device \cite{Hays2020,hays2021}, where post-selection was used to isolate the parity sector of interest. Furthermore, the desired parity sector can be initialized controllably by microwave pumping \cite{Wesdorp2023,Ackermann2023,Kurilovich2024,Zatsarynna2024}. In this approach, pumping is resonant with processes in a fixed-parity subspace, and the parity change subsequently occurs via spontaneous relaxation; parity polarization of 90\% can be obtained. 

\begin{figure}
\centering
    \includegraphics[width=0.5\textwidth]{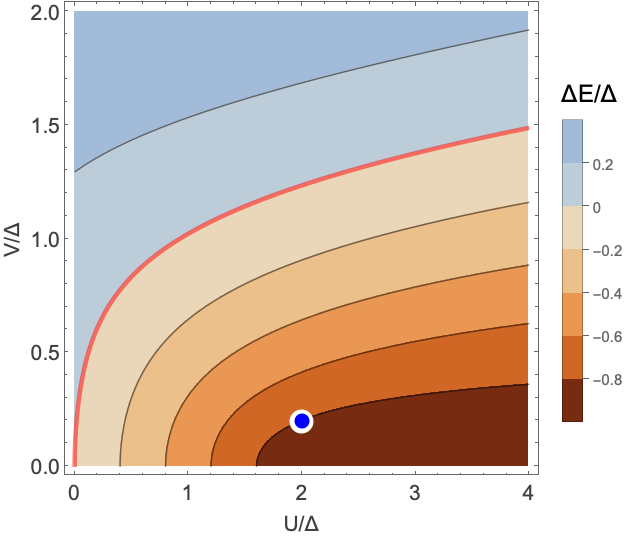}
\caption{The energy difference between the lowest-lying states in the odd and even parity sectors $\delta E = E_{G,o} - E_{G,e}$. Shades of blue (orange) represent the part of the phase space in which the ground state has even (odd) fermionic parity, and the crossover is indicated with the solid red line. Even in the region of odd-parity ground state, which includes the reference point, the even-parity sector is experimentally accessible using parity control (see main text).}
\label{transition}
\end{figure}

In general, finite $U$ favours an odd-parity absolute ground state. For $U \sim 2\Delta$ the ground state is indeed a spin-doublet unless $V$ is large. For completeness and for easier orientation in the parameter space, in Fig.~\ref{transition} we plot the energy difference between lowest-lying odd and even states $\delta E = E_{G,o} - E_{G,e}$ in the $(U,V)$ plane with all other parameters set to the reference values.
In the present work, however, we do not address state preparation or lifetime limitations and leave these issues for future study; see also Refs.~\cite{Ackermann2023,Kurilovich2024,Zatsarynna2024} for a broader discussion. We therefore restrict the analysis to the even-parity sector, irrespective of its energy offset relative to the lowest odd-parity state.

\section{Methods}

\subsection{Zero-bandwidth approximation}

We introduce the orbitals $f_{L\sigma}=\frac{1}{\sqrt{N}} \sum_k c_{Lk\sigma}$, $f_{R\sigma}=\frac{1}{\sqrt{N}} \sum_k c_{Rk\sigma}$, $N$ being the number of orbitals in each superconducting band. By rewriting the tunnelling part of the Hamiltonian in terms of these operators and dropping from consideration all other orbitals in the superconductors, we obtain the zero-bandwidth (ZBW) Hamiltonian \cite{vecino2003}. This approximation captures the essence of the problem and provides a convenient starting point for more controlled approaches \cite{Baran2023}. For the present problem, comparison with NRG (next subsection) for representative parameter sets shows that the ZBW method captures all qualitative features of interest and can therefore be used reliably for mapping out trends. The ZBW calculations reduce to diagonalisation of small matrices.

\subsection{Numerical renormalization group}

For a fully quantitative description, especially of phenomena near $U \approx 2\Delta$ where discrete states interact with the continuum (which is beyond the ZBW approximation), we employ the numerical renormalization group (NRG) \cite{wilson1975,bulla2008}. NRG relies on logarithmic discretization of the leads, mapping to a Wilson chain, and iterative diagonalization. In the present case, the method is particularly demanding due to the absence of symmetries beyond fermionic parity\footnote{For zero magnetic field, one can make use of the conserved $x$-component of total spin. However, magnetic field along the $z$-axis is required to extract the strength of the SOC effective field.}, which can be mitigated by using a coarser discretization parameter at the cost of somewhat larger systematic errors. Specifically, we use $\Lambda=8$ and we keep up to 3000 states per iteration. Past experience shows that using large values of $\Lambda$ does not lead to significant errors (in particular not to qualitative errors) \cite{zitko2009,zitko2010}.
Even with such large $\Lambda$ memory requirements are significant, on the order of 20 to 30GB. The calculation takes on the order of 3 to 4 hours for each parameter set when running on 8 cores of a modern CPU.

\subsection{Variational method}
\label{sec:var}

In Ref.~\cite{ilicin2025} we have recently proposed a variational Ansatz for solving the single-channel SC AIM for any value of $U$, from the ABS to the YSR limit. In this work, we generalize this approach to the two-channel case as appropriate for a Josephson junction. The Ansatz becomes 
$\ket{\Psi} = \sum_{s} \ket{\varphi^s} + \sum_{l} \ket{\psi_l} + \sum_{s,l,l'} \ket{\chi^s_{l,l'}}$, that contains wavefunctions with up to two QPs in the leads:
\begin{equation}
    \begin{split}
        \ket{\varphi^+} &= \frac{a^+}{\sqrt{2}} \left(1 + d^{\dag}_{\up} d^{\dag}_{\dw} \right)  \ket{\Phi_0} \>, \\ 
        \ket{\varphi^-} &= \frac{a^-}{\sqrt{2}} \left(1 - d^{\dag}_{\up} d^{\dag}_{\dw} \right)  \ket{\Phi_0} \>, \\ 
        \ket{\psi_{L}} &= \sum_{k} \frac{\alpha_{Lk}}{\sqrt{2}} \left( d^{\dag}_{\up} \gamma^{\dag}_{Lk\dw} - d^{\dag}_{\dw} \gamma^{\dag}_{Lk\up} \right) \ket{\Phi_0} \>, \\
        \ket{\psi_{R}} &= \sum_{k} \frac{\alpha_{Rk}}{\sqrt{2}} \left( d^{\dag}_{\up} \gamma^{\dag}_{Rk\dw} - d^{\dag}_{\dw} \gamma^{\dag}_{Rk\up} \right) \ket{\Phi_0} \>, \\
        \ket{\chi^{\pm}_{LL}} &= \sum_{k,q} \frac{\beta_{LLkq}^{\pm}}{2\sqrt{2}} \left(1 \pm d^{\dag}_{\up} d^{\dag}_{\dw} \right) \left( \gamma_{Lk\up}^\dag \gamma_{Lq\dw}^\dag - \gamma_{Lk\dw}^\dag \gamma_{Lq \up}^\dag \right) \ket{\Phi_0} \>, \\
        \ket{\chi^{\pm}_{RR}} &= \sum_{k,q} \frac{\beta_{RRkq}^{\pm}}{2\sqrt{2}} \left(1 \pm d^{\dag}_{\up} d^{\dag}_{\dw} \right) \left( \gamma_{Rk\up}^\dag \gamma_{Rq\dw}^\dag - \gamma_{Rk\dw}^\dag \gamma_{Rq \up}^\dag \right) \ket{\Phi_0} \>, \\
        \ket{\chi^{\pm}_{LR}} &= \sum_{k,q} \frac{\beta_{LRkq}^{\pm}}{2} \left(1 \pm d^{\dag}_{\up} d^{\dag}_{\dw} \right) \left( \gamma_{Lk\up}^\dag \gamma_{Rq\dw}^\dag - \gamma_{Lk\dw}^\dag \gamma_{Rq \up}^\dag \right) \ket{\Phi_0} \>.
    \end{split}
    \label{eq:ansatz}
\end{equation}
Here $\gamma_{lk\sigma}$ is the annihilation operator for a Bogoliubov QP in lead $l\in \{L, R\}$ with momentum $k$ and spin $\sigma$, defined in Appendix~\ref{app:var}. Here $\ket{\Phi_0} = \ket{\halfbox} \otimes \ket{\mathrm{BCS}}$ is the reference state with unfilled QD and BCS ground state in the leads.
Note that there are six independent two-QP states, corresponding to $(l,l') = (L,L),$ $(R,R), (L,R)$, each combination occurring with a $+1$ and $-1$ relative sign in the ABS part of the wavefunction. The numerical prefactors in the Ansatz have been chosen such that the wavefunction is normalized as
\begin{equation}
    |a^+|^2 + |a^-|^2 + \sum_{lk} \left( |\alpha_{lk}^+|^2+|\alpha_{lk}^-|^2 \right)  
    + \sum_{(ll')kk'} \left( |\beta_{ll'kk'}^+|^2 + |\beta_{ll'kk'}^-|^2 \right) = 1,
\end{equation}
where $(ll')$ corresponds to enumeration over the three combinations of lead indexes. The indexes $\beta_{ll}$ are symmetric in momentum indexes: $\beta_{llkq}=\beta_{llqk}$.
In the presence of spin-orbit coupling, the Ansatz needs to be extended with spin-triplet counterparts to $\ket{\Psi_L}$, $\ket{\Psi_R}$, $\ket{\chi^\pm_{ll'}}$. For the particular choice of spin-flip tunneling terms with SOC effective field direction along the $x$-axis, it is convenient to work with the eigenstates of the $x$-component of the total spin operator $S_x$. With our choice of the hopping terms, only the $S_x=0$ terms are required:
\begin{equation}
 \begin{split}
        \ket{\psi^{T}_{L}} &= \sum_{k} (+i) \frac{\alpha^{T}_{Lk}}{\sqrt{2}} \left( d^{\dag}_{\up} \gamma^{\dag}_{Lk\up} - d^{\dag}_{\dw} \gamma^{\dag}_{Lk\dw} \right) \ket{\Phi_0} \>,  \\
        \ket{\psi^{T}_{R}} &= \sum_{k} (-i) \frac{\alpha^{T}_{Rk}}{\sqrt{2}} \left( d^{\dag}_{\up} \gamma^{\dag}_{Rk\up} - d^{\dag}_{\dw} \gamma^{\dag}_{Rk\dw} \right) \ket{\Phi_0} \>, \\
         \ket{\chi^{\pm,T}_{ll}} &= \sum_{k,q} \frac{\xi^{\pm}_{llkq}}{2\sqrt{2}} \left(1 \pm d^{\dag}_{\up} d^{\dag}_{\dw} \right)
         \left( \gamma_{lk\up}^\dag \gamma_{lq\up}^\dag - \gamma_{lk\dw}^\dag \gamma_{lq \dw}^\dag \right) \ket{\Phi_0} \>, \\
         \ket{\chi^{\pm,T}_{LR}} &= \sum_{k,q} \frac{\xi_{LRkq}^{\pm}}{2} \left(1 \pm d^{\dag}_{\up} d^{\dag}_{\dw} \right) 
         \left( \gamma_{Lk\up}^\dag \gamma_{Rq\up}^\dag - \gamma_{Lk\dw}^\dag \gamma_{Rq \dw}^\dag \right) \ket{\Phi_0} \>,
    \end{split}
    \label{eq:ansatzSz}
\end{equation}
Imaginary factors $(+i)$ and $(-i)$ are included to simplify the variational equations and to better reveal the relation between normal and spin-flip tunneling processing.  Further details about the variational method are provided in Appendix~\ref{app:var}. We will mostly use the variational method for qualitative guidance: the results presented in the figures in the main text are obtained using the ZBW and NRG. In particular limits, the variational solution becomes simple, allowing us to transparently write the quantitatively correct wave-functions. These results are then used to provide further clarification of the qualitative behaviour observed in the numerical (ZBW and NRG) results. 

\section{General considerations}

\subsection{Symmetry properties}

Away from special points in the parameter space, only fermionic parity and (for zero magnetic field) the $x$ component of the total spin are conserved. The $y$ and $z$ spin components are not conserved because of the SOC, and for $\VD\neq0$ there is no particle-hole (p-h) symmetry, due to the absence of bipartite hopping structure.

In general, the results are symmetric with respect to $\lambda=1/2$, so only the interval $\lambda \in [0:1/2]$ needs to be considered. This follows from the fact that one can relate Hamiltonians with $\lambda$ and $1-\lambda$ through a transformation that maps spin-conserving tunneling terms onto the spin-flip tunneling terms and vice versa. In particular, the case of pure spin-flip scattering ($\lambda=1$) maps onto the case of pure normal scattering ($\lambda=0$). 

In the following subsections, we discuss specific sets of parameters, in which the model has increased symmetry.

\subsubsection{$X$-symmetry}

\begin{figure}
\centering
    \includegraphics[width=0.9\textwidth]{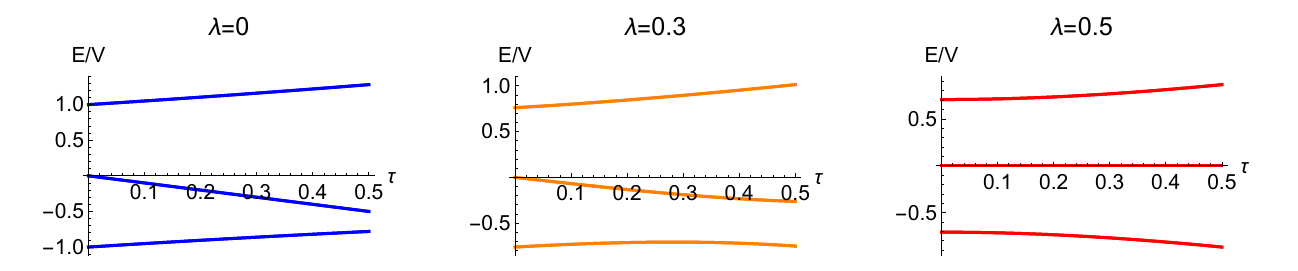}
    \includegraphics[width=0.9\textwidth]{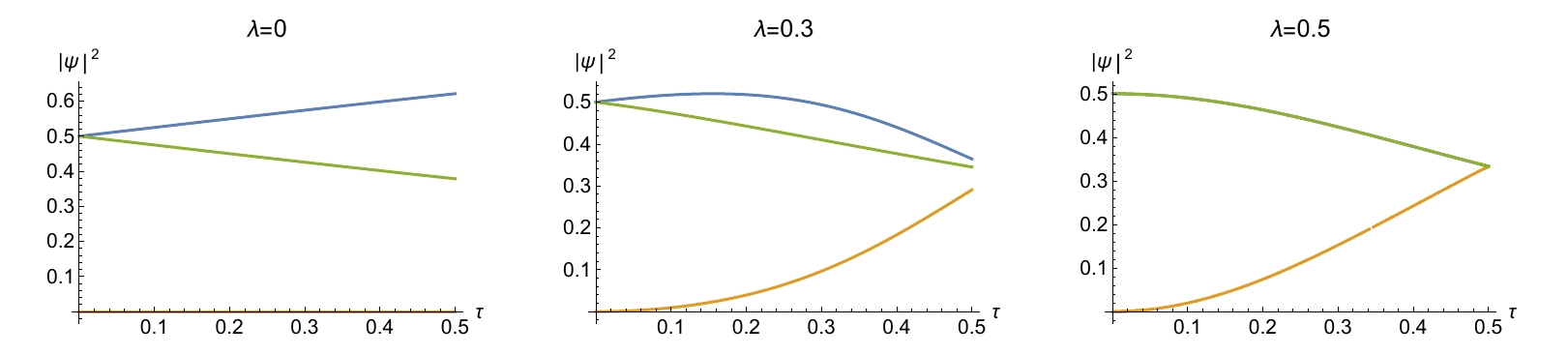}
\caption{Top: Eigenenergies of the normal-state ($\Delta=0$) single-electron part of the Hamiltonian  for three values of the dimensionless SOC parameter $\lambda$ as a function of the background tunnelling parameter $\tau=t/V$. Bottom: Corresponding probabilities for electron occupation on the QD orbital for lowest (blue), middle (orange) and
highest-energy state (green).}
\label{figt}
\end{figure}

In Fig.~\ref{figt} we plot the eigenvalues of the single-electron non-interacting Hamiltonian at $\Delta=0$ as a function of $\tau$ for several values of $\lambda$. In this simplified case, the Hamiltonian is a $6 \times 6$ matrix, and the eigenvalues come in Kramers pairs. Generally, there is no symmetry with respect to zero energy for $\tau\neq0$, except for the special case of $\lambda=1/2$, when the unitary transformation
\begin{equation}
    X := 
    \begin{cases}
        d^\dag_\sigma \to d_{\sigma}, \\
        c^\dag_{lk\sigma} \to i (-1)^l (-1)^\sigma c_{lk\bar\sigma},\\
    \end{cases}
\end{equation}
commutes with the Hamiltonian. In the definition, $(-1)^L=-1$ and $(-1)^R=1$, $\bar\sigma$ is spin inversion, and $(-1)^\uparrow=-1$, $(-1)^\downarrow=1$. The transformation maps spin-conserving tunneling term onto the spin-flip tunneling term and vice versa, and leaves the Hamiltonian invariant when the magnitudes of the two processes become equal. The symmetry is carried over to the full many-body Hamiltonian ($\Delta, U \neq 0$), under the conditions $\nu=1$, $\phi=\pi$. As we will see in the following, this leads to degeneracy of the Andreev pair qubit states. Consequently, system properties will be anomalous in vicinity of the point ($\lambda=1/2$, $\nu=1$, $\phi=\pi$). 

The existence of special symmetry for equal spin-conserving and spin-flipping tunneling rates is expected to be generic, and does not depend on the particular choice of SOC axis and parametrization, since an analogous mapping could be derived for other forms of SOC terms.

\subsubsection{$\eta$-symmetry}
\label{eta}

For $\nu=1$, $\VD=0$ and $\VN_L=\VN_R$, $\VS_L=\VS_R$ (i.e., for the standard left-right and particle-hole symmetric QD Josephson junction problem with no inter-lead tunneling, but with SOC), the Hamiltonian has a unitary symmetry (denoted by $\eta$) that simultaneously flips spin, mirror reflects the system, and exchanges particles and holes. It holds for arbitrary phase bias $\phi$. Even if the focus of this work is on the general situation with finite $\VD$, this symmetry represents a convenient reference point. Furthermore, the variational solution (Appendix B) can be most compactly expressed in terms of basis states that transform as $\eta$-invariants.

For this parameter tuning, the Hamiltonian has an anti-unitary particle-hole symmetry for any value of the phase-difference $\phi$, given by $\Pi K$, where:
\begin{equation}
    \Pi := \begin{cases}
        d_{\sigma} \rightarrow - d^{\dag}_{\bar{\sigma}} \\ 
        c_{lk\sigma} \rightarrow c^{\dag}_{l\bar{k}\bar{\sigma}}
    \end{cases} \>,
\end{equation}
and $K$ denotes complex conjugation. Here $\bar{\bullet}$ refers to the inversion of spin and the reflection of $k$ with respect to the Fermi surface. 
The Hamiltonian also has an anti-unitary mirror symmetry $MK$, where:
\begin{equation}
    M := c_{lk\sigma} \rightarrow c_{\bar{l}k\sigma} \>, 
\end{equation}
and $\bar{l}$ mirrors the channel (e.g. $\bar{L}=R$). The particular form of $MK$ makes use of the symmetric gauge, where conjugation is required to ''swap" the values of $\phi_{L}$ and $\phi_R$. Combining these two symmetries allows us to define a {\it unitary} transformation:
\begin{equation}
    \eta := \begin{cases}
        d_{\sigma} \rightarrow - d^{\dag}_{\bar{\sigma}} \\ 
        c_{lk\sigma} \rightarrow c^{\dag}_{\bar{l}\bar{k} \bar{\sigma}}
    \end{cases} \>,
\end{equation}
which leaves the Hamiltonian invariant. This symmetry exists for any value of $\phi$. 

Considering the variational states, we find $\eta \ket{\varphi^+} = +1 \ket{\varphi^+}$ and $\eta \ket{\varphi^-} =- \ket{\varphi^-}$. For other states in the Ansatz ~\eqref{eq:ansatz}, one can find specific linear combinations with definite parity. For example, $\ket{\psi^+} = \ket{\psi_L} + \ket{\psi_R}$, along with the condition $\alpha_{Lk} = \alpha_{R\bar{k}}$ defines an $\eta$-even state. Similar construction gives an odd single-QP state $\ket{\psi^-}$ provided that $\alpha_{Lk} = -\alpha_{R \bar{k}}$. The same considerations apply to the two-QP states, as well as to triplet states.

\subsection{Transition range between the ABS and YSR regimes}
\label{subtrans}

The $U=2\Delta$ special point corresponds to a quantum phase transition (level crossing) in the even-fermion-parity subspace. When $U$ exceeds $2 \Delta$ the lowest-energy state changes from the ABS-like to the YSR-like state, which can be represented in the atomic limit ($V\to0$) as 
\begin{align*}    
    \ket{\psi_{\text{ABS}}} = \frac{1}{\sqrt{2}} \left( 1 \pm d_{\up}^{\dag} d_{\dw}^{\dag} \right) \ket{\Phi_0} \longrightarrow \ket{\psi_{\text{YSR},l}} = \frac{1}{\sqrt{2}} \left(d_{\up}^{\dag} f_{l,\dw}^{\dag} - d_{\dw}^{\dag} f_{l,\up}^{\dag} \right) \ket{\Phi_0} \>. 
\end{align*}

For $\VD\neq0$, the direct tunneling constitutes a weak link between the superconductors, hence it generates a localized quasiparticle state with an energy level inside the gap \cite{zalom2025}. The creation of the YSR-type singlet $\ket{\psi_\mathrm{YSR}}$ becomes less energetically costly, because the quasiparticle can be generated in this new state instead of being delocalized further away in the leads. The centre of the cross-over region thus shifts toward lower $U$.

A quantitatively adequate approximative description of the shift of the transition point can be obtained from a single-QP mode Ansatz in the YSR regime. Neglecting the coupling with the dot level, the ''direct hopping'' matrix element
\begin{equation*}
    \mel{\psi_{\mathrm{YSR},L}}{H^{(D)}_c}{\psi_{\mathrm{YSR},R}} = V^{(D)} \sin{\frac{\phi}{2}} \>, 
\end{equation*} 
reduces the YSR state energy by:
\begin{equation}
    E_S = \Delta - V^{(D)} \sin{\frac{\phi}{2}} \>.
\end{equation}
As a consequence, the crossover point at finite $\tau$ shifts to
\begin{equation}
    U^* / 2 = \Delta - V^{(D)} \sin{\frac{\phi}{2}}.
\end{equation}

This discussion assumes other hopping coefficients ($\VN, \VS$) to be zero. When hopping to the dot is finite, there is an opposing effect, pushing the cross-over region towards larger values of $U$ \cite{ilicin2025}. For realistic parameters, the centre of ABS-YSR cross-over regime therefore always lies close to $U \approx 2 \Delta$.

\subsection{Emergence of local spin polarization}
\label{polariz}

We briefly discuss how local spin polarization along the $x$-axis arises in a model that conserves the $x$-component of the total spin $S^\mathrm{total}_x$.
We define an Ansatz defined in a two-orbital space (representing impurity, $d$, and bath, $f$) composed of an impurity local singlet ($L$), an inter-orbital singlet ($S$) and an inter-orbital triplet ($T$) with $S^\mathrm{total}_x=0$:
\begin{equation}
    \psi = 
    a_L d^\dag_\uparrow d^\dag_\downarrow 
    + 
    \frac{a_S}{\sqrt{2}} \left( d^\dag_\uparrow f^\dag_\downarrow - d^\dag_\downarrow f^\dag_\uparrow \right)
    + 
    \frac{a_T e^{i\phi_T}}{\sqrt{2}} \left( d^\dag_\uparrow f^\dag_\uparrow - d^\dag_\downarrow f^\dag_\downarrow \right),
\end{equation}
with $a_L$, $a_S$, $a_T$ real and satisfying normalization condition $a_L^2+a_S^2+a_T^2=1$, and a phase parameter $\phi_T$. The relative phase between $a_L$ and $a_S/a_T$ does not affect the argument, thus a single phase $\phi_T$ is sufficient in the Ansatz.
A quick calculations for the local spin polarization of the $d$ orbital yields $\langle \psi | S^d_x | \psi \rangle =-a_S a_T \cos\phi_T$. The other spin components $y$ and $z$ are strictly zero. We see that the local singlet is irrelevant, while the local spin polarization is generated by the interference of inter-orbital singlet and triplet components. It is maximal for equal strength of singlet and triplet components, $a_S=a_T=1/\sqrt{2}$ and, this is key, for coefficients that are in phase, $\phi_T=0$.
In this case, the wavefunction factorizes as $\psi=(1/2) \left( d^\dag_{\uparrow}-d^\dag_{\downarrow} \right) \left( f^\dag_\uparrow + f^\dag_\downarrow \right)$, corresponding to two spins polarized in opposite directions along the $x$-axis. This simple picture provides the underlying intuition for understanding the local spin polarization of subgap states due to a combination of singlet and triplet YSR components with an appropriate phase relation.

\section{Results}

To obtain a qualitative picture of the general behaviour in the typical operating range, we now perform a study using the computationally inexpensive ZBW approach, complemented by heuristic interpretations of key observations (Secs.~\ref{sec:local-moment-fraction} to \ref{sec:matel}). Later, we validate these results using the NRG calculations (Sec.~\ref{sec:nrg}). For quick reading, each section begins with a short paragraph summarizing the main observations and key physical mechanisms.

\subsection{Local-moment fraction}
\label{sec:local-moment-fraction}

In this section, we use the local-moment fraction $P_1=\langle n-2n_\uparrow n_\downarrow\rangle$ to track the crossover from ABS-like to YSR-like subgap states. The main mechanisms controlling state properties are the competition between pairing and Coulomb repulsion, which drives the ABS--YSR crossover, orthogonality constraints between the two subgap states, which can force them to evolve differently, and (near small phase bias) the influence of nearby continuum states, which can strongly reshape the excited state.

\begin{figure}
\centering
    \includegraphics[width=0.9\textwidth]{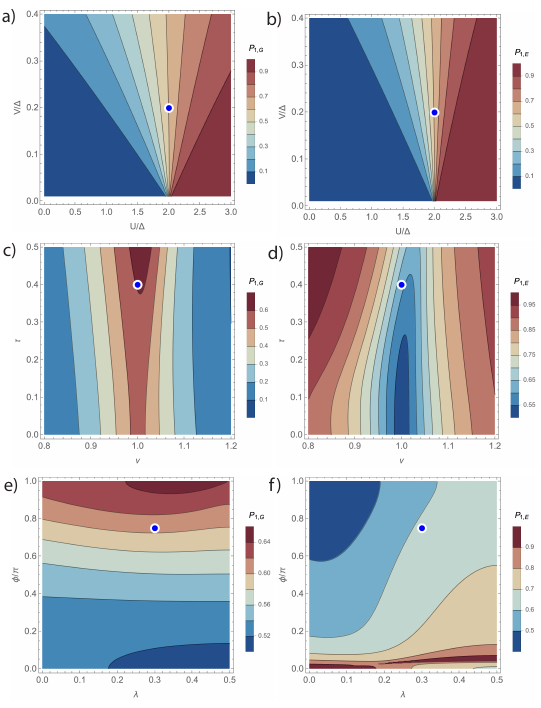}
\caption{Local-moment fraction, i.e., the probability of single-electron occupancy on the quantum dot, $P_1=\langle n-2n_\uparrow n_\downarrow \rangle$, in ground (G) and excited (E) subgap states; ZBW results. (a,b) Dependence on the electron--electron repulsion $U$ and the hybridization $V$. (c,d) Dependence on the displacement from half-filling, $\nu$, and the relative strength of background tunnelling through high-energy orbitals, $\tau$. (e,f) Dependence on the strength of spin-orbit coupling, $\lambda$, and the phase bias $\phi$.
Reference parameter values (Sec.~\ref{secparam}) are indicated by the blue markers. }
\label{fig1}
\end{figure}

The local-moment fraction, i.e., the probability of single-electron occupancy on the quantum dot, can be used to directly characterize the subgap states. From the perspective of the variational Ansatz, $P_1$ corresponds to contributions from $\ket{\psi_L}$ and $\ket{\psi_R}$ (and their spin-1 equivalents). This quantity evolves from 0 in the deep ABS limit to 1 in the deep YSR limit; the change is sudden in the atomic limit ($V \to 0$), but becomes a smooth crossover for finite $V$. The crossover region is narrower for the excited state than for the ground state, but generally they are quite similar, see Fig.~\ref{fig1}a,b. This generalizes the findings from Ref.~\cite{ilicin2025} which focused on the ground state G in the case of a single superconductor, which corresponds to the $\phi \to 0$ limit of the present work. For $\phi=0$, the excited state E merges with the continuum on the YSR side (see Fig.~6 in Ref.~\cite{ilicin2025}, as well as Fig.~\ref{fig4}a in this work). This is a peculiarity of the $\phi=0$ case that maps onto a single-channel Hamiltonian. For a general phase difference $\phi$ not too close to 0, there exist two subgap states for all values of $U$ and $V$, including in the YSR regime \cite{Kirsanskas2015}.

The lines of constant $P_1$ in the $(U,V)$ plane are almost linear (Fig.~\ref{fig1}a,b). For low $V$, this follows from the variational equations (Appendix \ref{appA1}) that describe linear mixing between the ABS and YSR terms, but such behaviour is seen to extend into the intermediate coupling regime, too. 
This scaling is, however, specific to $P_1$; other quantities tend to exhibit more complex dependencies, as will become clear in the following. We also note that the $P_1 = 0.5$ curve, which may be considered as the centre of the cross-over region, has a slight negative slope in the $(U,V)$ plane. This is in line with the observations made in Sec.~\ref{subtrans}.

The dependence in the $(\nu,\tau)$ plane is strikingly different for the ground and the excited state, see Fig.~\ref{fig1}c,d. The ground state G has maximal $P_1$ when $\nu$ is tuned near half filling, $\nu=1$, and it grows slightly with increasing $\tau$. The first observation is naturally expected, since it directly results from the behaviour of the YSR singlets for large $U$, where $\nu$ very directly tunes the average occupancy of the impurity orbital. The weak dependence on $\tau$ is only a small correction to this general trend; in the ZBW picture, it arises because the weight of the intermediate-energy non-interacting wavefunction on the QD orbital increases with increasing $\tau$ (see also Fig.~\ref{figt}, bottom panels, orange lines). 

Quite surprisingly, however, the local-moment fraction of the excited state E actually increases away from $\nu=1$, see Fig.~\ref{fig1}d. This can be understood from the requirement that the two subgap states remain orthogonal to each other. On the $\nu<1$ side, as the ground states becomes increasingly of pure $\ket{0}$ character with decreasing $\nu$, at the detriment of the $\ket{1}$ and $\ket{2}$ components, the lowest energy excited states maintaining orthogonality will need to have a sizeable $\ket{1}$ component, since the $\ket{2}$ part is too high in energy due to the relatively large value of the electron-electron repulsion $U$. Clearly, this behaviour is specific to the singular $U \sim 2\Delta$ cross-over region in vicinity of $\nu=1$; it is not relevant for low $U$ (where the two orthogonal states away from $\nu=1$ are simply of $\ket{0}$ and $\ket{2}$ type, respectively, and $\ket{1}$ is merely a small interaction-induced correction), nor for high $U$ (where the two subgap states are both YSR states with well defined local moment that is screened by two different linear combinations of Bogoliubov quasiparticles from the superconducting leads, as long as $\phi$ is not too close to 0 \cite{Kirsanskas2015,pavesic2024}). Fig.~\ref{fig1}d is a clear example of states in the crossover regime possessing unconventional properties, with the excited state displaying them most prominently.

Regarding the remaining parameters from the set, the SOC $\lambda$ and the phase $\phi$, for the ground state we find only a small variation of order 10\% across the full range of $\phi$ and a tiny dependence on $\lambda$, see Fig.~\ref{fig1}e. For the excited state, we  find moderate dependence on $\lambda$ and $\phi$, except for the region of very small $\phi$, see Fig.~\ref{fig1}f: this is where the excited state would enter the continuum in the full model, while in the ZBW approximation it mixes with a discrete state that has the character of a doublet state with an unbound Bogoliubov quasiparticle in the superconductor. This results in a rapid variation and anomalous behaviour (here for $\phi/\pi \lesssim 0.1$) that will also be visible in other quantities presented in the following. Since the focus of this work is in generic $\phi$, we will not analyse these peculiarities of the small $\phi$ range.

We note the opposite trend in $P_1$ for G and E states as regards their $\phi$ dependence, respectively increasing and decreasing, compare Fig.~\ref{fig1}e and \ref{fig1}f. This can be explained by the opposite dispersion (energy vs. bias $\phi$) of the two states: G increases with $\phi$ moving away from $\phi=0$, while E decreases, see Fig.~\ref{figdispersion}. Generally speaking, the vicinity to the continuum of excited states (which are essentially the doublet states with a decoupled Bogoliubov quasiparticle) tends to increase the single-occupancy component.

\subsection{Magnetic response}

In this section, we analyse the magnetic properties of the subgap states. We show that spin--orbit coupling, phase bias, and background tunnelling induce a small triplet admixture and spin polarization, with distinct behaviour in the ground and excited states. The main mechanisms are the spinful character inherited from the YSR sector and the effective spin--orbit field, which mixes spin components and selects a preferred spin orientation.

Since subgap states host local moment, they exhibit magnetic response. For $\phi \neq 0,\pi$, and in the simultaneous presence of spin–orbit coupling and background tunnelling, the states are spin-polarized in the absence of external magnetic field. The polarization occurs for essentially the same reasons as for the the doublet states in the odd-parity subspace in Andreev spin qubit (ASQ) devices \cite{Bargerbos2023}. The YSR components of states G and E simply inherit these properties.

\begin{figure}
\centering
    \includegraphics[width=0.9\textwidth]{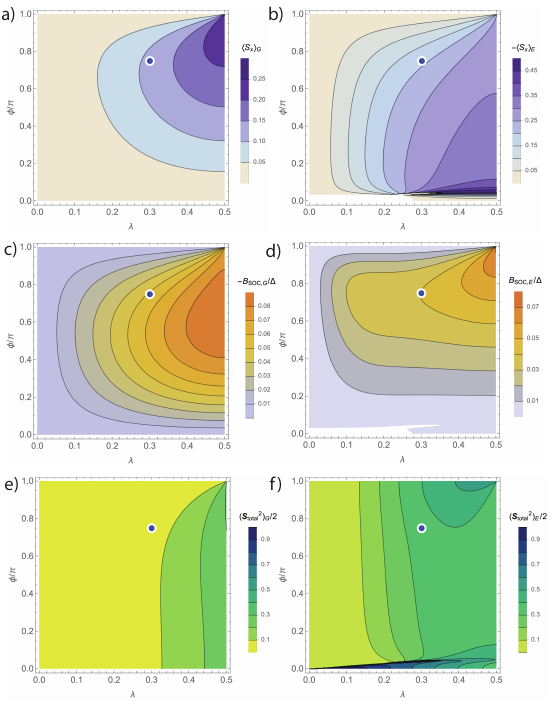}
\caption{Magnetic properties of the subgap states as a function of the SOC strength $\lambda$ and the phase bias $\phi$; ZBW results. (a,b) Spin polarization along the SOC axis ($x$); note the difference in signs. (c,d) Strength of the SOC effective magnetic field.
(e,f) Size of the YSR spin-triplet component quantified by $\langle \mathbf{S}_\mathrm{total}^2 \rangle$.
}
\label{fig2}
\end{figure}

The spin polarization $\langle S_x \rangle$ of G and E states points in the opposite directions, see Fig.~\ref{fig2}a,b. We plot the results in the $(\lambda,\phi)$ plane, as these two parameters determine the magnitude of the polarization. The odd-parity doublets in ASQs form degenerate Kramers pairs, unless all conditions for the spin splitting are present \cite{Beri2008,zazunov2009}; spin splitting and spin polarization occur concurrently. The even-parity states are in general non-degenerate, however $\langle S_x \rangle \neq 0$ only in the presence of the specified three conditions. Because of the "pre-existing energy splitting", the effective SOC field $B_\mathrm{SOC}$ is not the same for the G and E states, see Fig.~\ref{fig2}c,d. This quantity is extracted by applying a weak probe field along the transverse $z$ direction and observing the change of direction of the spin polarization, which gives
\begin{equation}
B_\mathrm{SOC} = \frac{\langle S_x \rangle}{\langle S_z \rangle} B_z.
\end{equation}

The spin polarization grows monotonously with $\lambda$, see Fig.~\ref{fig2}a,b. For the G state, the splitting is largest for intermediate values of $\phi$ close to $\pi/2$ (where time-reversal symmetry breaking is strongest) and it shifts toward $\phi=\pi$ as $\lambda$ increases. At the special point $\phi=\pi$, $\lambda=1/2$ the variation is singular, with a direction-dependent limiting value of the spin-polarization, i.e., the high-symmetry point behaves as a branch-point singularity.
\footnote{The singular structure in the results is not specific to ZBW, but is also present in NRG.}
For the E state, the trend as regards the $\phi$-dependence is, however, the opposite as for the G state: the maximum value moves from $\phi \sim \pi/2$ towards $\phi=0$ as $\lambda$ increases. 

Interestingly, the trends seen in $\langle S_x \rangle$ are not fully reflected in the behaviour of the effective SOC field. For the G state, the general trends in $\langle S_x \rangle$ and $B_\mathrm{SOC}$ are similar. For the E state, the peak in $B_\mathrm{SOC}$ as a function of $\phi$ moves toward $\phi=\pi$ with increasing $\lambda$, in distinction to the variation of $\langle S_x \rangle$. One should note that the spin polarization depends on two factors, the size of the effective field, as shown in Fig.~\ref{fig2}c,d, but also on $P_1$, on which we remarked an opposite $\phi$ dependence for G and E (see Fig.~\ref{fig1}e,f). This $P_1$ effect is strong enough to control the $\phi$-dependence of $\langle S_x \rangle$ as well.

For $\lambda\neq 0$, sub-gap states acquire an admixture of high-spin components, predominantly the YSR triplet, Eqs.~\eqref{eq:ansatzSz}, corresponding to a state in which the local moment on the QD aligns with the Bogoliubov quasiparticle spin. This is quantified through $\langle S_\mathrm{total}^2 \rangle$, which vanishes for spin singlets. Fig.~\ref{fig2}e,f reveal that the high-spin part develops with increasing $\lambda$, as expected, with weak dependence on $\phi$. The largest admixture of the YSR triplet is found for the state E at low values of $\phi$. This is expected, since this state is closest to the Bogoliubov continuum, where the YSR triplets are submerged in the absence of SOC.
We again observe singular behaviour in vicinity of $\lambda=1/2$, $\phi=\pi$.

\subsection{Pairing properties}
\label{sec:pairing}

In this section, we analyse the pairing properties of the subgap states through the $\ket{0}$ and $\ket{2}$ components. We show that Coulomb repulsion suppresses proximitized even-occupancy contributions by disfavouring charge fluctuations, while particle--hole asymmetry makes the $\ket{0}$ and $\ket{2}$ weights unequal, leading to deviations from half filling even at nominal $\nu=1$. The key sources of asymmetry are background tunnelling and spin--orbit coupling.

\begin{figure}
    \includegraphics[width=0.9\textwidth]{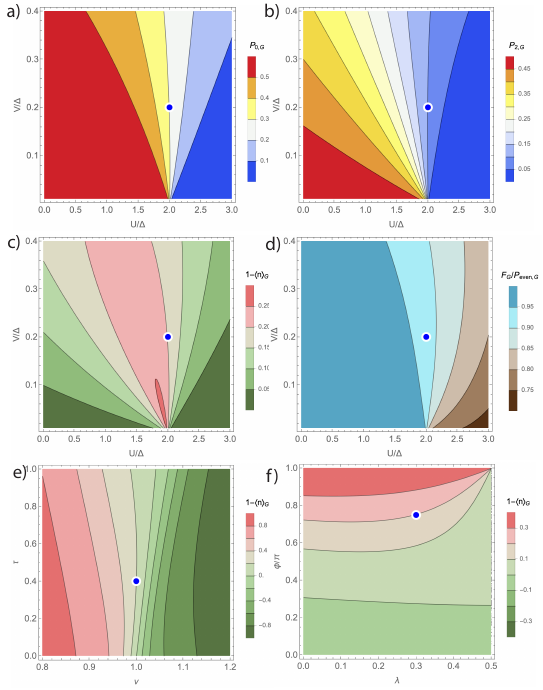}
\caption{Pairing properties of the state G; ZBW results.
(a,b) Probability of zero occupancy, $P_0$, and double occupancy, $P_2$, as a function of repulsion $U$ and hybridization $V$.
(c) Deviation from half-filling that results from these $P_0$ and $P_2$. (d) Overlap with the conventional ABS ground-state wavefunction $\psi=1/\sqrt{2}
\left( e^{i\phi/4} | 0 \rangle + e^{-i\phi/4} | 2 \rangle \right)$, normalized by the probability of even occupancy $P_e=P_0+P_2$. (e,f) Deviation from half-filling vs. $\nu$ and $\tau$, and vs. $\lambda$ and $\phi$. }
\label{fig3}
\end{figure}

We now focus on the $\ket{0}$ and $\ket{2}$ components associated with proximitized superconductivity. The lack of p-h symmetry for $\tau\neq0,\lambda\neq 1/2$ implies nonequal weights of $\ket{0}$ and $\ket{2}$ parts at the nominal half-filling point $\nu=1$. Fig.~\ref{fig3}a,b shows the $(U,V)$ dependence of the probabilities for zero and double occupancy, $P_0$ and $P_2$, in the state G. Both monotonically decrease with increasing $U$, but not at equal rates, creating imbalance between the two components. The excited state E shows the opposite pattern of $P_0$ and $P_2$ evolution, so that $P_{0,E} \sim P_{2,G}$ and $P_{2,E} \sim P_{0,G}$ (not shown). The occupancies, $\langle n \rangle = P_1  + 2 P_2$, for both states deviate from 1 at the nominal half-filling point $\nu=1$, see the case of $\langle n \rangle_G$ in Fig.~\ref{fig3}c, but so that $\langle n \rangle_G + \langle n \rangle_E \sim 2$ to a good approximation (yet not exactly).

In the absence of background tunnelling and SOC, the ground-state ABS wavefunction for low $U$ has the form
\begin{equation}
    \ket{\psi}_G = \frac{1}{\sqrt{2}} \left( e^{i\phi/4} \ket{0} + e^{-i\phi/4} \ket{2} \right).
\end{equation}
We can quantify the departure from this form by computing the expectation value of the projector to this state, $F=\langle P_\psi \rangle$. We furthermore take into account the emergence of single-electron component for finite $U$ by rescaling this quantity by $P_\text{even}=P_0+P_2$. We present a plot of $F/P_\mathrm{even}$ in Fig.~\ref{fig3}d for state G (the case of state E is similar). This quantity has a value not too far from 1 in a significant part of the $(U,V)$ plane, and even at $U\sim 2\Delta$ we still find $F/P_\mathrm{even} \approx 0.9$. This shows that despite the complex $(U,V)$ dependencies of various quantities at low $V$, and despite the lack of p-h symmetry, the nature of the state in the $\ket{0},\ket{2}$ subspace (i.e., disregarding the YSR admixture) is still essentially that of an ABS to a reasonable degree of approximation for much of the $U \lesssim 2\Delta$ range.

The line where $\langle n \rangle_G = 1$ remains parametrically close to $\nu=1$, see Fig.~\ref{fig3}e showing the deviation from 1 in the $(\nu,\tau)$ plane.
This deviation depends significantly on the phase-bias $\phi$ and somewhat more weakly on $\lambda$, see Fig.~\ref{fig3}f. In particular, the system can be tuned to half filling along a line in the $(\lambda,\phi)$ plane. For $\lambda=1/2$, the system becomes p-h symmetric. For $\lambda$ close to $1/2$, the deviation develops a singular structure: in most of the $\phi$ range it remains close to 0, but then increases suddenly as $\phi\to\pi$.

\subsection{Energy separation}

In this section, we examine when both subgap states exist within the gap and how their energy splitting evolves. The behaviour reflects the opening of a second scattering channel at finite phase bias, which stabilizes the excited state, the near-degeneracy around $U\sim2\Delta$, which enhances the splitting, detuning away from half filling, which separates the $\ket{0}$ and $\ket{2}$ sectors, and the enhanced symmetry near $\lambda=1/2,\phi=\pi$, which drives the states toward degeneracy.

\begin{figure}
\centering
    \includegraphics[width=0.9\textwidth]{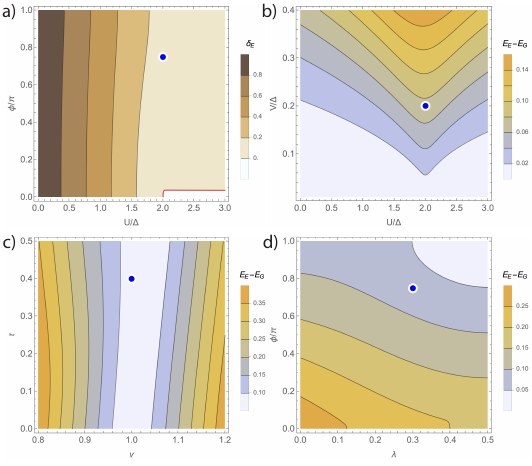}
\caption{Energies of the subgap states; ZBW results. (a) Binding energy of the excited subgap state, $\delta_E$, i.e., its position below the continuum of Bogoliubov excitation. Negative value indicates that the excited state is absorbed in the continuum (corresponding to a level crossing in ZBW approximation). The red curve indicates the locus of $\delta_E=0$. (b,c,d) Energy splitting between G and E states as a function of different parameter pairs.}
\label{fig4}
\end{figure}

Before addressing the energy separation between the states G and E, we first need to establish when both states actually exist within the gap. In the limit $\phi\to0$, this is the case only in the ABS regime. The excited state merges with the continuum of Bogoliubov states close to $U \sim 2\Delta$, see for example Fig.~6b in Ref.~\cite{ilicin2025}. For $\phi=0$, only a single linear combination of Bogoliubov states from the superconductors couples to the impurity orbital, while a second scattering channel emerges only at finite $\phi$. \footnote{
An arbitrary number of SC leads can be integrated out to define a single matrix-valued (Nambu) hybridization function that can be mapped to a single Wilson chain \cite{liu2016,Zalom2023}. In this sense, the SC-AIM is effectively a single-bath quantum impurity problem. Nevertheless, for a finite phase bias, the existence of the second scattering channel \cite{Kirsanskas2015} is encoded in the out-of-diagonal matrix elements being reduced compared to the $\phi=0$ limit by a factor of $\cos(\phi/2)$. In addition, in the presence of background tunnelling $\tau$, the hybridization matrix encodes sub-gap $\delta$-peak contributions corresponding to additional in-gap states \cite{zalom2025}.
}
The region where the E state does not exist is  rather small for generic parameters and confined to low values of $\phi$ in the $U>2\Delta$ range. For the reference values used in this work, it is restricted to $\phi/\pi \lesssim 0.05$, see Fig.~\ref{fig4}a (white region enclosed within the red curve). The same plot also reveals that the E state remains rather close to the gap edge for $U \gtrsim 2\Delta$. 

In Fig.~\ref{fig4}b,c,d we show the energy splitting $E_E-E_G$ along three cuts in the parameter space: $(U,V)$ plane, $(\nu,\tau)$ plane, and $(\lambda,\phi)$ plane for panels b,c,d, respectively. The splitting monotonously grows with $V$, with the largest splitting for a given value of $V$ occurring for $U \sim 2\Delta$  (see panel b): this directly reflects that fact that this region is controlled by the singularity associated with increased state degeneracy\cite{ilicin2025}, with two ABS states being resonant with the divergently dense Bogoliubov states at the band edge, a feature that is captured at qualitative level even within the ZBW approximation \cite{zitko2022}. The splitting grows as one moves away from half filling ($\nu=1$), since detuning enhances the energy difference beyond that induced by the proximity effect alone by altering the relative weights of the $\ket{0}$ and $\ket{2}$ components (see panel c). At the chosen reference point, the background tunnelling ratio $\tau$ has little effect on the energetics. Finally, the plot in panel d directly reveals the approach to the degeneracy point $E_G=E_E$ at $\lambda=1/2,\phi=\pi$ due to the enhanced symmetry. 

\subsection{Transition matrix elements}
\label{sec:matel}

In this section, we analyse transitions between the subgap states in the charge, spin, and current channels. We show that the corresponding matrix elements have distinct parameter dependences because they probe different aspects of the evolving subgap-state character. Charge transitions weaken toward the YSR regime, spin transitions grow with the emergence of spin polarization, and inductive transitions track the phase dispersion and level splitting of the states.

\begin{figure}
\centering
    \includegraphics[width=0.9\textwidth]{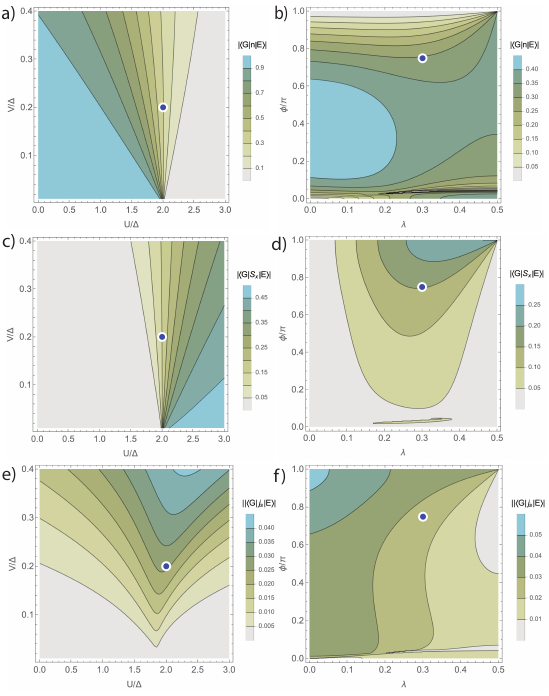}
\caption{Matrix elements for transitions between states G and E for the QD charge operator $n$, the QD spin operator $S_x$, and the Josephson supercurrent, here defined as the dimensionless quantity $j_s=\langle \partial H/\partial \phi \rangle /\Delta$; ZBW results. (a,c,e) Dependence on the repulsion $U$ and the hybridization $V$. (b,d,f) Dependence on the SOC parameter $\lambda$ and the phase bias $\phi$.}
\label{fig5}
\end{figure}

Fig.~\ref{fig5}a shows the expected trend towards insensitivity to gate-potential driven transitions as the system moves from the ABS to the YSR regime; this plot simply reflects the growth of single electron occupancy quantified by $P_{1,G}$ and $P_{1,E}$ and thus strongly resembles Fig.~\ref{fig1}a,b. Correspondingly, Fig.~\ref{fig5}c shows how spin transitions become possible for strong enough interaction $U$. This crossover in spin channel is more abrupt than the one in the charge channel. This can be ascribed to the properties of the state E: the expectation value $\langle S_x \rangle_E$ shows a similar threshold behaviour as a function of $U$, whereas $\langle S_x \rangle_G$ shows a smoother evolution (plots not shown). We note again that in the absence of SOC, background tunnelling, or finite $\phi$, this matrix element would be zero.
Fig.~\ref{fig5}e shows how for inductive transitions (flux driving \cite{Park2017,park2020,metzger2021,Fauvel2024}) the evolution in the $(U,V)$ plane reflects the dispersion of states which becomes larger as $V$ increases, especially in the $U \approx 2\Delta$ range; since strong dispersion also implies large level separation, this plot resembles that of $E_E-E_G$ in Fig.~\ref{fig4}b.

Fig.~\ref{fig5}b,d,f reveal that charge transitions are strong for intermediate values of $\phi \sim \pi/2$ with only a weak $\lambda$-dependence that slightly favours small SOC, magnetic transitions are favoured by strong SOC and phase bias close to $\pi$, while inductive transitions are favoured by weaker SOC and phase bias close to $\pi$. It should be noted that for $\phi=\pi$ the charge transitions are forbidden.
This is because $\ket{G}$ has only $\ket{0}$ component, while $\ket{E}$ has only $\ket{2}$ component in this limit, which occurs at $\phi=\pi$ as a result of cancellation of induced pairing on the QD dot site, thus any finite $\tau$ (or $\nu\neq1$) will split the degenerate subgap states into unmixed $\ket{0}$ and $\ket{2}$. 
At $\phi=\pi$, the relative strength of spin and inductive transitions depends on $\lambda$. Similarly, one can find parameter choices where either spin or inductive transitions become negligible. Finally, there are clearly regimes where all three couplings are large. These tuning capabilities might be of interest, for example, for applications in quantum transduction.

\subsection{Distinct crossovers of two subgap states}

In this section, we analyse how detuning affects ABS- and YSR-like states differently. By unbalancing the $\ket{0}$ and $\ket{2}$ components, it strongly shifts ABS-like states, while YSR-like states, dominated by single occupancy, are affected much less. This creates an intermediate region in which G and E have different character, with clear consequences for the level response and transition matrix elements.

Away from half filling, G and E states change their character from ABS-like to YSR-like at different values of $U$. This happens because for $\nu \neq 1$, either zero or double occupancy of the dot level is preferred, thus the energy balance of the ABS states will change significantly as $\nu$ is tuned away from 1 because they will have unequal $\ket{0}$ and $\ket{2}$ compositions, whereas the subgap states in the YSR regime that favour single occupancy will experience much smaller energy shifts. This implies that there is a range of $U$ values in vicinity of $U=2\Delta$ where the E and G states will have dissimilar characters ($G$ will be ABS-like and $E$ will be YSR-like), which has implications for inter-level transitions.

\begin{figure}
    \centering
    \includegraphics[width=0.75\linewidth]{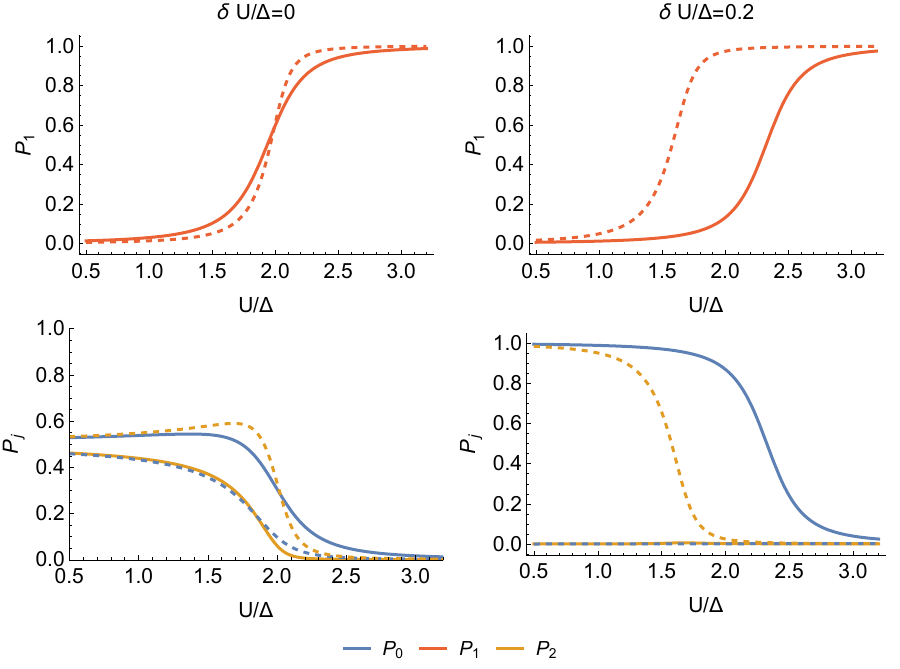}
    \caption{Probability of single (a, b), and zero/double dot level occupancy (c, d) in the G (solid) and E (dashed) states. Coupling strength is $V /\Delta = 0.1$, other parameters take reference values. Sufficiently away from half-filling (b,d), states are completely polarized, and the crossover between ABS and YSR regimes occurs at different values of $U$. The dimensionless parameter $\delta = \nu - 1$ quantifies the deviation from the particle-hole symmetry.}
    \label{fig:Pj_no-ph}
\end{figure}

In Fig.~\ref{fig:Pj_no-ph} we first compare the $U$-dependence of the local-moment fraction $P_1$, as well as the zero and double occupancy probabilities $P_0$ and $P_2$, for $\delta=0$ and $\delta\neq 0$, where $\delta = \nu - 1$. For $\delta \neq 0$, the excited state becomes YSR-like at significantly lower value of $U$, as a consequence of the large charge polarization ($P_0/P_2$ imbalance), see Fig~\ref{fig:Pj_no-ph}d. 

\begin{figure}
    \centering
    \includegraphics[height=0.37\linewidth]{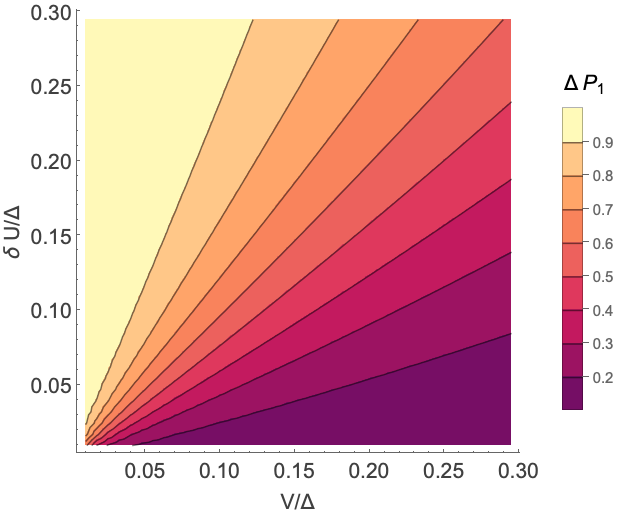}
    \includegraphics[height=0.37\linewidth]{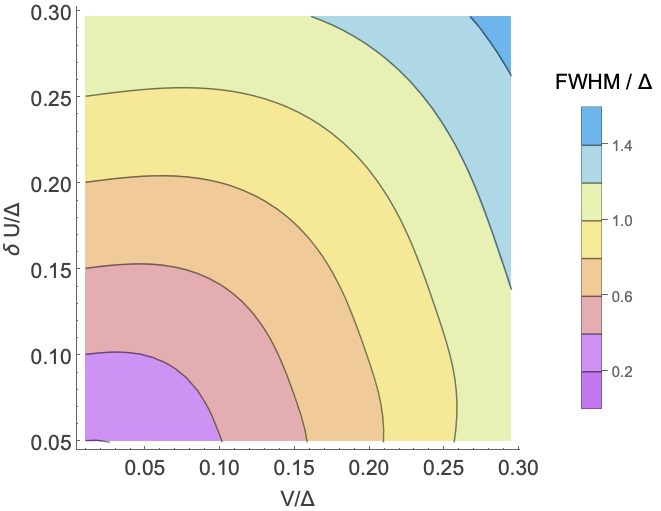}
    \caption{Characterisation of the difference in local moment fractions of $E$ and $G$ states, $\Delta P_1 (U) = P_{1,E} (U) - P_{1,G} (U)$, as function of $V$ and $\delta$. Other parameters are considered constant and take the reference values. a) Maximum value of $\Delta P_1 (U)$ (i.e. strongest difference of ABS vs. YSR character), and b) Full width at half maximum of $\Delta P_1 (U)$ (i.e., the width of the region where E and G have different character).}
    \label{fig:deltaP1}
\end{figure}

The difference of character of G and E states is the most pronounced at small values of $V$, when the crossover is more sudden for both states and the region in which the states have mixed character is narrower. In Fig.~\ref{fig:deltaP1}, we identify the existence and quantify the width of this region via $\Delta P_1 (U) = P_{1,E}(U) - P_{1,G} (U)$. In Fig~\ref{fig:deltaP1}a, we show the maximum deviation\footnote{Here, $P_1$ is treated as a function of $U$ only. The dependence on other parameters is weak, and they are considered constant when maxima and FWHM are considered.} of the local-moment fraction as a function of $\delta = \nu - 1$ and $V$. Not surprisingly, $\Delta P_1$ increases with increasing $\delta$. Furthermore, at large values of $V$, $\max\Delta P_1 (U) < 1$, since the crossover is more gradual, and both states have mixed character in the region of interest. We also show the full width at half maximum of $\Delta P_1$, which increases both with $\delta$ and $V$. The latter trend should be taken with a grain of salt, as at large $V$, gradual transition between ABS and YSR regimes destroys the distinction in the nature of $G$ and $E$ states.

\begin{figure}
    \centering
    \includegraphics[width=0.85\linewidth]{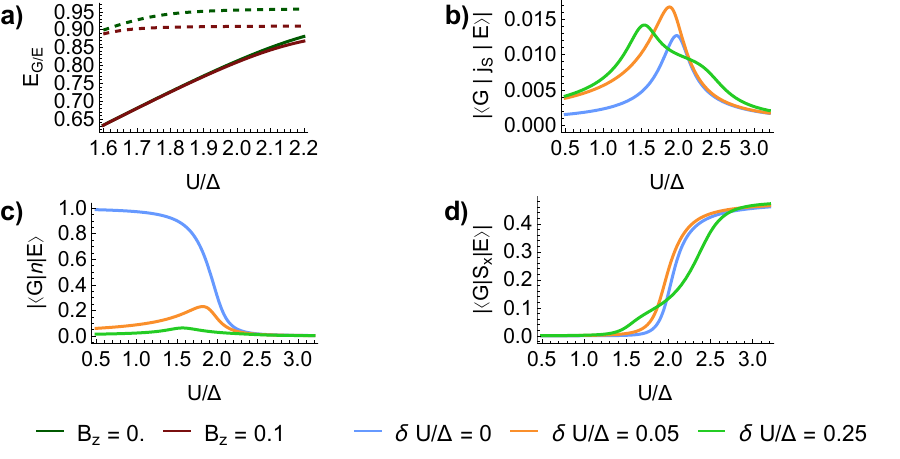}
    \caption{Consequences of distinct crossover regions of $G$ and $E$ states. a) Energy of $G$ (solid lines) and $E$ (dashed lines) at (non)zero external magnetic field. For values of $U$ on the plot, $E$ has significantly higher $P_1$ resulting in high effective $g$ factor (stronger splitting) relative to $G$. (b), (c), (d) Transition matrix elements for operators $j_S, n$ and $S_x$. At large $\delta$ different characters of $G$ and $E$ result in qualitatively different behaviour of transition matrix elements around $U=2\Delta$. In all plots, $V=0.1 \Delta$, and in (a) $\delta U =0.15 \Delta$.}
    \label{fig:mels_nophi}
\end{figure}

The consequences of the distinct crossovers are shown in Fig.~\ref{fig:mels_nophi}. At large values of $\delta$, the matrix elements of the charge operator vanishes in the ABS region due to (nearly) total charge-polarization of the states. In the intermediate region, the same matrix elements are suppressed because of the different natures of $G$ and $E$. For the same reason, matrix elements of $j_S$ and $S_x$ show different behaviour far away from nominal half-filling ($\delta U= 0.25 \Delta$). Finally, when $\Delta P_1 (U)$ is high, the response to external magnetic fields of $E$ can be significant, while $G$ can remain more or less insensitive (see Fig.~\ref{fig:mels_nophi}a). 

\subsection{NRG validation}
\label{sec:nrg}

In this section, we compare the ZBW results with NRG calculations for the full continuum model. We show that ZBW captures the main qualitative features of the ABS--YSR crossover because much of the relevant physics is controlled by local charge and occupancy effects

While NRG calculations for the full model with SOC and two SC leads with a phase bias are numerically very costly, we have been able to perform a systematic study that supports the qualitative correctness of the results obtained using the ZBW approximation in the proceeding subsections. Because the superconductors are now described using continuum Hamiltonians, the hybridization and background tunnelling terms are defined differently: the hybridization will be expressed in terms of the hybridisation strength $\Gamma=2\pi\rho V^2$, where $\rho$ is the normal-state density of states of each lead, while the background tunnelling will be quantified through $\tau'$ defined as $V^{(D)}=\tau' \sqrt{\Gamma D}$, where $D$ is the half-bandwidth of the bath. Other parameters are defined as before and we used the same values as in the reference parameter set for the ZBW approximation in the earlier sections. 

\begin{figure}[htbp]
\centering
     \includegraphics[width=\textwidth]{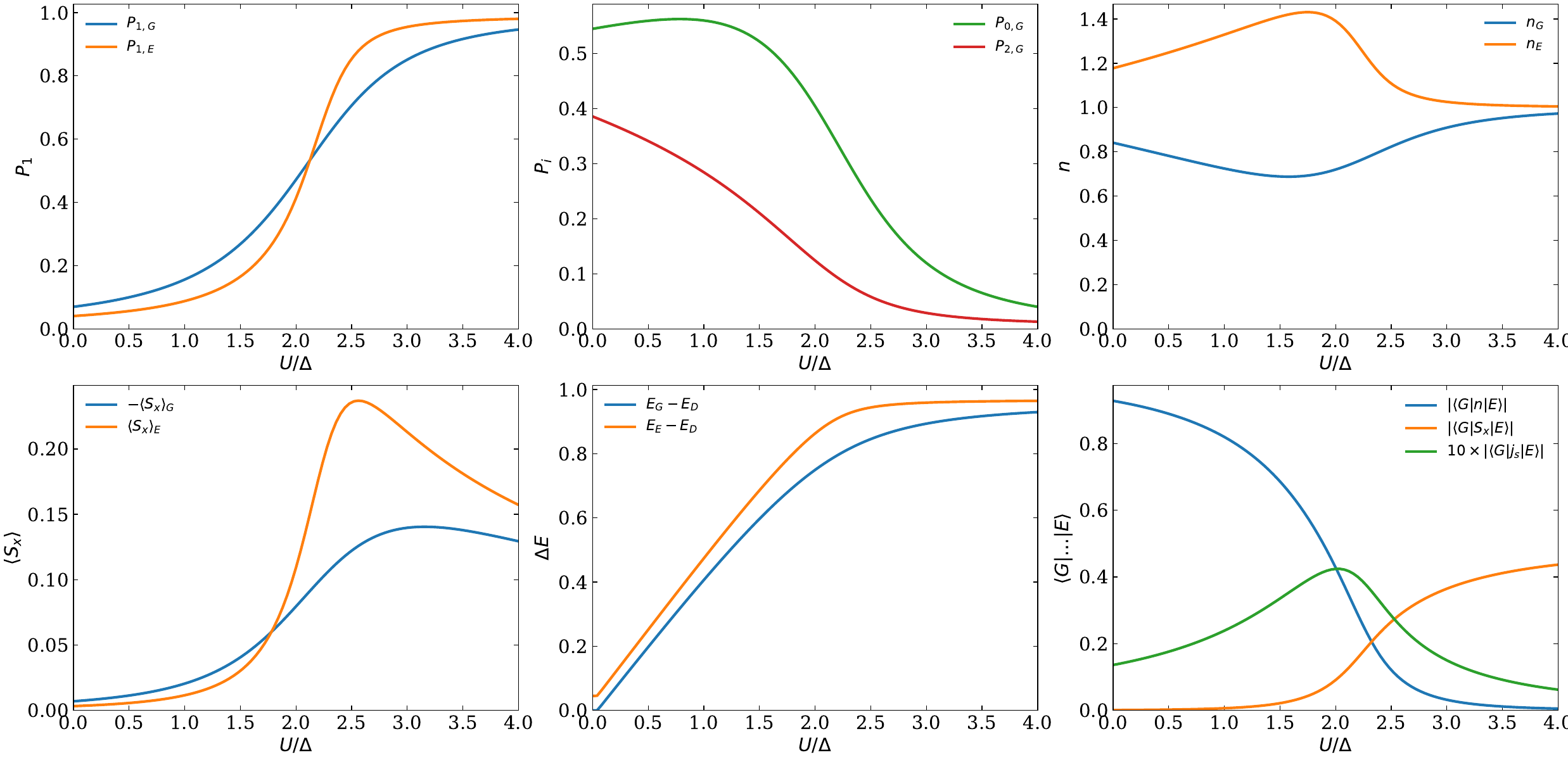}
\caption{NRG results: $U$-dependence of key properties for $\Gamma/\Delta=0.1$, $\nu=1$, $\phi=0.75\pi$, $\lambda=0.3$, and $\tau'=3$.}
\label{figA}
\end{figure}

In Fig.~\ref{figA} we plot the $U$-dependence of key quantities. This calculation is performed for intermediately strong coupling $\Gamma=0.1\Delta$; this value is typical of hybrid nanowire devices \cite{coulomb2}. The local-moment fraction $P_1$ evolves from values close to 0 to values close to 1 as the system evolves from the ABS to YSR regime (panel a); as in Fig.~\ref{fig1}(a,b) we find that the state E crosses over faster than state G. The projectors $P_{0,G}$ and $P_{2,G}$ (panel b) also generally follow the behaviour seen for the ZBW in Fig.~\ref{fig3}(a,b). Likewise, the occupancy in state G is suppressed for $U \lesssim 2\Delta$, with a local minimum at $U \approx 1.5\Delta$ (panel c), similar to what is seen in Fig.~\ref{fig3}(c); the state E shows increased local occupancy in this range. The spin polarization (panel d) increases for $U \sim 2\Delta$ and it peaks somewhere in the $U>2\Delta$ range; this is particularly pronounced for the state E. The same behaviour is found in the ZBW model (results not shown).
From the energy positions with respect to the ground state (panel e) we read off that the maximum level separation occurs for $U \gtrsim 2\Delta$, in agreement with Fig.~\ref{fig4}(b). Finally, the matrix elements (panel f) show the same trends as the ZBW prediction in Fig.~\ref{fig5}(a,c,e).

\begin{figure}
\centering
    \includegraphics[width=0.7\textwidth]{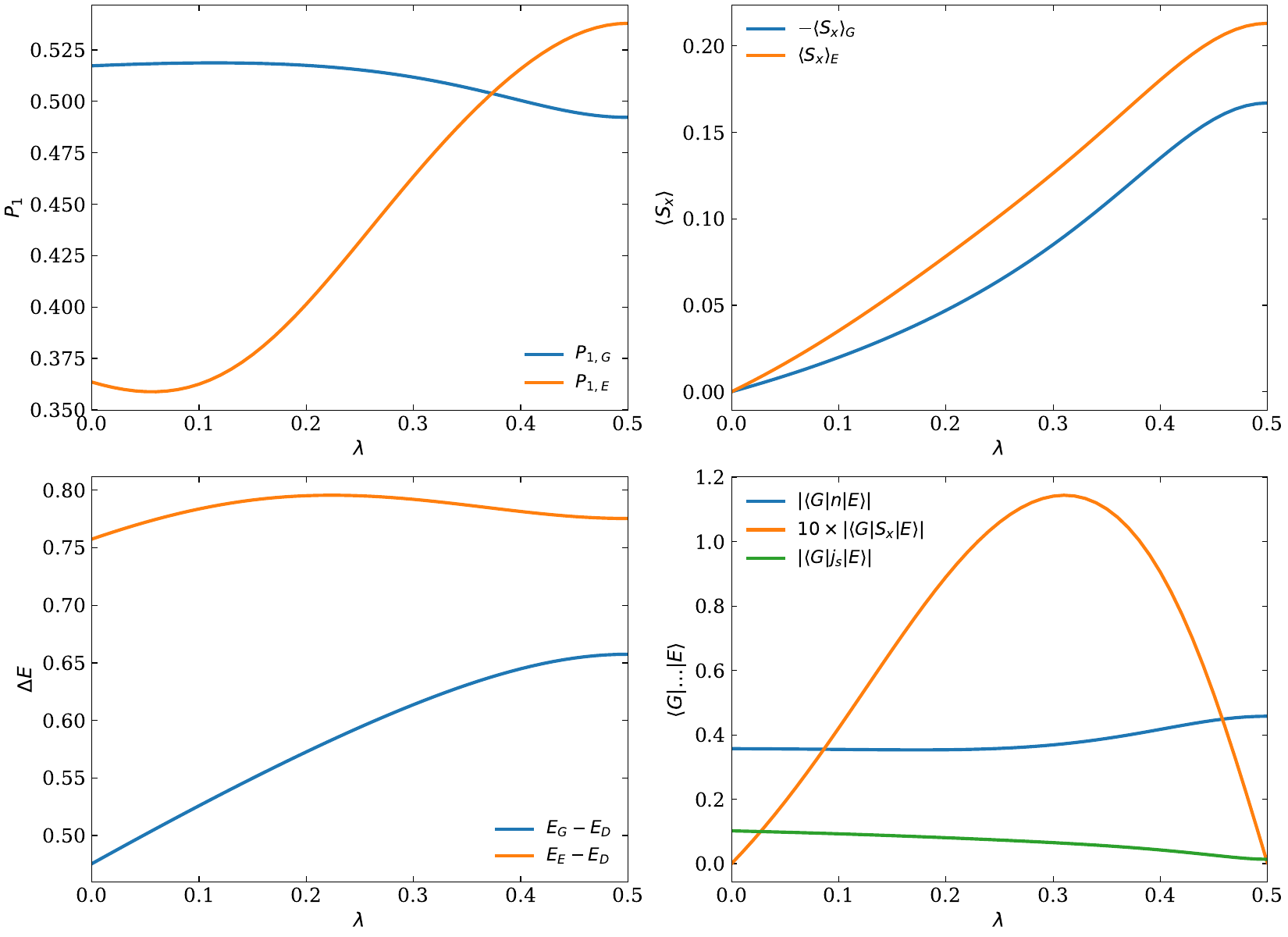}
\caption{NRG results: $\lambda$-dependence for $U/\Delta=2$, $\Gamma/\Delta=0.2$, $\nu=1$, $\phi=0.75\pi$, and $\tau'=3$.}
\label{figB3}
\end{figure}

In Figs.~\ref{figB3} we address the SOC dependence. We find the same trends as in ZBW, compare with Figs.~\ref{fig2}(a,b), \ref{fig4}(d) and \ref{fig5}(b,d,f), down to certain details, for example $|\langle G|n|e \rangle|$ first being relatively flat, and then increasing, which is indeed the behaviour seen in ZBW model in the same range of $\phi$.

\begin{figure}[htbp]
    \centering
    \includegraphics[width=0.65\linewidth]{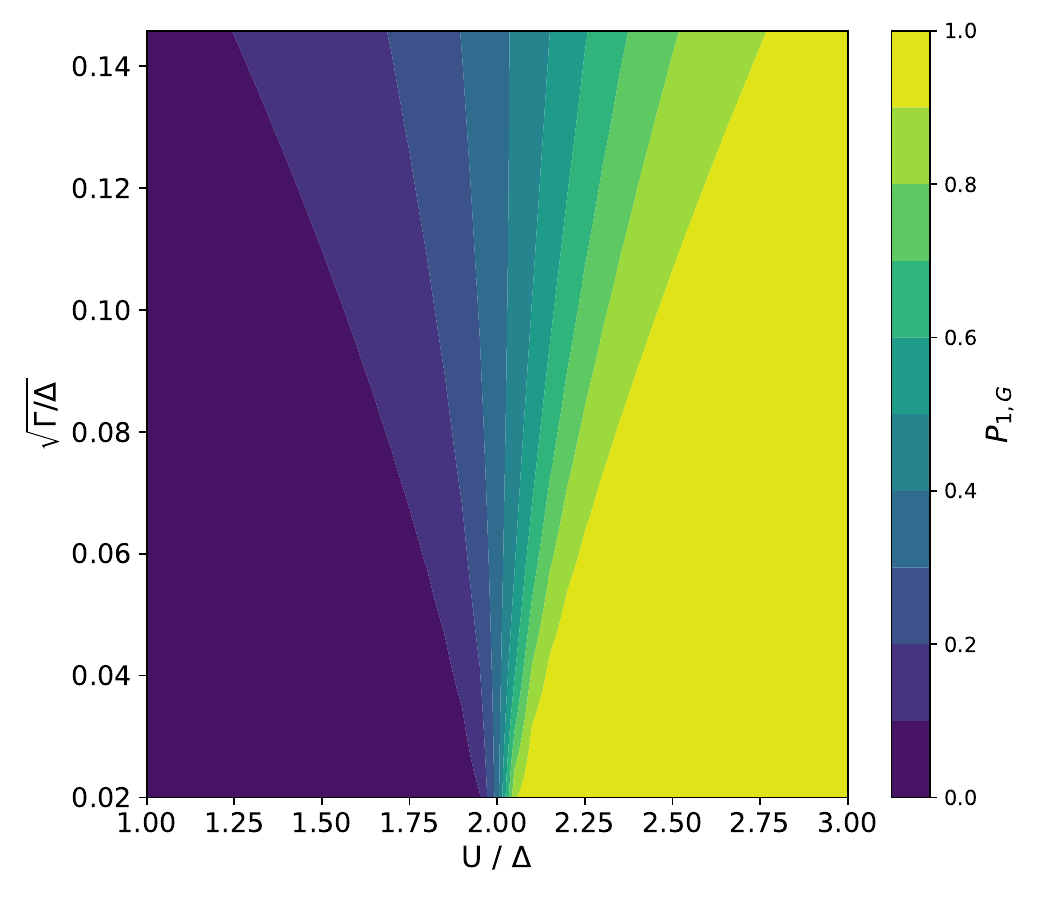}
    \caption{NRG results: local-moment fraction $P_{1,G}$ in the $(U,\sqrt{\Gamma})$ plane at half-filling, $\nu=1$, and no SOC, $\lambda=0$. The single-channel case is shown, corresponding to $\tau, \phi = 0$. Lines of constant $P_{1,G}$ are almost linear in the range shown, qualitatively matching the ZBW results (see Fig.~\ref{fig2}).}
    \label{fig:P1G_NRG}
\end{figure}
In Fig.~\ref{fig:P1G_NRG} we show $P_{1, G}$ in the $(U,\sqrt{\Gamma})$ plane for the single-channel ($\phi, \tau=0$) case with no SOC, $\lambda =0$. By comparing the results with Fig.~\ref{fig1}, we conclude that the qualitative behaviour in the $(U,\sqrt{\Gamma})$ plane can be correctly captured by the ZBW approximation. Most features from Fig.~\ref{fig:P1G_NRG} extend to the general two-channel case.

\begin{figure}  
\centering
    \includegraphics[width=0.9\textwidth]{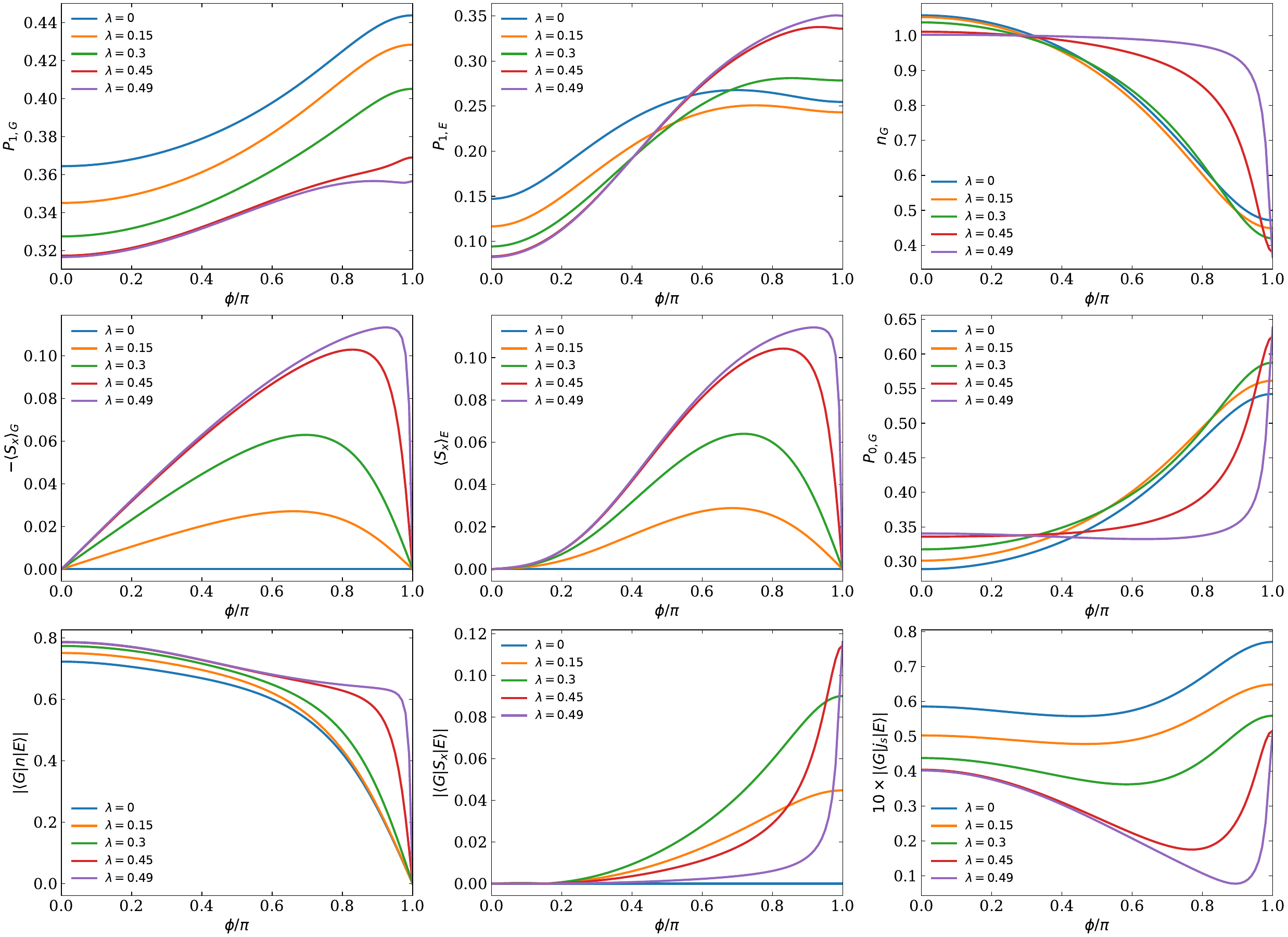}
\caption{NRG results: $\phi$-dependence for $U/\Delta=2$, $\Gamma/\Delta=0.1$, $\nu=1$, $\lambda=0.3$, and $\tau'=3$.}
\label{figphi}
\end{figure}

In Fig.~\ref{figphi} we plot phase-bias dependences for a range of $\lambda$ that also includes some values close to $\lambda^*=1/2$. Particularly noteworthy are the results for the occupancy, $n_G$, whose drop to small values becomes particularly sharp as $\lambda \to \lambda^*$. This explains anomalies seen in all other quantities, for instance the drop in spin polarization and matrix element for the charge operator.

\begin{figure}
\centering
    \includegraphics[width=0.65\textwidth]{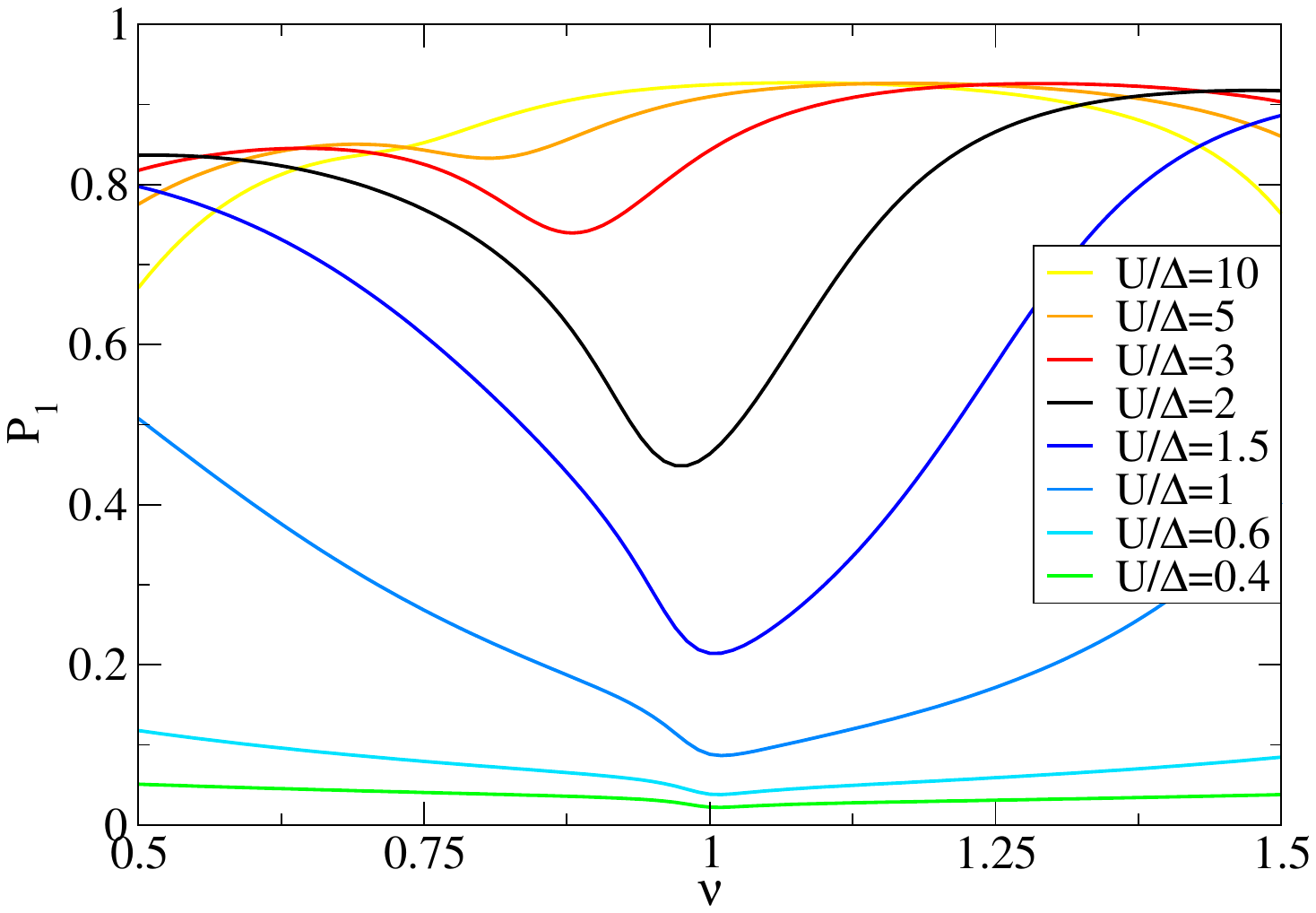}
\caption{NRG results: $\nu$-dependence of the local-moment fraction $P_1$ in the excited state E for a range of $U/\Delta$, showing a {\it minimum} for $\nu \sim 1$ in the cross-over range $U/\Delta \sim 2$. Other parameters are fixed: $\Gamma/\Delta=0.1$, $\phi=0.75\pi$, $\lambda=0.3$, and $\tau'=3$.}
\label{figeps}
\end{figure}

Finally, we return to the gate-voltage dependence of the local-moment fraction $P_1$, where we remarked unusual behaviour for the state E. We systematically investigated this quantity for a range of $U/\Delta$ ratios, see Fig.~\ref{figeps}. The results show that in a range of $U/\Delta$ spanning the ABS-YSR crossover region this quantity has a minimum value in vicinity of $\nu=1$, where a maximum value is generally expected for low-energy states. This minimum becomes well developed with increasing $U$ as soon as $P_1$ starts to become appreciable, for $U/\Delta \sim 1$. It fades away deeper in the YSR regime; it is hardly visible at $U/\Delta \sim 5$. This is thus indeed a particularity of the cross-over region.

Taken together, these results establish that the ZBW correctly reproduces all key features of the ABS-YSR crossover at the qualitative level. This is due to charge physics and level occupancies being controlled by the high energy scales and local effects, which are captured adequately even by truncated-basis models. Nevertheless, we note that the use of the NRG is essential both in obtaining quantitatively correct results, and in capturing all qualitative features of the model. Since the continuum effects are not included, phenomena like the anomalous $\Gamma^{2/3}$ scaling of binding energies \cite{ilicin2025} cannot be produced within the ZBW framework. In such cases, either NRG or a variational calculation (see Appendix~\ref{app:var} and Ref.~\cite{ilicin2025}) is required.

\section{Experimental observability and design implications}

The reference parameter set is broadly consistent with typical values for InAs/Al nanowires embedded in transmon circuits in Refs.~\cite{bargerbos2022,Bargerbos2023,pitavidal2023}. Our results therefore constitute a prediction of physical effects that could, in principle, be observed in the same or similar devices, provided the system is initialized in the even-parity subspace. In such devices, transitions can be driven in two ways: via microwave excitation through the feedline, or by direct modulation of the QD gate \cite{Bargerbos2023}. Feedline-driven excitations couple to matrix elements of the Josephson current, i.e. to phase-dependent transitions, while gate driving can act through several channels, including spin via electric-dipole spin resonance (EDSR) \cite{Rashba2003,Flindt2006,Golovach2006,Nowack2007}. At present, however, the relative importance of these mechanisms remains unclear. Further controlled experiments, combined with detailed theoretical modelling, could help identify the relevant processes.

To experimentally distinguish between ABS and YSR states, and detect the ABS-YSR mixing effects in the intermediate region, one should measure the local charge susceptibility (quantum capacitance). This is possible using RF reflectometry in dispersive gate-sensing configuration \cite{vanVeen2019,malinowski2022}.

The results presented in the previous section also have some clear design implications. In particular, for qubits optimized for charge coupling with minimal magnetic sensitivity, it is crucial to remain in the ABS-dominated regime with $U \ll 2\Delta$, with small or moderate $\lambda$. For devices aiming to exploit spin-selective transitions, the crossover regime $U \approx 2\Delta$ with moderate $\lambda$ and $\tau$ is optimal. In fact, device platforms that are suitable for building ASQs would also be appropriate for these applications.  It should be noted, of course, that in this parameter regime, the overall ground state of the system is in the odd-parity sector (spin-doublet).
A long parity lifetime is thus required.

\section{Conclusion}

We have studied the superconducting Anderson impurity model for an interacting quantum dot Josephson junction with spin-orbit coupling. We uncovered some noteworthy properties: 1) mixing of ABS and YSR leads to properties that are not typical for pure charge qubit states, because the even-parity states carry local moments; 2) in the presence of SOC, both subgap states are spin polarized in the absence of magnetic field; 3) matrix elements are large in charge (capacitive), spin (magnetic), and current (inductive) channels, with the possibility of tuning the relative values.

An important next step is to extend this analysis to the multi-orbital junctions with several "active" levels, which are directly relevant for devices exhibiting several ABS pairs in the relevant frequency window.

\section*{Acknowledgments}

\paragraph{Funding information}
We acknowledge the support of the Slovenian Research and Innovation Agency (ARIS) under P1-0416 and J1-3008, and discussions with Don Rolih and Peter Zalom, in particular on the topic of symmetry properties of impurity Hamiltonians.

\begin{appendix}
\numberwithin{equation}{section}

\section{Estimate of interlead hopping parameter $\VD$}
\label{app:VD}

The effective direct tunneling between the superconducting leads originates from higher-energy inactive orbitals of the quantum dot. The amplitude $\VD$ can be estimated using second-order perturbation theory \cite{Bargerbos2023}:
\begin{equation}
    \VD \approx \sum_i \frac{|V_i|^2}{\delta_i},
\end{equation}
where $\delta_i = E_i - \epsilon_0$ is the energy of the $i$-th orbital, measured from the energy of the active channel, and $V_i$ is the tunneling amplitude from the leads (assumed equal for $L$ and $R$) to the orbital in question. To estimate $\VD$,  we assume $V_i = V$, neglect spin-flip processes, and model the QD as a particle in a box. The direct-hopping amplitude is then approximated as:
\begin{equation}
    \VD = \frac{|V|^2}{\epsilon_{b}} \sum_{n_x,n_y,n_z\neq(1,1,1)} \frac{1}{n_x^2 + n_y^2 + n_z^2 - 3} \>, 
    \label{eq:appVD}
\end{equation}
where $\epsilon_b = \hbar^2 \pi^2 / 2m^* L^2$, and the $3$ comes from the energy of the ground state. We assume cubic geometry with side $L \approx 100 \mathrm{nm}$ and take $m^* = 0.023 m_e$, the effective mass of electrons in bulk InAs, which yields $\epsilon_b \approx 1.6 \mathrm{meV}$. The sum in Eq.~\eqref{eq:appVD} does not converge, so we chose a cut-off at $n_{x,y,z} \leq 15$ that corresponds to the factor $\sim 20$. The cut-off is chosen so that the energy of the highest level taken into account ($n_x=n_y=n_z=5$) has a value $\sim 1 \mathrm{eV}$, comparable to the bandwidth of the conduction band in InAs. Finally, taking the reference value $V = 0.2 \Delta \approx 0.05 \mathrm{meV}$, we arrive at the estimate $\tau \approx 0.7$, close to the chosen value for the reference set (0.4). The actual magnitude of $\tau$ will of course strongly depend on many factors, such as the specific shape of the dot, and the concrete values of tunneling amplitudes for each channel. Therefore, the value obtained here should be understood as a rough estimate of the order of $\tau$, and the main takeaway of this calculation is that $\VD$ can be comparable to $V$, even when level separation of dot orbitals is the largest energy scale in the problem ($ \epsilon_b > 5 \Delta$). We also note that an interorbital Hartree term would lead to an additional correlated direct hopping term in the effective Hamiltonian. However, quantitative details related to these mechanisms are beyond the scope of this work.

\section{Variational approach}
\label{app:var}

\subsection{Variational equations for conventional quantum-dot Josephson junction}
\label{appA1}

We first consider the conventional quantum-dot Josephson junction problem with no SOC ($\lambda=0$) and no direct tunneling ($\tau=0)$.
To make contact with our previous work on the single-channel impurity problem  \cite{ilicin2025}, in the first step we address the half-filling case ($\nu=1$). Making use of symmetry, we will reduce the minimization problem to a single simple transcendental equation for the energy similar to that in the single-channel case.

The Bogoliubov quasiparticle operators in the Ansatz from Sec.~\ref{sec:var} are defined as
\begin{align*}
    \gamma_{lk\up} &= u_{lk} c_{lk\up} - v_{lk} c_{l-k\dw}^\dag \\
    \gamma_{l-k\dw}^\dag &= v_{lk}^* c_{lk\up} + u_{lk}^* c_{l-k\dw}^\dag \>,
\end{align*}
with symmetric phases in the BCS coherence factors to simplify subsequent expressions:
\begin{align*}
    u_{lk} &= e^{-i \phi_l / 2} \sqrt{\frac{1}{2} \left( 1 + \frac{\epsilon_k}{E_k} \right)}, \\ 
    v_{lk} &= e^{i \phi_l /2} \sqrt{\frac{1}{2} \left( 1 - \frac{\epsilon_k}{E_k} \right)} \>.
\end{align*}
Here, $E_k = \sqrt{\Delta^2 + \epsilon_k ^2}$ is the dispersion of Bogoliubons. 
We choose a symmetric gauge $\phi_R = - \phi_L = \phi / 2$.

Minimization of the functional $f \equiv \expval{H}{\Psi} - \epsilon \braket{\Psi}$ yields the following variational equations for the singlet-state energy $\epsilon$:
\begin{align}
    \left( \epsilon - \frac{U}{2} \right) a^\pm &= \frac{\VN_L}{\sqrt{N}} \sum_{k} \alpha_{Lk} (v_{Lk}^* \pm u_{Lk}^*) + \frac{\VN_R}{\sqrt{N}} \sum_{k} \alpha_{Rk} (v_{Rk}^* \pm u_{Rk}^*) \>, \label{eq:minimization1} \\ 
    \left( \epsilon - E_{Lk} \right) \alpha_{Lk} &= \frac{\VN_L}{\sqrt{N}} a^{+} (v_{Lk} + u_{Lk})
    + \frac{\VN_L}{\sqrt{N}} a^{-} (v_{Lk} - u_{Lk}) \nonumber \\  
    &+ \frac{\VN_L}{\sqrt{N}} \sum_{k'} (u_{Lk'}^* - v_{Lk'}^*) \beta_{LLkk'}^+ + \frac{\VN_L}{\sqrt{N}} \sum_{k'} (u_{Lk'}^* + v_{Lk'}^*) \beta_{LLkk'}^- \nonumber \\ 
    &+  \frac{\VN_R}{\sqrt{2N}} \sum_{k'} (u_{R k'}^* - v_{R k'}^*) \beta_{LRkk'}^+ + \frac{\VN_R}{\sqrt{2N}} \sum_{k'} (u_{Rk'}^* + v_{Rk'}^*) \beta_{LRkk'}^- 
    \label{eq:minimization2} \>,\\     
    \left( \epsilon - E_{Rk} \right) \alpha_{Rk} &= \frac{\VN_R}{\sqrt{N}} a^{+} (v_{Rk} + u_{Rk})
    + \frac{\VN_R}{\sqrt{N}} a^{-} (v_{Rk} - u_{Rk}) \nonumber \\  
    &+ \frac{\VN_R}{\sqrt{N}} \sum_{k'} (u_{Rk'}^* - v_{Rk'}^*) \beta_{RRkk'}^+ + \frac{\VN_R}{\sqrt{N}} \sum_{k'} (u_{Rk'}^* + v_{Rk'}^*) \beta_{RRkk'}^- \nonumber \\ 
    &+  \frac{\VN_L}{\sqrt{2N}} \sum_{k'} (u_{L k'}^* - v_{L k'}^*) \beta_{LRk'k}^+ + \frac{\VN_L}{\sqrt{2N}} \sum_{k'} (u_{Lk'}^* + v_{Lk'}^*) \beta_{LRk'k}^- 
    \label{eq:minimization2RR} \>
\end{align}
\begin{align}
    \left( \epsilon - \frac{U}{2} - (E_{lk} + E_{{lk'}}) \right) \beta_{llkk'}^\pm &= 
     \frac{\VN_l}{2\sqrt{N}} \alpha_{lk} (u_{lk'} \mp  v_{lk'}) + \frac{\VN_l}{2\sqrt{N}} (u_{lk} \mp v_{lk}) \alpha_{lk'} 
     \label{eq:minimization3} \>, \\
    \left( \epsilon - \frac{U}{2} - (E_{Lk} + E_{{Rk'}}) \right) \beta_{LRkk'}^\pm &= 
     \frac{\VN_R}{\sqrt{2N}} \alpha_{Lk} (u_{Rk'} \mp  v_{Rk'}) + \frac{\VN_L}{\sqrt{2N}} (u_{Lk} \mp v_{Lk}) \alpha_{Rk'} 
     \label{eq:minimization4} \>.
\end{align}

By substituting the variational coefficients $a^\pm$ and $\beta$ from Eqs.~\eqref{eq:minimization1},\eqref{eq:minimization3},~\eqref{eq:minimization4} we can obtain equations for $\alpha_{Lk}, \alpha_{Rk}$ alone. To simplify notation, we switch to continuous energy domain ($\epsilon_k \rightarrow x$), and assume constant density of states $\rho = 1 / 2D$. We introduce functions $s_l(x) = v_l(x) + u_l(x)$ and  $a_l(x) = v_l(x) - u_l(x)$ for (anti)symmetric linear combination of coherence factors. We set $\VN_L=\VN_R \equiv V$ and define the total hybridization strength as $\Gamma=2\pi \rho V^2$.

The functions $s_l(x)$ are even with respect to the $\eta$ symmetry: $s_R(-x) = s_L(x)$, while $a_l(x)$ are odd: $a_R(-x)=-a_L(x)$. The transformation acts on the coherence factors by inverting the energy ($x\rightarrow -x$) and changing the label $L$ to $R$ and vice-versa. The symmetry relations are hence a direct consequence of $v_l(x) = u_l(-x)$ and $\phi_L=-\phi_R$. In practice, the symmetry can be utilized to reduce the system to separate equations for $\eta$-even and $\eta$-odd functions 
\begin{equation}
\alpha_e(x) = \alpha_L(x) + \alpha_R(-x)     
\end{equation}
and 
\begin{equation}
\alpha_o(x) = \alpha_L(x) -  \alpha_R(-x).
\end{equation}
As in Ref.~\cite{ilicin2025} we introduce the function $N_1(x)=\epsilon - E(x) - (\Gamma / \pi) I_2(x, \epsilon)$ with the leading two-QP energy contribution 
\begin{equation*}
    I_2(x, \epsilon) = \int \frac{\mathrm{d}x'}{\epsilon - U/2 - (E(x) + E(x'))},
\end{equation*}
as well as $N_2(x,x')=\epsilon-U/2-\left[E(x)+E(x')\right]$. Finally, we define the integrals
\begin{equation}
    C_{e} = \int s_L^*(y) \alpha_{e}(y) \mathrm{d}y,
    \quad
    C_{o} = \int a_L^*(y) \alpha_{o}(y) \mathrm{d}y.
\end{equation}
The resulting equations can now be written as
\begin{align}
   N_1(x) \alpha_e(x) &=
   \frac{\Gamma}{\pi} s_L(x)  \frac{C_e}{\epsilon - U / 2} +
   \frac{\Gamma}{2\pi} s_L(x) \int \frac{\alpha_e(x') s_L^*(x') \mathrm{d}x'}{N_2(x,x')}, 
      \label{eq:alpha_L} \\
   N_1(x) \alpha_o(x) &=
   \frac{\Gamma}{\pi} a_L(x)  \frac{C_o}{\epsilon - U / 2} + 
   \frac{\Gamma}{2\pi} a_L(x)  \int \frac{\alpha_o(x') a_L^*(x') \mathrm{d}x'}{N_2(x,x')}.
   \label{eq:alpha_R} 
\end{align}
In the following, we drop the subleading terms that correspond to small quantitative corrections which do not significantly affect the range of validity of the variational solution \cite{ilicin2025}, and the equations reduce to
\begin{align}
   N_1(x) \alpha_e(x) &= \frac{\Gamma s_L(x) C_e}{2\pi (\epsilon - U / 2)}, \\
   N_1(x) \alpha_o(x) &= \frac{\Gamma a_L(x) C_o}{2\pi (\epsilon - U / 2)}.
\end{align}

\begin{figure}[htpb]
    \centering
    \includegraphics[width=\linewidth]{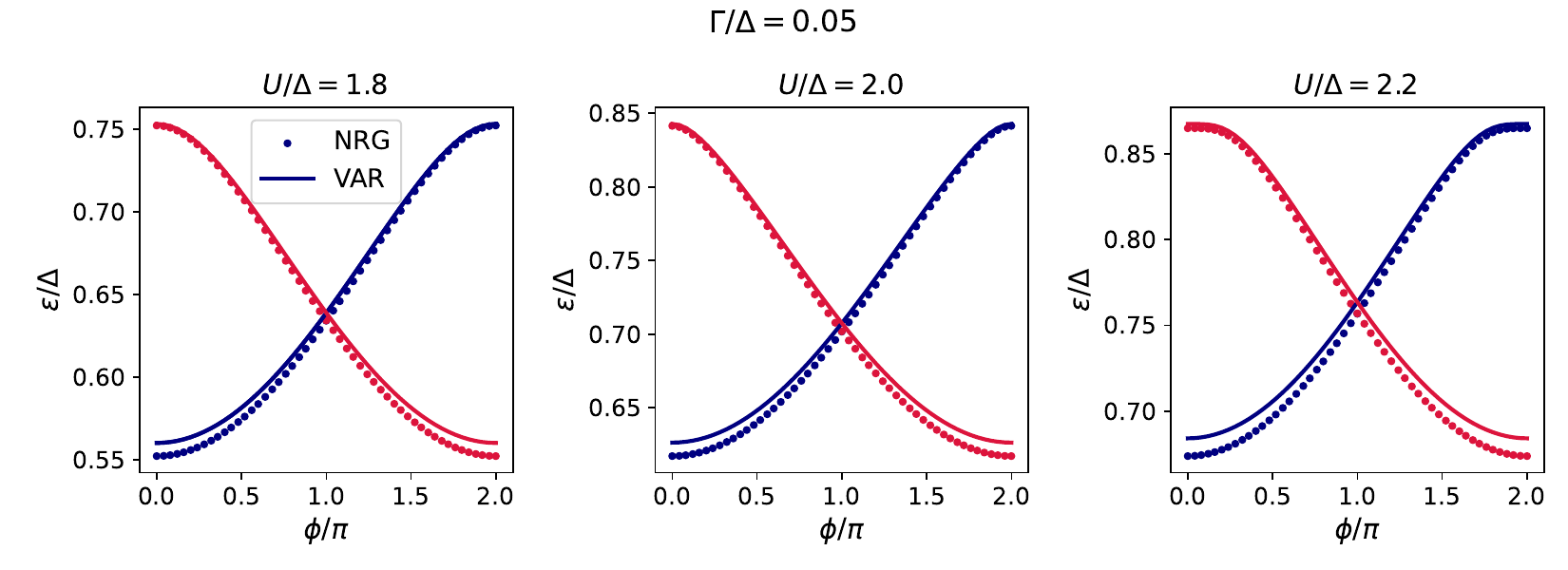}
    \includegraphics[width=\linewidth]{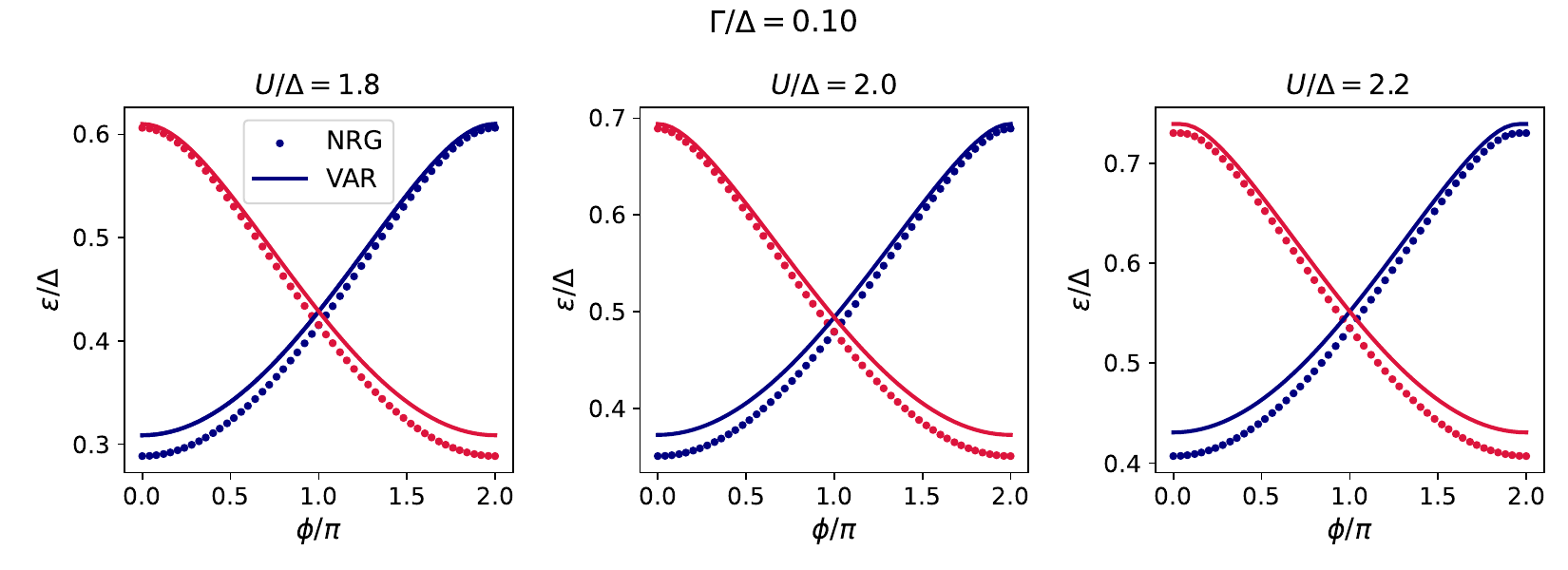}
    \caption{Variational estimates of the subgap energies for a range of $\Gamma$ and $U$. The $\eta$-even ($\eta$-odd) variational solution is shown with a solid blue (red) line. The corresponding NRG data (dots) is shown as a reference. In the NRG calculations, we used the discretization parameter $\Lambda = 8$, and $z$-averaging over four values ($z=0.25, 0.5, 0.75, 1$).
    }
    \label{fig:VAR_phi_sweeep}
\end{figure}

For the $\eta$-even case, we multiply the expression with $s_L^*(x)$, integrate with respect to $x$, and make use of $|s_L(x)|^2=1+2|u(x)||v(x)|\cos(\varphi/2)$ to get a transcendental equation for the energy of the even state:
\begin{equation}
    \epsilon_e = U / 2 + \frac{\Gamma}{\pi} \int \mathrm{d}x \frac{1 + 2 |u(x)| |v(x)| \cos{\left(\varphi/2\right)}}{\epsilon_e - E(x) - (\Gamma / \pi) I_2(x, \epsilon_e)} \>. 
    \label{eq:eps_plus}
\end{equation}
Note the correspondence of this equation with Eq.~(26) in Ref.~\cite{ilicin2025}, up to the $\cos(\phi/2)$ factor. This is the same factor that appears in the anomalous part of the hybridization function for the corresponding quantum impurity problem. 
Similar procedure with $\alpha_o$ leads to the equation for the $\eta$-odd state:
\begin{equation}
    \epsilon_o = U / 2 + \frac{\Gamma}{\pi} \int \mathrm{d}x \frac{1 - 2 |u(x)| |v(x)| \cos{\left(\varphi/2\right)}}{\epsilon_o - E(x) - (\Gamma / \pi) I_2(x, \epsilon_o)} \>.
    \label{eq:eps_minus}
\end{equation}
The roots of these eigenvalue equations can be obtained numerically. From previous work we know that they correctly capture all spectral features in the low-$\Gamma$ regime, including the anomalous $\Gamma^{2/3}$ scaling \cite{ilicin2025}. In Fig.~\ref{fig:VAR_phi_sweeep} we compare the variational solutions with the NRG data, finding good agreement at low to intermediate values of the hybridization strength $\Gamma$.

\subsection{Away from half filling}
\label{phasym}

We proceed by generalizing the variational solution to the case with no p-h symmetry (still assuming no SOC and no direct tunneling). First, we write the Hamiltonian as a sum of the p-h symmetric and antisymmetric terms: 
\begin{equation}
    H = H_s + H_a \>.
\end{equation}
We parametrize the deviation from the p-h symmetric point as $\nu = 1 + \delta$ and write (up to a constant):
\begin{equation}
    H_a = - U \delta \sum_{\sigma} \left( n_{\sigma} - \frac{1}{2} \right) \>.
\end{equation}
Non-zero $\delta$ results in additional terms in the variational equations.
\begin{equation}
\begin{split}
    \ref{eq:minimization1} \quad &\mathrel{+}= a^\mp U \delta, \\
    \ref{eq:minimization2} \quad &\mathrel{+}= 0, \\
    \ref{eq:minimization2RR} \quad &\mathrel{+}= 0, \\
    \ref{eq:minimization3} \quad &\mathrel{+}= \beta^\mp_{llkk'} U \delta, \\   
    \ref{eq:minimization4} \quad &\mathrel{+}= \beta^\mp_{LRkk'} U \delta. 
\end{split}
\end{equation}
These terms break the $\eta$-symmetry and result in coupling between the subgap states from different $\eta$-parity sectors. In practice, this means that variational solutions cannot be expressed in terms of two separate eigenvalue equations like in Eqs.~\eqref{eq:eps_plus},~\eqref{eq:eps_minus}, but rather as solutions to a $2\times2$ non-linear eigenvalue problem. The eigenvalue equations this becomes 
\begin{align}
\begin{split}
    N_{1\delta}(x) \alpha_{e}(x) &= \frac{\Gamma}{\pi} \frac{2C_e(U-2\epsilon)+4C_o U \delta}{-U_\delta} s_L(x) \\
&+\frac{\Gamma}{2\pi}
\left(
\int \frac{s_L^*(x') N_2(x,x') \alpha_{e}(x') \mathrm{d}x'}{N_2(x,x')^2-U^2\delta^2}
+
U \delta 
\int \frac{a_L^*(x')\alpha_{o}(x') \mathrm{d}x'}{N_2(x,x')^2-U^2\delta^2}
\right) s_L(x),
\end{split}\\
\begin{split}
    N_{1\delta}(x) \alpha_{o}(x) &= \frac{\Gamma}{\pi} \frac{2C_o(U-2\epsilon)+4C_e U \delta}{-U_\delta} a_L(x) \\
&+\frac{\Gamma}{2\pi}
\left(
\int \frac{a_L^*(x') N_2(x,x') \alpha_{o}(x') \mathrm{d}x'}{N_2(x,x')^2-U^2\delta^2}
+
U \delta 
\int \frac{s_L^*(x')\alpha_{e}(x') \mathrm{d}x'}{N_2(x,x')^2-U^2\delta^2}
\right) a_L(x),
\end{split}
\end{align}
Here
\begin{equation}
\begin{split}
    N_{1\delta}(x) &= \epsilon-E(x)-(\Gamma/\pi) I_{2\delta}(x,\epsilon), \\
    I_{2\delta}(x,\epsilon) &= \int \frac{N_2(x,x')\mathrm{d}x'}{N_2(x,x')^2-U^2\delta^2}, \\
    N_2(x,x') &= \epsilon-U/2-\left( E(x) + E(x') \right), \\
    U_\delta &= (U-2\epsilon)^2 -4 U^2\delta^2= (U-2\epsilon-2U\delta)(U-2\epsilon+2U\delta).
\end{split}
\end{equation}

In the $\delta\to0$ limit, the auxiliary quantities behave as $N_{1\delta} \to N_1$, $I_{2\delta}\to I_2$, and $U_\delta \to (U-2\epsilon)^2$, the two equations decouple into separate equations for the $\eta$-even and $\eta$-odd sectors, and we recover the known expressions previously found for the p-h symmetric case. We also note that in this limit, $U_\delta$ behaves for small $\Gamma$ as $U_\delta \sim (U-2\Delta)^2$ on the YSR side ($U>2\Delta$), and as $U_\delta \sim (g_\mathrm{ABS}/\pi)^2 \Gamma^2$ on the ABS side ($U<2\Delta$), with $g_\mathrm{ABS}=4\sqrt{-(2+\theta)/\theta}\arctan\sqrt{-(2+\theta)/\theta}$ with the parameter $\theta$ defined as $U=2\Delta(1+\theta)$ \cite{ilicin2025}.

For finite $\delta$, we drop the two-QP contributions (they represent only a small correction), and find compact equations:
\begin{align}
    C_e &= \frac{2C_e(U-2\epsilon)+4C_o U \delta}{-U_\delta} \Gamma K_e, \label{Ce} \\
    C_o &= \frac{2C_o(U-2\epsilon)+4C_e U \delta}{-U_\delta} \Gamma K_o, \label{Co}
\end{align}
with
\begin{align}
    K_e &= \frac{1}{\pi} \int \mathrm{d}x \frac{1 + 2 |u(x)| |v(x)| \cos{\left(\varphi/2\right)}}{\epsilon - E(x) - (\Gamma / \pi) I_{2\delta}(x, \epsilon)}, \\
    K_o &= \frac{1}{\pi} \int \mathrm{d}x \frac{1 - 2 |u(x)| |v(x)| \cos{\left(\varphi/2\right)}}{\epsilon - E(x) - (\Gamma / \pi) I_{2\delta}(x, \epsilon)}.
\end{align}
The characteristic equation of the eigenvalue problem is
\begin{equation}
    \left[ 2\Gamma K_e \left(U-2\epsilon\right) + U_\delta \right]
    \left[ 2\Gamma K_o \left(U-2\epsilon\right) + U_\delta \right] - 16\Gamma^2 K_e K_o U^2 \delta^2 = 0,
\end{equation}
or, making use of the definition of $U_\delta$,
\begin{equation}
    \left( \epsilon - U/2 - \Gamma K_e \right) 
    \left( \epsilon - U/2 - \Gamma K_o \right) 
     = U^2\delta^2.
\end{equation}
The roots are
\begin{equation}
  \epsilon = \frac{U}{2} + \frac{\Gamma}{2} \left[ K_e + K_o \pm \sqrt{ \Gamma^2 (K_e-K_o)^2 + 4 U^2 \delta^2 }\right].
\end{equation}
For small $\delta$, this becomes
\begin{equation}
    \epsilon = \frac{U}{2} + \Gamma K_{e/o} + \frac{U^2 \delta^2}{\Gamma (K_{e/o}-K_{o/e})}.
\end{equation}
We note that the eigenvalue $\epsilon$ appears in $I_{2\delta}$ and thus in $K_e$ and $K_o$. For each eigenvalue, the implicit equation for $\epsilon$ can be solved numerically by iteration, as in the single-channel case \cite{ilicin2025}.

\subsection{Spin-orbit coupling}

We now consider the spin-flip tunneling terms. The equations for the singlet terms are extended as follows:
\begin{align}
    \ref{eq:minimization1} \quad &\mathrel{+}= 
    \frac{\VS_L}{\sqrt{N}} \sum_k \alpha^T_{Lk} \left( v_{Lk}^* \pm u_{Lk}^* \right)
    +  \frac{\VS_R}{\sqrt{N}} \sum_k \alpha^T_{Rk} \left( v_{Rk}^* \pm u_{Rk}^* \right), \\
    \ref{eq:minimization2} \quad &\mathrel{+}= 
    - i\frac{\VS_L}{\sqrt{N}} \sum_{k'} \left( u_{Lk'}^* - v_{Lk'}^* \right) \xi_{LLkk'}^+ 
    - i\frac{\VS_L}{\sqrt{N}} \sum_{k'} \left( u_{Lk'}^* + v_{Lk'}^* \right) \xi_{LLkk'}^- \nonumber \\
    &\quad\quad +i \frac{\VS_R}{\sqrt{2N}} \sum_{k'} \left( u_{Rk'}^* - v_{Rk'}^* \right) \xi_{LRkk'}^+ 
    +i \frac{\VS_R}{\sqrt{2N}} \sum_{k'} \left( u_{Rk'}^* + v_{Rk'}^* \right) \xi_{LRkk'}^-,  \\
        \ref{eq:minimization2RR} \quad &\mathrel{+}= 
    i \frac{\VS_R}{\sqrt{N}} \sum_{k'} \left( u_{Rk'}^* - v_{Rk'}^* \right) \xi_{RRkk'}^+ 
    +i \frac{\VS_R}{\sqrt{N}} \sum_{k'} \left( u_{Rk'}^* + v_{Rk'}^* \right) \xi_{RRkk'}^- \nonumber \\
    &\quad\quad +i \frac{\VS_L}{\sqrt{2N}} \sum_{k'} \left( u_{Lk'}^* - v_{Lk'}^* \right) \xi_{LRk'k}^+ 
    +i \frac{\VS_L}{\sqrt{2N}} \sum_{k'} \left( u_{Lk'}^* + v_{Lk'}^* \right) \xi_{LRk'k}^-,  \\
    \ref{eq:minimization3} \quad &\mathrel{+}= \frac{\VS_l}{2\sqrt{N}} \alpha^T_{lk} \left( u_{lk'} \mp v_{lk'} \right)
    + \frac{\VS_l}{2\sqrt{N}} \alpha^T_{lk'} \left( u_{lk} \mp v_{lk} \right), \\
    \ref{eq:minimization4} \quad &\mathrel{+}= 
   - \frac{\VS_R}{\sqrt{2N}} \alpha^T_{Lk}  \left( u_{Rk'} \mp v_{Rk'} \right)
   - \frac{\VS_L}{\sqrt{2N}} \alpha^T_{Rk'} \left( u_{Lk}  \mp v_{Lk} \right).
\end{align}
The new equations for the triplet terms are
\begin{equation}
\label{soc1}
\begin{split}
    \left( \epsilon - E_{Lk} \right) \alpha^T_{Lk} &=
    \frac{\VS_L}{\sqrt{N}} a^+ \left( v_{Lk} + u_{Lk} \right) 
  + \frac{\VS_L}{\sqrt{N}} a^- \left( v_{Lk} - u_{Lk} \right) \\
 &+i \frac{\VN_L}{\sqrt{N}}  \sum_{k'} \left( u_{Lk'}^* - v_{Lk'}^* \right) \xi_{LLkk'}^+
  +i \frac{\VN_L}{\sqrt{N}}  \sum_{k'} \left( u_{Lk'}^* + v_{Lk'}^* \right) \xi_{LLkk'}^- \\
 &+i \frac{\VN_R}{\sqrt{2N}} \sum_{k'} \left( u_{Rk'}^* - v_{Rk'}^* \right) \xi_{LRkk'}^+
  +i \frac{\VN_R}{\sqrt{2N}} \sum_{k'} \left( u_{Rk'}^* + v_{Rk'}^* \right) \xi_{LRkk'}^- \\
 &+ \frac{\VS_L}{\sqrt{N}}  \sum_{k'} \left( u_{Lk'}^* - v_{Lk'}^* \right) \beta_{LLkk'}^+
  + \frac{\VS_L}{\sqrt{N}}  \sum_{k'} \left( u_{Lk'}^* + v_{Lk'}^* \right) \beta_{LLkk'}^- \\
 &- \frac{\VS_R}{\sqrt{2N}} \sum_{k'} \left( u_{Rk'}^* - v_{Rk'}^* \right) \beta_{LRkk'}^+ 
  - \frac{\VS_R}{\sqrt{2N}} \sum_{k'} \left( u_{Rk'}^* + v_{Rk'}^* \right) \beta_{LRkk'}^-,
\end{split}
\end{equation}
\begin{equation}
\label{soc2}
\begin{split}
    \left( \epsilon - E_{Rk} \right) \alpha^T_{Rk} &=
    \frac{\VS_R}{\sqrt{N}} a^+ \left( v_{Rk} + u_{Rk} \right) 
  + \frac{\VS_R}{\sqrt{N}} a^- \left( v_{Rk} - u_{Rk} \right) \\
 &-i \frac{\VN_R}{\sqrt{N}}  \sum_{k'} \left( u_{Rk'}^* - v_{Rk'}^* \right) \xi_{RRkk'}^+
  -i \frac{\VN_R}{\sqrt{N}}  \sum_{k'} \left( u_{Rk'}^* + v_{Rk'}^* \right) \xi_{RRkk'}^- \\
 &+i \frac{\VN_L}{\sqrt{2N}} \sum_{k'} \left( u_{Lk'}^* - v_{Lk'}^* \right) \xi_{LRk'k}^+
  +i \frac{\VN_L}{\sqrt{2N}} \sum_{k'} \left( u_{Lk'}^* + v_{Lk'}^* \right) \xi_{LRk'k}^- \\
 &+ \frac{\VS_R}{\sqrt{N}}  \sum_{k'} \left( u_{Rk'}^* - v_{Rk'}^* \right) \beta_{RRkk'}^+
  + \frac{\VS_R}{\sqrt{N}}  \sum_{k'} \left( u_{Rk'}^* + v_{Rk'}^* \right) \beta_{RRkk'}^- \\
 &- \frac{\VS_L}{\sqrt{2N}} \sum_{k'} \left( u_{Lk'}^* - v_{Lk'}^* \right) \beta_{LRk'k}^+ 
  - \frac{\VS_L}{\sqrt{2N}} \sum_{k'} \left( u_{Lk'}^* + v_{Lk'}^* \right) \beta_{LRk'k}^-,
\end{split}
\end{equation}
\begin{equation}
\label{soc3}
\begin{split}
\left( \epsilon - \frac{U}{2} - \left( E_{Lk} + E_{Lk'} \right) \right) \xi^\pm_{LLkk'} &= 
-i\frac{\VN_L}{2\sqrt{N}} \alpha^T_{Lk}  \left( u_{Lk'} \mp v_{Lk'} \right)
+i\frac{\VN_L}{2\sqrt{N}} \alpha^T_{Lk'} \left( u_{Lk}  \mp v_{Lk}  \right) \\
&
+i\frac{\VS_L}{2\sqrt{N}} \alpha_{Lk}    \left( u_{Lk'} \mp v_{Lk'} \right)
-i\frac{\VS_L}{2\sqrt{N}} \alpha_{Lk'}   \left( u_{Lk}  \mp v_{Lk}  \right) \\
&
+U \delta \xi^\mp_{LLkk'},
\end{split}
\end{equation}
\begin{equation}
\label{soc4}
\begin{split}
\left( \epsilon - \frac{U}{2} - \left( E_{Rk} + E_{Rk'} \right) \right) \xi^\pm_{RRkk'} &= 
+i\frac{\VN_R}{2\sqrt{N}} \alpha^T_{Rk}  \left( u_{Rk'} \mp v_{Rk'} \right)
-i\frac{\VN_R}{2\sqrt{N}} \alpha^T_{Rk'} \left( u_{Rk}  \mp v_{Rk}  \right) \\
&
-i\frac{\VS_R}{2\sqrt{N}} \alpha_{Rk}    \left( u_{Rk'} \mp v_{Rk'} \right)
+i\frac{\VS_R}{2\sqrt{N}} \alpha_{Rk'}   \left( u_{Rk}  \mp v_{Lk}  \right) \\
&
+U \delta \xi^\mp_{RRkk'},
\end{split}
\end{equation}
\begin{equation}
\label{soc5}
\begin{split}
\left( \epsilon - \frac{U}{2} - \left( E_{Lk} + E_{Rk'} \right) \right) \xi^\pm_{LRkk'} &= 
-i \frac{\VN_L}{\sqrt{2N}} \alpha^T_{Rk'} \left( u_{Lk}  \mp v_{Lk} \right)
-i\frac{\VN_R}{\sqrt{2N}} \alpha^T_{Lk}  \left( u_{Rk'} \mp v_{Rk'}\right) \\
&
-i\frac{\VS_L}{\sqrt{2N}} \alpha_{Rk'}   \left( u_{Lk}  \mp v_{Lk} \right)
-i\frac{\VS_R}{\sqrt{2N}} \alpha_{Lk}    \left( u_{Rk'} \mp v_{Rk'}  \right) \\
&+U \delta  \xi^\mp_{LRkk'}.
\end{split}
\end{equation}
The set of equations can be reduced to four equations for $\alpha_{L,R}$ and $\alpha^T_{L,R}$ by substituting coefficients $a^\pm$, as well as $\beta$ and $\xi$ functions. Dropping the 2QP corrections\footnote{The 2QP corrections due to spin-flip processes are obtained by taking the expressions for 2QP corrections due to spin-conserving processes and replacing $V \to \VS$ and $\alpha_{e,o} \to \alpha_{e,o}^T$.}, we find
\begin{align}
\label{q1}
N_{1\delta}(x) \alpha_e(x) &= 4V
\frac{(C_e V + C_e^T \VS)(U-2\epsilon) + 2U\delta (C_o V + C_o^T \VS)}{-U_\delta }s_L(x),  \\
\label{q2}
N_{1\delta}(x) \alpha_o(x) &= 4V
\frac{(C_o V + C_o^T \VS)(U-2\epsilon) + 2U\delta (C_e V + C_e^T \VS)}{-U_\delta }a_L(x), \\
\label{q3}
N_{1\delta}(x) \alpha_e^T(x) &= 4\VS
\frac{(C_e V + C_e^T \VS)(U-2\epsilon) + 2U\delta (C_o V + C_o^T \VS)}{-U_\delta }s_L(x), \\
\label{q4}
N_{1\delta}(x) \alpha_o^T(x) &= 4\VS
\frac{(C_o V + C_o^T \VS)(U-2\epsilon) + 2U\delta (C_e V + C_e^T \VS)}{-U_\delta }a_L(x).
\end{align}
Here $N_{1\delta}(x,\epsilon) = \epsilon-E(x)-\Gamma_T I_{2\delta}(x,\epsilon)$, where now $\Gamma_T$ is the total hybridisation strength, with contributions from both spin-conserving and spin-flipping processes:
\begin{equation}
   \Gamma_T = \left( 2V^2+2({\VS})^2 \right) \pi \rho. 
\end{equation}
We have also introduced
\begin{equation}
    C^T_{e} = \int s_L^*(y) \alpha^T_{e}(y) \mathrm{d}y,
    \quad
    C^T_{o} = \int a_L^*(y) \alpha^T_{o}(y) \mathrm{d}y.
\end{equation}

One then derives a characteristic equation for this $4\times 4$ eigenvalue problem, which turns out to take a very simple form:
\newcommand{\GN}{\Gamma^{\mathrm{(N)}}}
\newcommand{\GS}{\Gamma^{\mathrm{(SF}}}
\begin{equation}
    \left( \epsilon-U/2- \Gamma_T K_e \right)
    \left( \epsilon-U/2- \Gamma_T K_o \right)
    = U^2 \delta^2.
\end{equation}
The expression has the same structure as in the absence of SOC, which implies that the singlet and triplet channels are basically decoupled and their effects are additive. This is more clearly seen by considering Eqs.~\eqref{q1}-\eqref{q4} for the p-h symmetric case, in the limit $\delta \rightarrow 0$. Since SOC does not break the $\eta$-symmetry, eigenenergy in each of the sectors is given by a single transcendental equation analogous to Eqs.~\eqref{eq:eps_plus} and  \eqref{eq:eps_minus}. For the even sector, the minimization equations reduce to 
\begin{align}
     \left(\epsilon - U / 2 \right) \alpha_e (x) &= 2V \left(C_e V + C_e^T \VS \right) \frac{s_L(x)}{N_1(x)}     
     \label{eq:T_even_1} \\
     \left(\epsilon - U / 2 \right) \alpha_e^T(x) &= 2 \VS \left(C_e V + C_e^T \VS \right) \frac{s_L(x)}{N_1(x)} \>.
    \label{eq:T_even_2}
\end{align}
By summing Eq.~\eqref{eq:T_even_1} weighted by $V$ and Eq.~\eqref{eq:T_even_2} weighted by $\VS$, multiplying both sides of the resulting expression by $s_L^*(x)$, integrating and cancelling the common factor, one finds an equation of the same form as Eq.~\eqref{eq:eps_plus}, but involving the total hybridization strength $\Gamma_T$. The same holds in the odd $\eta$-parity case. 

More generally, even when $\delta\neq0$, Eq.~\eqref{q1} and Eq.~\eqref{q3} have the same complex coefficients, and $\alpha_{e}$ and $\alpha_e^T$ have the same complex phase. The same holds for $\alpha_o$ and $\alpha_o^T$. Singlet and triplet coefficients are thus in phase, therefore according to the general considerations from Sec.~\ref{polariz}, there is no local spin polarization on the QD site, even if a spin-triplet component is present. In order to have finite $\langle S_x \rangle$, it is thus necessary to break the phase relation by including a direct hopping term (i.e. background tunneling processes).

\subsection{Background tunneling processes}

Finally, we take into consideration the background tunneling processes between the SC leads. The following additional terms need to be added to the variational equations:
\begin{align}
\label{E1}
    \ref{eq:minimization1} \quad &\mathrel{+}= \frac{\sqrt{2}\VD}{N} \sum_{k,k'}  \beta_{LRkk'}^\pm \left( v^*_{Lk}u^*_{Rk'} + u^*_{Lk} v^*_{Rk'} \right) \\
\label{D1}
    \ref{eq:minimization2} \quad &\mathrel{+}= \frac{\VD}{N} \sum_{k'} \alpha_{Rk'} \left( u_{Lk} u^*_{Rk'}
     - v_{Lk} v^*_{Rk'} \right) \\
\label{D2}
         \ref{eq:minimization2RR} \quad &\mathrel{+}= \frac{\VD}{N} \sum_{k'} \alpha_{Lk'} \left( u_{Rk} u^*_{Lk'}
     - v_{Rk} v^*_{Lk'} \right)  \\
\label{S1}
    \ref{eq:minimization3} \quad &\mathrel{+}= \frac{\VD}{\sqrt{2}N} \sum_q
    \left[ 
    \beta^\pm_{LRk'q} 
    \left( u_{Lk} u_{Rq}^* - v_{Lk}v_{Rq}^* \right)
    +
    \beta^\pm_{LRkq}
    \left( u_{Lk'} u_{Rq}^* - v_{Lk'}v_{Rq}^* \right)
    \right]\,\text{ for }l=L, \\
\label{S2}
\ref{eq:minimization3} \quad &\mathrel{+}= \frac{\VD}{\sqrt{2}N} \sum_q
    \left[ 
    \beta^\pm_{LRqk'} 
    \left( u_{Rk} u_{Lq}^* - v_{Rk}v_{Lq}^* \right)
    +
    \beta^\pm_{LRqk}
    \left( u_{Rk'} u_{Lq}^* - v_{Rk'}v_{Lq}^* \right)
    \right]
    \,\text{ for }l=R, \\
\label{E5}
    \ref{eq:minimization4} \quad &\mathrel{+}= \frac{\sqrt{2}\VD}{N} \Biggl[ \sum_q
    \left\{
    \beta^\pm_{LLqk} \left( u_{Rk'}u_{Lq}^*-v_{Rk'}v_{Lq}^* \right)
    +
    \beta^\pm_{RRqk'} \left( u_{Lk} u_{Rq}^*-v_{Lk}v_{Rq}^* \right) \right\} \nonumber \\
    & \quad\quad\quad\quad\quad\quad+
    a^\pm \left( v_{Lk}u_{Rk'} + u_{Lk} v_{Rk'} \right)    
    \Biggr].
\end{align}
as well as
\begin{align}
\label{D3}
\ref{soc1} \quad &\mathrel{+}= -\frac{\VD}{N} \sum_{k'} \alpha^T_{Rk'} \left( u_{Lk} u^*_{Rk'} - v_{Lk} v^*_{Rk'} \right), \\
\label{D4}
\ref{soc2} \quad &\mathrel{+}= -\frac{\VD}{N} \sum_{k'} \alpha^T_{Lk'} \left( u_{Rk} u^*_{Lk'} - v_{Rk} v^*_{Lk'} \right), \\
\ref{soc3} \quad &\mathrel{+}= \frac{\VD}{\sqrt{2}N} \sum_q
\Bigl[ 
- \xi^\pm_{LRk'q} \left( u_{Lk} u^*_{Rq} - v_{Lk} v^*_{Rq} \right)
+ 
\xi^\pm_{LRkq} \left( u_{Lk'} v^*_{rq} - v_{Lk'} v^*_{Rq} \right)
\Bigr], \\
\ref{soc4} \quad &\mathrel{+}= \frac{\VD}{\sqrt{2}N} \sum_q
\Bigl[ 
 \xi^\pm_{LRqk'} \left( u_{Rk} u^*_{Lq} - v_{Rk} v_{Lq}^*  \right)
 -
 \xi^\pm_{LRqk} \left( u_{Rk'} u^*_{Lq}  -  v_{Rk'} v_{Lq}^* \right)
\Bigr], \\
\ref{soc5} \quad &\mathrel{+}= \frac{\sqrt{2} \VD}{N} \sum_q
\Bigl[ 
-\xi^\pm_{LLqk}  \left( u_{Rk'} u_{Lq}^*  - v_{Rk'} v_{Lq}^*  \right) 
+
\xi^\pm_{RRqk'}  \left( u_{Lk} u_{Rq}^*  - v_{Lk}  v_{Rq}^*  \right) 
\Bigr].
\end{align}
Concerning the phase dependence, it should be noted that 
\begin{align}
   v_L u_R + u_L v_R & \sim \cos(\phi/2), \\
   u_L u_R^*-v_L v_R^*  & \sim i\sin(\phi/2), \\
   v_L^* u_R^* + u_L^* v_R^* &\sim \cos(\phi/2).
\end{align}

The inter-lead tunneling processes induce three effects that we describe in more detail below: 1) mixing between L and R 1QP terms ($\alpha$),
2) mixing between LL, RR, and LR 2QP terms ($\beta$ and $\xi$), 3) Cooper pair splitting and recombination (coupling between $a_\pm$ and $\beta_{LR}$).

Eqs.~\eqref{D1} and \eqref{D2}, as well as \eqref{D3} and \eqref{D4}, describe mixing between the L and R combinations of the singlet and triplet YSR states. These processes explicitly break the symmetry of the coefficients $\alpha_{L/R}$ and $\alpha_{L/R}^T$ with respect to the Fermi level. They thus embody the breaking of p-h symmetry by direct tunneling. Such processes are present for all values of the phase bias $\phi$ different from 0 and $\pi$, due to the $\sin(\phi/2)$ dependence. Importantly, compared to the effects of departure from the p-h symmetric point due to potential scattering ($\delta$ term), which were found to be the same for the singlet and triplet states, here the mixing leads to different signs in the singlet and triplet state because of the different phase structure of the wavefunction in these subspaces. If one only retains these $\VD$ terms, the variational equations take the following form (again we drop the 2QP terms):
\begin{align}
\label{w1}
N_{1\delta}(x) \alpha_e(x) &= \left[ 4V
\frac{(C_e V + C_e^T \VS)(U-2\epsilon) + 2U\delta (C_o V + C_o^T \VS)}{-U_\delta}
+ \frac{C_o \VD}{2}
\right]
s_L(x),  \\
\label{w2}
N_{1\delta}(x) \alpha_o(x) &= \left[ 4V
\frac{(C_o V + C_o^T \VS)(U-2\epsilon) + 2U\delta (C_e V + C_e^T \VS)}{-U_\delta }
+ \frac{C_e \VD}{2}
\right]
a_L(x), \\
\label{w3}
N_{1\delta}(x) \alpha_e^T(x) &= \left[ 4\VS
\frac{(C_e V + C_e^T \VS)(U-2\epsilon) + 2U\delta (C_o V + C_o^T \VS)}{-U_\delta }
- \frac{C^T_o \VD}{2}
\right]
s_L(x), \\
\label{w4}
N_{1\delta}(x) \alpha_o^T(x) &= \left[ 4\VS
\frac{(C_o V + C_o^T \VS)(U-2\epsilon) + 2U\delta (C_e V + C_e^T \VS)}{-U_\delta }
- \frac{C^T_e \VD}{2}
\right]a_L(x).
\end{align}
One consequence of the mixing between $\alpha_e$ and $\alpha_o$, and $\alpha_e^T$ and $\alpha_o^T$ is that there is no apparent way to express eigenenergies via a single transcendental equation, even when $\delta = 0$. This comes from the fact that the direct tunneling term breaks the p-h, and thus the $\eta$-symmetry.
Since there is different mixing between $\alpha_e$ and $\alpha_o$, and between $\alpha_e^T$ and $\alpha_o^T$, the singlet and the triplet amplitudes can have different complex phases. Following Sec.~\ref{polariz}, this leads to non-zero $\langle S_x \rangle$. This observation is one of the key results of our variational analysis.

The three-way mixing of 2QP states is described by Eqs.~\eqref{S1},\eqref{S2} and the first part of Eq.~\eqref{E5}, involving LL, RR, and LR configurations, and analogous structure for the triplet counterparts. 2QP components are generally only a minor correction, thus the effects of this mixing is not expected to lead to any important effects.

Finally, Eq.~\eqref{E1} and the last term in Eq.~\eqref{E5} show that zero-QP and two-QP states (LR) are coupled via inter-lead tunneling. This corresponds to Cooper pair breaking and recombination processes, which occur for all $\phi \neq \pi$. Because Cooper pairs are singlets, there is no analogous coupling that involves the LR two-QP spin triplets ($\xi_\mathrm{LR}^\pm$ terms). These terms change the mathematical structure of the problem so that it is no longer possible to explicitly express $a^\pm$ and $\beta_\mathrm{LR}$ in terms of $\alpha_{L/R}$, because the relations are not algebraic but rather take the form of integral equations. For this reason, it is not possible to fully reduce the eigenproblem to equations for $\alpha_{L/R}$ alone for $\VD \neq 0$. A detailed analysis of the effect thus seems difficult in this formalism.

\end{appendix}

\bibliography{main}

@article{ilicin2025,
  title = {Variational solution of the superconducting {Anderson} impurity model and the band-edge singularity phenomena},
  volume = {19},
  ISSN = {2542-4653},
  url = {http://dx.doi.org/10.21468/SciPostPhys.19.1.006},
  DOI = {10.21468/scipostphys.19.1.006},
  number = {1},
  journal = {SciPost Physics},
  publisher = {Stichting SciPost},
  author = {Iličin,  Teodor and Žitko,  Rok},
  year = {2025},
  month = jul 
}

@article{Kirsanskas2015,
doi = {10.1103/physrevb.92.235422},
url = {https://doi.org/10.1103/physrevb.92.235422},
year = {2015},
month = dec,
publisher = {American Physical Society ({APS})},
volume = {92},
number = {23},
pages = { 235422 },
author = {Gediminas Kir{\v{s}}anskas and Moshe
Goldstein and Karsten Flensberg and Leonid I. Glazman
and Jens Paaske},
title = {{Yu-Shiba-Rusinov} states in phase-biased
superconductor-quantum dot-superconductor junctions},
journal = {Physical Review B}
}

@article{zalom2025,
  title = { {Andreev} bound state spectroscopy of a quantum-dot-based
  {Aharonov-Bohm} interferometer with superconducting terminals },
  author = { P. Zalom and D. Rolih and R. \v{Z}itko },
  volume = { 113 },
  pages = {075130},
  year = { 2026 },
  journal = { Phys. Rev. B }
  }

@article{kurilovich2021,
doi = {10.1103/physrevb.104.174517},
url = {https://doi.org/10.1103/physrevb.104.174517},
year = {2021},
month = nov,
publisher = {American Physical Society ({APS})},
volume = {104},
number = {17},
pages = { 174517 },
author = {Pavel D. Kurilovich and Vladislav D.
Kurilovich and Valla Fatemi and Michel H. Devoret and
Leonid I. Glazman},
title = {Microwave response of an {Andreev} bound state},
journal = {Physical Review B}
}

@article{PitaVidal2025,
  title = {Blueprint for All-to-All-Connected Superconducting Spin
  Qubits},
  volume = {6},
  ISSN = {2691-3399},
  url = {http://dx.doi.org/10.1103/PRXQuantum.6.010308},
  DOI = {10.1103/prxquantum.6.010308},
  number = {1},
  journal = {PRX Quantum},
  publisher = {American Physical Society (APS)},
  author = {Pita-Vidal,  Marta and Wesdorp,  Jaap J.
  and Andersen,  Christian Kraglund},
  year = {2025},
  month = jan,
  pages = { 010308 },
  }

@article{Zatsarynna2024,
  title = {Many-body quantum dynamics of spin-orbit coupled {Andreev}
  states in a {Zeeman} field},
  volume = {109},
  ISSN = {2469-9969},
  url = {http://dx.doi.org/10.1103/PhysRevB.109.214505},
  DOI = {10.1103/physrevb.109.214505},
  number = {21},
  journal = {Physical Review B},
  publisher = {American Physical Society (APS)},
  author = {Zatsarynna,  Kateryna and Nava,  Andrea
  and Zazunov,  Alex and Egger,  Reinhold},
  year = {2024},
  month = jun,
  pages = { 214505 },
  }

@article{Fatemi2022,
  title = {Microwave Susceptibility Observation of Interacting
  Many-Body {Andreev} States},
  volume = {129},
  ISSN = {1079-7114},
  url = {http://dx.doi.org/10.1103/PhysRevLett.129.227701},
  DOI = {10.1103/physrevlett.129.227701},
  number = {22},
  journal = {Physical Review Letters},
  publisher = {American Physical Society (APS)},
  author = {Fatemi,  V. and Kurilovich,  P. D. and
  Hays,  M. and Bouman,  D. and Connolly,  T. and
  Diamond,  S. and Frattini,  N. E. and Kurilovich,
  V. D. and Krogstrup,  P. and Nygård,  J. and
  Geresdi,  A. and Glazman,  L. I. and Devoret,  M. H.},
  year = {2022},
  month = nov,
  pages = { 227701 },
  }

@article{Wesdorp2024,
  title = {Microwave spectroscopy of interacting {Andreev} spins},
  volume = {109},
  ISSN = {2469-9969},
  url = {http://dx.doi.org/10.1103/PhysRevB.109.045302},
  DOI = {10.1103/physrevb.109.045302},
  number = {4},
  journal = {Physical Review B},
  publisher = {American Physical Society (APS)},
  author = {Wesdorp,  J. J. and Matute-Cañadas,  F. J.
  and Vaartjes,  A. and Gr\"{u}nhaupt,  L. and Laeven,
  T. and Roelofs,  S. and Splitthoff,  L. J. and
  Pita-Vidal,  M. and Bargerbos,  A. and van Woerkom,
  D. J. and Krogstrup,  P. and Kouwenhoven,  L. P. and
  Andersen,  C. K. and Yeyati,  A. Levy and van Heck,
  B. and de Lange,  G.},
  year = {2024},
  month = jan,
  pages = { 045302 },
  }

@article{hays2021,
year = {2021},
month = jul,
publisher = {American Association for the Advancement of
Science ({AAAS})},
volume = {373},
number = {6553},
pages = {430--433},
author = {M. Hays and V. Fatemi and D. Bouman and J.
Cerrillo and S. Diamond and K. Serniak and T.
Connolly and P. Krogstrup and J. Nyg{\aa}rd and A.
Levy Yeyati and A. Geresdi and M. H. Devoret},
title = {Coherent manipulation of an {Andreev} spin
qubit},
journal = {Science}
}

@article{chtchelkatchev2003,
year = {2003},
month = jun,
publisher = {American Physical Society ({APS})},
volume = {90},
number = {22},
author = {Nikolai M. Chtchelkatchev and Yu. V. Nazarov},
title = {{Andreev} Quantum Dots for Spin Manipulation},
journal = {Physical Review Letters}
}

@article{deLange2015,
year = {2015},
month = sep,
publisher = {American Physical Society ({APS})},
volume = {115},
number = {12},
author = {G. de Lange and B. van Heck and A. Bruno and
D. J. van Woerkom and A. Geresdi and
S. R. Plissard and
E. P. A. M. Bakkers and A. R. Akhmerov and
L. DiCarlo},
title = {Realization of Microwave Quantum Circuits
Using Hybrid Superconducting-Semiconducting Nanowire
{Josephson} Elements},
journal = {Physical Review Letters}
}

@article{Zazunov2003,
year = {2003},
month = feb,
publisher = {American Physical Society ({APS})},
volume = {90},
number = {8},
author = {A. Zazunov and V. S. Shumeiko and E. N.
Bratus' and J. Lantz and G. Wendin},
title = {{Andreev} Level Qubit},
journal = {Physical Review Letters}
}

@article{Beri2008,
year = {2008},
month = jan,
publisher = {American Physical Society ({APS})},
volume = {77},
number = {4},
author = {B. B{\'{e}}ri and J. H. Bardarson and C. W. J. Beenakker},
title = {Splitting of {Andreev} levels in a {Josephson}
junction by spin-orbit coupling},
journal = {Physical Review B}
}

@article{pitavidal2023,
  doi = {10.1038/s41567-023-02071-x},
  url = {https://doi.org/10.1038/s41567-023-02071-x},
  year = {2023},
  month = may,
  publisher = {Springer Science and Business Media {LLC}},
  author = {Marta Pita-Vidal and Arno Bargerbos and Rok {\v{Z}}itko and Lukas J. Splitthoff and Lukas Gr\"{u}nhaupt
  and Jaap J. Wesdorp and Yu Liu and Leo P. Kouwenhoven and Ram{\'{o}}n Aguado and Bernard van Heck and Angela Kou
  and Christian Kraglund Andersen},
  title = {Direct manipulation of a superconducting spin qubit strongly coupled to a transmon qubit},
  journal = {Nature Physics}
}

@article{tosi2019,
year = {2019},
month = jan,
publisher = {American Physical Society ({APS})},
volume = {9},
number = {1},
pages = { 011010 },
author = {L. Tosi and C. Metzger and
M. F. Goffman and C. Urbina and H. Pothier and Sunghun Park and A. Levy Yeyati and
J. Nyg{\aa}rd and P. Krogstrup},
title = {Spin-Orbit Splitting of {Andreev} States Revealed by Microwave Spectroscopy},
journal = {Physical Review X}
}

@article{padurariu2010,
year = {2010},
month = apr,
publisher = {American Physical Society ({APS})},
volume = {81},
number = {14},
pages = { 144519 },
author = {C. Padurariu and Yu. V. Nazarov},
title = {Theoretical proposal for superconducting spin qubits},
journal = {Physical Review B}
}

@article{bargerbos2022,
  doi = {10.1103/prxquantum.3.030311},
  url = {https://doi.org/10.1103/prxquantum.3.030311},
  year = {2022},
  month = jul,
  publisher = {American Physical Society ({APS})},
  volume = {3},
  number = {3},
  author = {Arno Bargerbos and Marta Pita-Vidal and Rok {\v{Z}}itko and Jes{\'{u}}s
  {\'{A}}vila and Lukas J. Splitthoff and Lukas Gr\"{u}nhaupt and Jaap J. Wesdorp
  and Christian K. Andersen and Yu Liu and Leo P. Kouwenhoven and Ram{\'{o}}n Aguado
  and Angela Kou and Bernard van Heck},
  title = {Singlet-Doublet Transitions of a Quantum Dot {Josephson} Junction
  Detected in a Transmon Circuit},
  journal = {{PRX} Quantum}
}

@article{Bargerbos2023,
   title = {Spectroscopy of Spin-Split {Andreev} Levels in a Quantum Dot
   with Superconducting Leads},
   volume = {131},
   ISSN = {1079-7114},
   url = {http://dx.doi.org/10.1103/PhysRevLett.131.097001},
   DOI = {10.1103/physrevlett.131.097001},
   number = {9},
   journal = {Physical Review Letters},
   publisher = {American Physical Society (APS)},
   author = {Bargerbos,  Arno and Pita-Vidal,  Marta
   and Žitko,  Rok and Splitthoff,  Lukas J. and
   Gr\"{u}nhaupt,  Lukas and Wesdorp,  Jaap J. and
   Liu,  Yu and Kouwenhoven,  Leo P. and Aguado,
   Ramón and Andersen,  Christian Kraglund and Kou,
   Angela and van Heck,  Bernard},
   year = {2023},
   month = aug,
   pages = { 097001 },
   }

@article{zazunov2009,
  doi = {10.1103/physrevlett.103.147004},
  url = {https://doi.org/10.1103/physrevlett.103.147004},
  year = {2009},
  month = oct,
  publisher = {American Physical Society ({APS})},
  volume = {103},
  number = {14},
  pages = { 147004 },
  author = {A. Zazunov and R. Egger and T. Jonckheere
  and T. Martin},
  title = {Anomalous {Josephson} Current through a
  Spin-Orbit Coupled Quantum Dot},
  journal = {Physical Review Letters}
  }

@article{Fauvel2024,
  title = {Opportunities for the direct manipulation of a phase-driven
  {Andreev} spin qubit},
  volume = {109},
  ISSN = {2469-9969},
  url = {http://dx.doi.org/10.1103/PhysRevB.109.184515},
  DOI = {10.1103/physrevb.109.184515},
  number = {18},
  journal = {Physical Review B},
  publisher = {American Physical Society (APS)},
  author = {Fauvel,  Yoan and Meyer,  Julia S. and
  Houzet,  Manuel},
  year = {2024},
  month = may,
  pages = { 184515 },
  }

@article{Park2017,
year = {2017},
month = sep,
publisher = {American Physical Society ({APS})},
volume = {96},
number = {12},
author = {Sunghun Park and A. Levy Yeyati},
title = {Andreev spin qubits in multichannel {Rashba}
nanowires},
journal = {Physical Review B},
pages = { 125416 },
}

@article{metzger2021,
year = {2021},
month = jan,
publisher = {American Physical Society ({APS})},
volume = {3},
number = {1},
author = {C. Metzger and Sunghun Park and L. Tosi and
C. Janvier and A. A. Reynoso and M. F. Goffman and C.
Urbina and A. Levy Yeyati and H. Pothier},
title = {Circuit-{QED} with phase-biased {Josephson}
weak links},
journal = {Physical Review Research},
pages = { 013036 },
}

@article{park2020,
doi = {10.1103/physrevlett.125.077701},
url = {https://doi.org/10.1103/physrevlett.125.077701},
year = {2020},
month = aug,
publisher = {American Physical Society ({APS})},
volume = {125},
number = {7},
pages = { 077701 },
author = {Sunghun Park and C. Metzger and L. Tosi and M. F. Goffman
and C. Urbina and H. Pothier and A. Levy Yeyati},
title = {From Adiabatic to Dispersive Readout of Quantum Circuits},
journal = {Physical Review Letters}
}

@article{hybrid2010,
  title = { Hybrid superconductor-quantum dot devices },
  author = { Silvano De Franceschi and Leo Kouwenhoven and
  Christian Sch\"onenberger and Wolfgang Wernsdorfer },
  journal = { Nat. Nanotechnology },
  volume = { 5 },
  pages = { 703 },
  year = { 2010 },
  }

@article{Baran2023,
  title = {Surrogate model solver for impurity-induced superconducting
  subgap states},
  volume = {108},
  ISSN = {2469-9969},
  url = {http://dx.doi.org/10.1103/PhysRevB.108.L220506},
  DOI = {10.1103/physrevb.108.l220506},
  number = {22},
  journal = {Physical Review B},
  publisher = {American Physical Society (APS)},
  author = {Baran,  Virgil V. and Frost,  Emil J. P.
  and Paaske,  Jens},
  year = {2023},
  month = dec,
  pages = { L220506 },
  }

@article{vecino2003,
author = {Vecino, E and Mart{\'\i}n-Rodero, A and Yeyati, A},
title = { {Josephson} current through a correlated quantum level:
{Andreev} states and $\pi$ junction behavior },
journal = {Phys. Rev. B},
year = {2003},
volume = {68},
pages = {035105},
}

@article{janvier2015,
author = {Janvier, C. and Tosi, L. and Bretheau, L. and Girit, C. O.
and Stern, M. and Bertet, P. and Joyez, P. and Vion, D. and Esteve, D.
and Goffman, M. F. and Pothier, H. and Urbina, C.},
title = {Coherent manipulation of {Andreev} states in superconducting
atomic contacts},
volume = {349},
pages = {1199},
year = {2015},
journal = {Science}
}

@article{Cheung2024,
title = {Photon-mediated long-range coupling of two {Andreev} pair
qubits},
volume = {20},
ISSN = {1745-2481},
url = {http://dx.doi.org/10.1038/s41567-024-02630-w},
DOI = {10.1038/s41567-024-02630-w},
number = {11},
journal = {Nature Physics},
publisher = {Springer Science and Business Media LLC},
author = {Cheung,  L. Y. and Haller,  R. and
Kononov,  A. and Ciaccia,  C. and Ungerer,  J. H.
and Kanne,  T. and Nygård,  J. and Winkel,  P. and
Reisinger,  T. and Pop,  I. M. and Baumgartner,  A.
and Sch\"{o}nenberger,  C.},
year = {2024},
month = oct,
pages = {1793–1797}
}

@article{Reynoso2012,
  title = {Spin-orbit-induced chirality of {Andreev} states in {Josephson}
  junctions},
  volume = {86},
  ISSN = {1550-235X},
  url = {http://dx.doi.org/10.1103/PhysRevB.86.214519},
  DOI = {10.1103/physrevb.86.214519},
  number = {21},
  journal = {Physical Review B},
  publisher = {American Physical Society (APS)},
  author = {Reynoso,  Andres A. and Usaj,  Gonzalo and
  Balseiro,  C. A. and Feinberg,  D. and Avignon,  M.},
  year = {2012},
  month = dec,
  pages = { 214519 },
  }

@article{meden2019review,
  title = { The {Anderson}-{Josephson} quantum dot -- A theory
  perspective },
  author = { V. Meden },
  journal = { J. Phys.: Condens. Matter },
  volume = { 31 },
  pages = { 163001 },
  year = { 2019 },
  }

@article{MatuteCanadas2022,
  title = {Signatures of Interactions in the {Andreev} Spectrum of
  Nanowire {Josephson} Junctions},
  volume = {128},
  ISSN = {1079-7114},
  url = {http://dx.doi.org/10.1103/PhysRevLett.128.197702},
  DOI = {10.1103/physrevlett.128.197702},
  number = {19},
  journal = {Physical Review Letters},
  publisher = {American Physical Society (APS)},
  author = {Matute-Cañadas,  F. J. and Metzger,  C.
  and Park,  Sunghun and Tosi,  L. and Krogstrup,  P.
  and Nygård,  J. and Goffman,  M. F. and Urbina,  C.
  and Pothier,  H. and Yeyati,  A. Levy},
  year = {2022},
  month = may,
  pages = { 197702 },
  }

@article{Hays2020,
  doi = {10.1038/s41567-020-0952-3},
  url = {https://doi.org/10.1038/s41567-020-0952-3},
  year = {2020},
  month = jul,
  publisher = {Springer Science and Business Media {LLC}},
  volume = {16},
  number = {11},
  pages = {1103--1107},
  author = {M. Hays and V. Fatemi and K. Serniak and D. Bouman and S. Diamond
and G. de Lange and P. Krogstrup and J. Nyg{\aa}rd and A. Geresdi and M. H. Devoret},
  title = {Continuous monitoring of a trapped superconducting spin},
  journal = {Nature Physics}
}

@article{bretheau2013prx,
  title = {Supercurrent Spectroscopy of {Andreev} States},
  author = {Bretheau, L. and Girit, C. O. and Urbina, C. and Esteve, D. and Pothier, H.},
  journal = {Phys. Rev. X},
  volume = {3},
  pages = {041034},
  year = {2013},
}

@article{bretheau2013nature,
  title = {Exciting {Andreev} pairs in a superconducting atomic contact},
  volume = {499},
  ISSN = {1476-4687},
  url = {http://dx.doi.org/10.1038/nature12315},
  DOI = {10.1038/nature12315},
  number = {7458},
  journal = {Nature},
  publisher = {Springer Science and Business Media LLC},
  author = {Bretheau,  L. and Girit,  C.
  O. and Pothier,  H. and Esteve,  D. and Urbina,
  C.},
  year = {2013},
  month = jul,
  pages = {312–315}
  }

@article{Hays2018,
year = {2018},
month = jul,
publisher = {American Physical Society ({APS})},
volume = {121},
number = {4},
author = {M. Hays and G. de Lange and K. Serniak and
D. J. van Woerkom and D. Bouman and
P. Krogstrup and J. Nyg{\aa}rd and A. Geresdi and
M. H. Devoret},
title = {Direct Microwave Measurement of
{Andreev}-Bound-State Dynamics in a
Semiconductor-Nanowire {Josephson} Junction},
journal = {Physical Review Letters}
}

@article{jellinggaard2016,
  title = { Tuning {Yu-Shiba-Rusinov }states in a quantum dot },
  author = { A. Jellinggaard and K. Grove-Rasmussen
  and M. H. Madsen and J. Nyg{\aa}rd },
  journal = { Phys. Rev. B },
  volume = { 94 },
  pages = { 064520 },
  year = { 2016 },
  }

@misc{shvetsov2025,
  title = { Approaching the ultrastrong coupling regime between an
  {Andreev} level and a microwave resonator },
  author = {
  O.O. Shvetsov and A. Khola and V. Buccheri and I.P.C. Cools and N.
  Trnjanin and A. Geresd and T. Kanne and J. Nyg\o ard },
  howpublished = { 2502.09243 },
  year = { 2025 },
  }

@article{Zellekens2022,
  title = {Microwave spectroscopy of {Andreev} states in {InAs}
  nanowire-based hybrid junctions using a flip-chip layout},
  volume = {5},
  ISSN = {2399-3650},
  url = {http://dx.doi.org/10.1038/s42005-022-01035-6},
  DOI = {10.1038/s42005-022-01035-6},
  number = {1},
  journal = {Communications Physics},
  publisher = {Springer Science and Business Media LLC},
  author = {Zellekens,  Patrick and Deacon,  Russell
  S. and Perla,  Pujitha and Gr\"{u}tzmacher,  Detlev
  and Lepsa,  Mihail Ion and Sch\"{a}pers,  Thomas and
  Ishibashi,  Koji},
  year = {2022},
  month = oct,
  pages = { 267 },
  }

@article{wilson1975,
  title = { The renormalization group: {Critical} phenomena and the
  {Kondo} problem },
  author = { K. G. Wilson },
  journal = { Rev. Mod. Phys. },
  year = { 1975 },
  pages = { 773 },
  volume = { 47 },
}

@article{bulla2008,
  title = { The numerical renormalization group method for quantum
  impurity systems },
  author = { R. Bulla and T. A. Costi and Th. Pruschke },
  journal = { Rev. Mod. Phys. },
  volume = { 80 },
  pages = { 395 },
  year = { 2008 },
  }

@article{zitko2022,
  title = {{Yu-Shiba-Rusinov} states, {BCS-BEC} crossover, and exact
  solution in the flat-band limit},
    volume = {106},
    number = {2},
    journal = {Physical Review B},
    publisher = {American Physical Society (APS)},
    author = {Žitko,  R. and Pavešić,  L.},
    year = {2022},
    month = jul,
    pages = { 024513 }
    }

@article{Larsen2015,
year = {2015},
month = sep,
publisher = {American Physical Society ({APS})},
volume = {115},
number = {12},
author = {T. W. Larsen and
K. D. Petersson and F. Kuemmeth and
T. S Jespersen and P. Krogstrup and
J. Nyg{\aa}rd and C. M. Marcus},
title = {Semiconductor-Nanowire-Based Superconducting Qubit},
journal = {Physical Review Letters}
}

@article{pavesic2024,
  title = {Strong-coupling theory of quantum-dot {Josephson} junctions:
  Role of a residual quasiparticle},
    volume = {109},
    ISSN = {2469-9969},
    url = {http://dx.doi.org/10.1103/PhysRevB.109.125131},
    DOI = {10.1103/physrevb.109.125131},
    number = {12},
    journal = {Physical Review B},
    publisher = {American Physical Society (APS)},
    author = {Pavešić,  Luka and Aguado,  Ramón and
    Žitko,  Rok},
    year = {2024},
    month = mar,
    pages = { 125131 },
    }

@article{pillet2013,
  title = { Tunneling spectroscopy of a single quantum dot coupled to
  a superconductor: From {Kondo} ridge to {Andreev} bound states  },
  journal = { Phys. Rev. B },
  volume = { 88 },
  pages = { 045101 },
  year = { 2013 },
  author = { J.-D. Pillet and P. Joyez and R. \v{Z}itko and M. F.
  Goffman },
  }

@article{lee2017prb,
  title = { Scaling of subgap excitations in a
  superconductor-semiconductor nanowire quantum dot },
  author = { E. J. H. Lee and X. Jiang and R. \v{Z}itko and R. Aguado
  and C. M. Lieber and De Franceschi, S. },
  journal = { Phys. Rev. B },
  volume = { 95 },
  pages = { 180502(R) },
  year = { 2017 },
  }

@article{liu2016,
  title = { Quantum impurities in channel mixing baths },
  journal = { Phys. Rev. B },
  volume = { 93 },
  pages = { 035102 },
  year = { 2016 },
  author = { Jin-Guo Liu, Da Wang, Qiang-Hua Wang },
  }

@article{Zalom2023,
  title = {Rigorous {Wilsonian} renormalization group for impurity models with a spectral gap},
  volume = {108},
  ISSN = {2469-9969},
  url = {http://dx.doi.org/10.1103/PhysRevB.108.195123},
  DOI = {10.1103/physrevb.108.195123},
  number = {19},
  journal = {Physical Review B},
  publisher = {American Physical Society (APS)},
  author = {Zalom,  Peter},
  year = {2023},
  month = nov 
}

@article{Kurilovich2021MicrowaveResponseABS,
  author       = {Kurilovich, Pavel D. and Kurilovich, Vladislav D. and Fatemi, V. and Devoret, Michel H. and Glazman, Leonid I.},
  title        = {Microwave response of an {Andreev} bound state},
  journal      = {Physical Review B},
  volume       = {104},
  number       = {17},
  pages        = {174517},
  year         = {2021},
  doi          = {10.1103/PhysRevB.104.174517},
  url          = {https://doi.org/10.1103/PhysRevB.104.174517}
}

@article{tenKate2025FiniteLengthGeJJ,
  author       = {ten Kate, S. C. and Ohnmacht, D. C. and Coraiola, M. and Antonelli, T. and Paredes, S. and Schupp, F. J. and Hinderling, M. and Bedell, S. W. and Belzig, W. and Cuevas, J. C. and Svetogorov, A. E. and Nichele, F. and Sabonis, D.},
  title        = {Finite Length Effects and {Coulomb} Interaction in {Ge} Quantum Well-Based {Josephson} Junctions Probed with Microwave Spectroscopy},
  journal      = {Physical Review Applied},
  volume       = {24},
  number       = {6},
  pages        = {064005},
  year         = {2025},
  doi          = {10.1103/hb1h-8jn9},
  url          = {https://doi.org/10.1103/hb1h-8jn9}
}

@article{Bordin2025ImpactABSLeads,
  title         = {Impact of {Andreev} Bound States within the Leads of a Quantum Dot {Josephson} Junction},
  author        = {Bordin, Alberto and Evertsz, Florian J. Bennebroek and Steffensen, Gorm O. and Dvir, Tom and Mazur, Grzegorz P. and van Driel, David and van Loo, Nick and Wolff, Jan Cornelis and Bakkers, Erik P. A. M. and Yeyati, Alfredo Levy and Kouwenhoven, Leo P.},
  journal       = {Physical Review X},
  volume        = {15},
  number        = {1},
  pages         = {011046},
  year          = {2025},
  publisher     = {American Physical Society},
  doi           = {10.1103/PhysRevX.15.011046},
  url           = {https://doi.org/10.1103/PhysRevX.15.011046}
}

@article{Pavesic2023ImpurityKnightShift,
  title   = {Impurity {Knight} shift in quantum dot {Josephson} junctions},
  author  = {Pave{\v{s}}i{\'c}, Luka and Pita Vidal, Marta and Bargerbos, Arno and {\v{Z}}itko, Rok},
  journal = {SciPost Phys.},
  volume  = {15},
  pages   = {070},
  year    = {2023},
  publisher = {SciPost},
  doi     = {10.21468/SciPostPhys.15.2.070},
  url     = {https://scipost.org/10.21468/SciPostPhys.15.2.070}
}

@article{zitko2009,
  title = { Energy resolution and discretization artefacts in the
  numerical renormalization group },
  author = { Rok \v{Z}itko and Thomas Pruschke },
  year = { 2009 },
  journal = { Phys. Rev. B },
  volume = { 79 },
  pages = { 085106 },
  }

@article{zitko2010,
 title = { Josephson current in strongly correlated double quantum
 dots  },
 volume = { 105 },
 pages = { 116803 },
 year = { 2010 },
 author = { R. \v{Z}itko and M. Lee and R. Lopez and R. Aguado
 and M.-S. Choi },
 journal = { Phys. Rev. Lett. },
 }

@article{coulomb2,
  author = { J. C. Estrada Salda{\~n}a and A. Vekris and L. Pave\v{s}i\'{c} and
  P. Krogstrup and R. \v{Z}itko and K. Grove-Rasmussen and J. Nyg{\aa}rd },
  title = { Excitations in a superconducting {Coulombic} energy gap },
  year = { 2022 },
  journal = { Nat. Commun. },
  volume = { 13 },
  pages = { 2243 },
  }

@article{Karrasch2009Supercurrent,
  author    = {Karrasch, C. and Meden, V.},
  title     = {Supercurrent and multiple singlet-doublet phase transitions of a quantum dot {Josephson} junction inside an {Aharonov--Bohm} ring},
  journal   = {Phys. Rev. B},
  volume    = {79},
  number    = {4},
  pages     = {045110},
  year      = {2009},
  month     = jan,
  publisher = {American Physical Society},
  doi       = {10.1103/PhysRevB.79.045110},
  url       = {https://link.aps.org/doi/10.1103/PhysRevB.79.045110}
}

@article{Ivanov1999,
  title = {Two-level {Hamiltonian} of a superconducting quantum point contact},
  volume = {59},
  ISSN = {1095-3795},
  url = {http://dx.doi.org/10.1103/PhysRevB.59.8444},
  DOI = {10.1103/physrevb.59.8444},
  number = {13},
  journal = {Physical Review B},
  publisher = {American Physical Society (APS)},
  author = {Ivanov,  D. A. and Feigel’man,  M. V.},
  year = {1999},
  month = apr,
  pages = {8444–8446}
}

@article{Desposito2001,
  title = {Controlled dephasing of {Andreev} states in superconducting quantum point contacts},
  volume = {64},
  ISSN = {1095-3795},
  url = {http://dx.doi.org/10.1103/PhysRevB.64.140511},
  DOI = {10.1103/physrevb.64.140511},
  number = {14},
  journal = {Physical Review B},
  publisher = {American Physical Society (APS)},
  author = {Despósito,  M. A. and Levy Yeyati,  A.},
  year = {2001},
  month = sep 
}

@article{Lantz2002,
  title = {Flux qubit with a quantum point contact},
  volume = {368},
  ISSN = {0921-4534},
  url = {http://dx.doi.org/10.1016/S0921-4534(01)01188-1},
  DOI = {10.1016/s0921-4534(01)01188-1},
  number = {1–4},
  journal = {Physica C: Superconductivity},
  publisher = {Elsevier BV},
  author = {Lantz,  J. and Shumeiko,  V.S. and Bratus,  E. and Wendin,  G.},
  year = {2002},
  month = mar,
  pages = {315–319}
}

@article{Zazunov2005,
  title = {Dynamics and phonon-induced decoherence of {Andreev} level qubit},
  volume = {71},
  ISSN = {1550-235X},
  url = {http://dx.doi.org/10.1103/PhysRevB.71.214505},
  DOI = {10.1103/physrevb.71.214505},
  number = {21},
  journal = {Physical Review B},
  publisher = {American Physical Society (APS)},
  author = {Zazunov,  A. and Shumeiko,  V. S. and Wendin,  G. and Bratus’,  E. N.},
  year = {2005},
  month = jun 
}

@article{Bretheau2014,
  title = {Theory of microwave spectroscopy of {Andreev} bound states with a {Josephson} junction},
  volume = {90},
  ISSN = {1550-235X},
  url = {http://dx.doi.org/10.1103/PhysRevB.90.134506},
  DOI = {10.1103/physrevb.90.134506},
  number = {13},
  journal = {Physical Review B},
  publisher = {American Physical Society (APS)},
  author = {Bretheau,  L. and Girit,  C. \"{O}. and Houzet,  M. and Pothier,  H. and Esteve,  D. and Urbina,  C.},
  year = {2014},
  month = oct 
}

@article{Wesdorp2023,
  title = {Dynamical Polarization of the Fermion Parity in a Nanowire {Josephson} Junction},
  volume = {131},
  ISSN = {1079-7114},
  url = {http://dx.doi.org/10.1103/PhysRevLett.131.117001},
  DOI = {10.1103/physrevlett.131.117001},
  number = {11},
  journal = {Physical Review Letters},
  publisher = {American Physical Society (APS)},
  author = {Wesdorp,  J. J. and Gr\"{u}nhaupt,  L. and Vaartjes,  A. and Pita-Vidal,  M. and Bargerbos,  A. and Splitthoff,  L. J. and Krogstrup,  P. and van Heck,  B. and de Lange,  G.},
  year = {2023},
  month = sep 
}

@article{Kurilovich2024,
  title = {On-demand population of {Andreev} levels by their ionization in the presence of {Coulomb} blockade},
  volume = {110},
  ISSN = {2469-9969},
  url = {http://dx.doi.org/10.1103/PhysRevB.110.184508},
  DOI = {10.1103/physrevb.110.184508},
  number = {18},
  journal = {Physical Review B},
  publisher = {American Physical Society (APS)},
  author = {Kurilovich,  Pavel D. and Kurilovich,  Vladislav D. and Svetogorov,  Aleksandr E. and Belzig,  Wolfgang and Devoret,  Michel H. and Glazman,  Leonid I.},
  year = {2024},
  month = nov 
}

@article{Ackermann2023,
  title = {Dynamical parity selection in superconducting weak links},
  volume = {107},
  ISSN = {2469-9969},
  url = {http://dx.doi.org/10.1103/PhysRevB.107.214515},
  DOI = {10.1103/physrevb.107.214515},
  number = {21},
  journal = {Physical Review B},
  publisher = {American Physical Society (APS)},
  author = {Ackermann,  Nico and Zazunov,  Alex and Park,  Sunghun and Egger,  Reinhold and Yeyati,  Alfredo Levy},
  year = {2023},
  month = jun 
}

@article{Petersson2010,
  title = {Charge and Spin State Readout of a Double Quantum Dot Coupled to a Resonator},
  volume = {10},
  ISSN = {1530-6992},
  url = {http://dx.doi.org/10.1021/nl100663w},
  DOI = {10.1021/nl100663w},
  number = {8},
  journal = {Nano Letters},
  publisher = {American Chemical Society (ACS)},
  author = {Petersson,  K. D. and Smith,  C. G. and Anderson,  D. and Atkinson,  P. and Jones,  G. A. C. and Ritchie,  D. A.},
  year = {2010},
  month = jul,
  pages = {2789–2793}
}

@article{aghaee2025,
  title = {Interferometric single-shot parity measurement in {InAs–Al} hybrid devices},
  volume = {638},
  ISSN = {1476-4687},
  url = {http://dx.doi.org/10.1038/s41586-024-08445-2},
  DOI = {10.1038/s41586-024-08445-2},
  number = {8051},
  journal = {Nature},
  publisher = {Springer Science and Business Media LLC},
  author = {Aghaee,  Morteza and Alcaraz Ramirez,  Alejandro and Alam,  Zulfi and Ali,  Rizwan and Andrzejczuk,  Mariusz and Antipov,  Andrey and Astafev,  Mikhail and Barzegar,  Amin and Bauer,  Bela and Becker,  Jonathan and Bhaskar,  Umesh Kumar and Bocharov,  Alex and Boddapati,  Srini and Bohn,  David and Bommer,  Jouri and Bourdet,  Leo and Bousquet,  Arnaud and Boutin,  Samuel and Casparis,  Lucas and Chapman,  Benjamin J. and Chatoor,  Sohail and Christensen,  Anna Wulff and Chua,  Cassandra and Codd,  Patrick and Cole,  William and Cooper,  Paul and Corsetti,  Fabiano and Cui,  Ajuan and Dalpasso,  Paolo and Dehollain,  Juan Pablo and de Lange,  Gijs and de Moor,  Michiel and Ekefj\"{a}rd,  Andreas and El Dandachi,  Tareq and Estrada Saldaña,  Juan Carlos and Fallahi,  Saeed and Galletti,  Luca and Gardner,  Geoff and Govender,  Deshan and Griggio,  Flavio and Grigoryan,  Ruben and Grijalva,  Sebastian and Gronin,  Sergei and Gukelberger,  Jan and Hamdast,  Marzie and Hamze,  Firas and Hansen,  Esben Bork and Heedt,  Sebastian and Heidarnia,  Zahra and Herranz Zamorano,  Jesús and Ho,  Samantha and Holgaard,  Laurens and Hornibrook,  John and Indrapiromkul,  Jinnapat and Ingerslev,  Henrik and Ivancevic,  Lovro and Jensen,  Thomas and Jhoja,  Jaspreet and Jones,  Jeffrey and Kalashnikov,  Konstantin V. and Kallaher,  Ray and Kalra,  Rachpon and Karimi,  Farhad and Karzig,  Torsten and King,  Evelyn and Kloster,  Maren Elisabeth and Knapp,  Christina and Kocon,  Dariusz and Koski,  Jonne V. and Kostamo,  Pasi and Kumar,  Mahesh and Laeven,  Tom and Larsen,  Thorvald and Lee,  Jason and Lee,  Kyunghoon and Leum,  Grant and Li,  Kongyi and Lindemann,  Tyler and Looij,  Matthew and Love,  Julie and Lucas,  Marijn and Lutchyn,  Roman and Madsen,  Morten Hannibal and Madulid,  Nash and Malmros,  Albert and Manfra,  Michael and Mantri,  Devashish and Markussen,  Signe Brynold and Martinez,  Esteban and Mattila,  Marco and McNeil,  Robert and Mei,  Antonio B. and Mishmash,  Ryan V. and Mohandas,  Gopakumar and Mollgaard,  Christian and Morgan,  Trevor and Moussa,  George and Nayak,  Chetan and Nielsen,  Jens Hedegaard and Nielsen,  Jens Munk and Nielsen,  William Hvidtfelt Padkar and Nijholt,  Bas and Nystrom,  Mike and O’Farrell,  Eoin and Ohki,  Thomas and Otani,  Keita and Paquelet W\"{u}tz,  Brian and Pauka,  Sebastian and Petersson,  Karl and Petit,  Luca and Pikulin,  Dima and Prawiroatmodjo,  Guen and Preiss,  Frank and Puchol Morejon,  Eduardo and Rajpalke,  Mohana and Ranta,  Craig and Rasmussen,  Katrine and Razmadze,  David and Reentila,  Outi and Reilly,  David J. and Ren,  Yuan and Reneris,  Ken and Rouse,  Richard and Sadovskyy,  Ivan and Sainiemi,  Lauri and Sanlorenzo,  Irene and Schmidgall,  Emma and Sfiligoj,  Cristina and Shah,  Mustafeez Bashir and Simoes,  Kevin and Singh,  Shilpi and Sinha,  Sarat and Soerensen,  Thomas and Sohr,  Patrick and Stankevic,  Tomas and Stek,  Lieuwe and Stuppard,  Eric and Suominen,  Henri and Suter,  Judith and Teicher,  Sam and Thiyagarajah,  Nivetha and Tholapi,  Raj and Thomas,  Mason and Toomey,  Emily and Tracy,  Josh and Turley,  Michelle and Upadhyay,  Shivendra and Urban,  Ivan and Van Hoogdalem,  Kevin and Van Woerkom,  David J. and Viazmitinov,  Dmitrii V. and Vogel,  Dominik and Watson,  John and Webster,  Alex and Weston,  Joseph and Winkler,  Georg W. and Xu,  Di and Yang,  Chung Kai and Yucelen,  Emrah and Zeisel,  Roland and Zheng,  Guoji and Zilke,  Justin},
  year = {2025},
  month = feb,
  pages = {651–655}
}

@article{Colless2013,
  title = {Dispersive Readout of a Few-Electron Double Quantum Dot with Fast rf Gate Sensors},
  volume = {110},
  ISSN = {1079-7114},
  url = {http://dx.doi.org/10.1103/PhysRevLett.110.046805},
  DOI = {10.1103/physrevlett.110.046805},
  number = {4},
  journal = {Physical Review Letters},
  publisher = {American Physical Society (APS)},
  author = {Colless,  J. I. and Mahoney,  A. C. and Hornibrook,  J. M. and Doherty,  A. C. and Lu,  H. and Gossard,  A. C. and Reilly,  D. J.},
  year = {2013},
  month = jan 
}

@article{GonzalezZalba2015,
  title = {Probing the limits of gate-based charge sensing},
  volume = {6},
  ISSN = {2041-1723},
  url = {http://dx.doi.org/10.1038/ncomms7084},
  DOI = {10.1038/ncomms7084},
  number = {1},
  journal = {Nature Communications},
  publisher = {Springer Science and Business Media LLC},
  author = {Gonzalez-Zalba,  M. F. and Barraud,  S. and Ferguson,  A. J. and Betz,  A. C.},
  year = {2015},
  month = jan 
}

@article{Crippa2019,
  title = {Gate-reflectometry dispersive readout and coherent control of a spin qubit in silicon},
  volume = {10},
  ISSN = {2041-1723},
  url = {http://dx.doi.org/10.1038/s41467-019-10848-z},
  DOI = {10.1038/s41467-019-10848-z},
  number = {1},
  journal = {Nature Communications},
  publisher = {Springer Science and Business Media LLC},
  author = {Crippa,  A. and Ezzouch,  R. and Aprá,  A. and Amisse,  A. and Laviéville,  R. and Hutin,  L. and Bertrand,  B. and Vinet,  M. and Urdampilleta,  M. and Meunier,  T. and Sanquer,  M. and Jehl,  X. and Maurand,  R. and De Franceschi,  S.},
  year = {2019},
  month = jul 
}

@article{Blais2004,
  title = {Cavity quantum electrodynamics for superconducting electrical circuits: An architecture for quantum computation},
  volume = {69},
  ISSN = {1094-1622},
  url = {http://dx.doi.org/10.1103/PhysRevA.69.062320},
  DOI = {10.1103/physreva.69.062320},
  number = {6},
  journal = {Physical Review A},
  publisher = {American Physical Society (APS)},
  author = {Blais,  Alexandre and Huang,  Ren-Shou and Wallraff,  Andreas and Girvin,  S. M. and Schoelkopf,  R. J.},
  year = {2004},
  month = jun 
}

@article{Schuster2005,
  title = {ac {Stark} Shift and Dephasing of a Superconducting Qubit Strongly Coupled to a Cavity Field},
  volume = {94},
  ISSN = {1079-7114},
  url = {http://dx.doi.org/10.1103/PhysRevLett.94.123602},
  DOI = {10.1103/physrevlett.94.123602},
  number = {12},
  journal = {Physical Review Letters},
  publisher = {American Physical Society (APS)},
  author = {Schuster,  D. I. and Wallraff,  A. and Blais,  A. and Frunzio,  L. and Huang,  R.-S. and Majer,  J. and Girvin,  S. M. and Schoelkopf,  R. J.},
  year = {2005},
  month = mar 
}

@article{Hinderling2023,
  title = {Flip-Chip-Based Microwave Spectroscopy of {Andreev} Bound States in a Planar {Josephson} Junction},
  volume = {19},
  ISSN = {2331-7019},
  url = {http://dx.doi.org/10.1103/PhysRevApplied.19.054026},
  DOI = {10.1103/physrevapplied.19.054026},
  number = {5},
  journal = {Physical Review Applied},
  publisher = {American Physical Society (APS)},
  author = {Hinderling,  M. and Sabonis,  D. and Paredes,  S. and Haxell,  D.Z. and Coraiola,  M. and ten Kate,  S.C. and Cheah,  E. and Krizek,  F. and Schott,  R. and Wegscheider,  W. and Nichele,  F.},
  year = {2023},
  month = may 
}

@article{Vigneau2023,
  title = {Probing quantum devices with radio-frequency reflectometry},
  volume = {10},
  ISSN = {1931-9401},
  url = {http://dx.doi.org/10.1063/5.0088229},
  DOI = {10.1063/5.0088229},
  number = {2},
  journal = {Applied Physics Reviews},
  publisher = {AIP Publishing},
  author = {Vigneau,  Florian and Fedele,  Federico and Chatterjee,  Anasua and Reilly,  David and Kuemmeth,  Ferdinand and Gonzalez-Zalba,  M. Fernando and Laird,  Edward and Ares,  Natalia},
  year = {2023},
  month = feb 
}

@article{vanVeen2019,
  title = {Revealing charge-tunneling processes between a quantum dot and a superconducting island through gate sensing},
  volume = {100},
  ISSN = {2469-9969},
  url = {http://dx.doi.org/10.1103/PhysRevB.100.174508},
  DOI = {10.1103/physrevb.100.174508},
  number = {17},
  journal = {Physical Review B},
  publisher = {American Physical Society (APS)},
  author = {van Veen,  Jasper and de Jong,  Damaz and Han,  Lin and Prosko,  Christian and Krogstrup,  Peter and Watson,  John D. and Kouwenhoven,  Leo P. and Pfaff,  Wolfgang},
  year = {2019},
  month = nov 
}

@article{Razmadze2019,
  title = {Radio-Frequency Methods for {Majorana}-Based Quantum Devices: Fast Charge Sensing and Phase-Diagram Mapping},
  volume = {11},
  ISSN = {2331-7019},
  url = {http://dx.doi.org/10.1103/PhysRevApplied.11.064011},
  DOI = {10.1103/physrevapplied.11.064011},
  number = {6},
  journal = {Physical Review Applied},
  publisher = {American Physical Society (APS)},
  author = {Razmadze,  Davydas and Sabonis,  Deividas and Malinowski,  Filip K. and Ménard,  Gerbold C. and Pauka,  Sebastian and Nguyen,  Hung and van Zanten,  David M.T. and O'Farrell,  Eoin C.T. and Suter,  Judith and Krogstrup,  Peter and Kuemmeth,  Ferdinand and Marcus,  Charles M.},
  year = {2019},
  month = jun 
}

@article{Menard2019,
  title = {Suppressing quasiparticle poisoning with a voltage-controlled filter},
  volume = {100},
  ISSN = {2469-9969},
  url = {http://dx.doi.org/10.1103/PhysRevB.100.165307},
  DOI = {10.1103/physrevb.100.165307},
  number = {16},
  journal = {Physical Review B},
  publisher = {American Physical Society (APS)},
  author = {Ménard,  Gerbold C. and Malinowski,  Filip K. and Puglia,  Denise and Pikulin,  Dmitry I. and Karzig,  Torsten and Bauer,  Bela and Krogstrup,  Peter and Marcus,  Charles M.},
  year = {2019},
  month = oct 
}

@article{Lu2025,
  title = {Andreev spin relaxation time in a shadow-evaporated {InAs} weak link},
  volume = {24},
  ISSN = {2331-7019},
  url = {http://dx.doi.org/10.1103/v3lq-t5z8},
  DOI = {10.1103/v3lq-t5z8},
  number = {2},
  journal = {Physical Review Applied},
  publisher = {American Physical Society (APS)},
  author = {Lu,  Haoran and Bofill,  David F. and Sun,  Zhenhai and Kanne,  Thomas and Nygård,  Jesper and Kjaergaard,  Morten and Fatemi,  Valla},
  year = {2025},
  month = aug 
}

@misc{pitavidal2025novel,
      title={Novel qubits in hybrid semiconductor-superconductor nanostructures}, 
      author={Marta Pita-Vidal and Rubén Seoane Souto and Srijit Goswami and Christian Kraglund Andersen and Georgios Katsaros and Javad Shabani and Ramón Aguado},
      year={2025},
      eprint={2512.23336},
      archivePrefix={arXiv},
      primaryClass={cond-mat.mes-hall},
      url={https://arxiv.org/abs/2512.23336}, 
}

@article{Rashba2003,
  title = {Orbital Mechanisms of Electron-Spin Manipulation by an Electric Field},
  volume = {91},
  ISSN = {1079-7114},
  url = {http://dx.doi.org/10.1103/PhysRevLett.91.126405},
  DOI = {10.1103/physrevlett.91.126405},
  number = {12},
  journal = {Physical Review Letters},
  publisher = {American Physical Society (APS)},
  author = {Rashba,  E. I. and Efros,  Al. L.},
  year = {2003},
  month = sep 
}

@article{Flindt2006,
  title = {Spin-Orbit Mediated Control of Spin Qubits},
  volume = {97},
  ISSN = {1079-7114},
  url = {http://dx.doi.org/10.1103/PhysRevLett.97.240501},
  DOI = {10.1103/physrevlett.97.240501},
  number = {24},
  journal = {Physical Review Letters},
  publisher = {American Physical Society (APS)},
  author = {Flindt,  Christian and Sørensen,  Anders S. and Flensberg,  Karsten},
  year = {2006},
  month = dec 
}

@article{Golovach2006,
  title = {Electric-dipole-induced spin resonance in quantum dots},
  volume = {74},
  ISSN = {1550-235X},
  url = {http://dx.doi.org/10.1103/PhysRevB.74.165319},
  DOI = {10.1103/physrevb.74.165319},
  number = {16},
  journal = {Physical Review B},
  publisher = {American Physical Society (APS)},
  author = {Golovach,  Vitaly N. and Borhani,  Massoud and Loss,  Daniel},
  year = {2006},
  month = oct 
}

@article{Nowack2007,
  title = {Coherent Control of a Single Electron Spin with Electric Fields},
  volume = {318},
  ISSN = {1095-9203},
  url = {http://dx.doi.org/10.1126/science.1148092},
  DOI = {10.1126/science.1148092},
  number = {5855},
  journal = {Science},
  publisher = {American Association for the Advancement of Science (AAAS)},
  author = {Nowack,  K. C. and Koppens,  F. H. L. and Nazarov,  Yu. V. and Vandersypen,  L. M. K.},
  year = {2007},
  month = nov,
  pages = {1430–1433}
}

@misc{malinowski2022,
      title={Quantum capacitance of a superconducting subgap state in an electrostatically floating dot-island}, 
      author={Filip K. Malinowski and R. K. Rupesh and Luka Pavešić and Zoltán Guba and Damaz de Jong and Lin Han and Christian G. Prosko and Michael Chan and Yu Liu and Peter Krogstrup and András Pályi and Rok Žitko and Jonne V. Koski},
      year={2022},
      eprint={2210.01519},
      archivePrefix={arXiv},
      primaryClass={cond-mat.mes-hall},
      url={https://arxiv.org/abs/2210.01519}, 
}

\end{document}